\documentclass[11pt]{iopart}
\usepackage{iopams}  
\usepackage{soul,color,xcolor}
\usepackage{bm}
\usepackage{amsthm}
\usepackage[ruled]{algorithm2e}
\usepackage{algorithmic}
\usepackage{enumitem}
\usepackage{multirow}
\usepackage{graphicx}
\usepackage{subcaption}
\usepackage{booktabs}
\usepackage{array}
\expandafter\let\csname equation*\endcsname=\relax
\expandafter\let\csname endequation*\endcsname=\relax
\usepackage{amsthm}
\usepackage{amsmath,amssymb}
\usepackage{url}
\usepackage{ulem}

\usepackage[letterpaper,centering, top=1.5in, bottom=1.5in, left=1.2in, right=1.2in]{geometry}

\newtheorem{theorem}{Theorem}[section]
\newtheorem{lemma}[theorem]{Lemma}
\newtheorem{proposition}[theorem]{Proposition}

\usepackage{enumitem}
\setlist[enumerate]{leftmargin=.5in}
\setlist[itemize]{leftmargin=.5in}

\begin{document}
\title[]{Fused $L_{1/2}$ prior for large scale linear inverse problem with Gibbs bouncy particle sampler}

\author{Xiongwen Ke$^1$, Yanan Fan$^2$, Qingping Zhou$^{1,*}$}

\address{$^1$ School of Mathematics and Statistics, Central South University, 932 South Lushan Rd, Hunan 410083, China\\
$^2$ Data61, CSIRO Sydney 2015, Australia}

\ead{qpzhou@csu.edu.cn}
\vspace{10pt}

\begin{abstract}
In this paper, we study Bayesian approach for solving large scale linear inverse problems arising in various scientific and engineering fields. We propose a fused $L_{1/2}$ prior with edge-preserving and sparsity-promoting properties and show that it can be formulated as a Gaussian mixture Markov random field. Since the density function of this family of prior is neither log-concave nor Lipschitz, gradient-based Markov chain Monte Carlo methods can not be applied to sample the posterior. Thus, we present a Gibbs sampler in which all the conditional posteriors involved have closed form expressions. The Gibbs sampler works well for small size problems but it is computationally intractable for large scale problems due to the need for sample high dimensional Gaussian distribution. To reduce the computation burden, we construct a Gibbs bouncy particle sampler (Gibbs-BPS) based on a piecewise deterministic Markov process. This new sampler combines elements of Gibbs sampler with bouncy particle sampler and its computation complexity is an order of magnitude smaller. We show that the new sampler converges to the target distribution. With computed tomography examples, we demonstrate that the proposed method shows competitive performance with existing popular Bayesian methods and is highly efficient in large scale problems.
\end{abstract}
 
\vspace{1em}
\noindent{\it Keywords\/}:
Bayesian inverse problem, bouncy particle sampler, global-local shrinkage prior, Gaussian mixture Markov random fields, Gibbs sampler

\maketitle

\newpage
\section{Introduction}
\label{sec:intro}
 Inverse problems are encountered in many fields, such as medical images, radar, geophysics and oceanography, where the unknown must be estimated from noisy, incomplete, and indirect measurements. Although inverse problems can be solved using optimization approaches, Bayesian approaches are particularly attractive, as it offers a coherent framework for uncertainty quantification. To carry out Bayesian inference, it is often required to perform Markov chain Monte Carlo (MCMC) simulations. However, obtaining accurate, efficient and reliable Bayesian solutions becomes significantly more challenging when dealing with ultra high dimensional problems. In this situation, the computational demands 
 become prohibitively expensive and time consuming. Consequently, the design of an efficient sampler under specific prior distribution becomes critically important, which is the central task of large scale inverse problems~\cite{lucka2012fast,zhou2020bayesian,durmus2022proximal}.

The prior plays a critical role in inverse problems, as this type of problems are typically ill-posed, leading to noisy, unstable estimates. Regularisation techniques have proven to be useful in such cases \cite{bardsley2018computational}. In the Bayesian setting, regularisation can be deployed via appropriate prior setting.
Popular priors used in Bayesian inverse problem includes $\mathrm{L}_1$-type prior~\cite{lucka2012fast},  total variation prior~\cite{lassas2004can}, Besov space priors~\cite{lassas2009discretization,dashti2012besov} and Markov random field (MRF) priors~\cite{bardsley2018computational} (Laplace MRF~\cite{bardsley2012laplace}, Cauchy MRF~\cite{suuronen2022cauchy}).
The global-local shrinkage family of priors, which has heavy tail and put sufficient probability mass around $0$, has become popular in high dimensional statistic due to its superior theoretical properties~\cite{carvalho2010horseshoe,song2023nearly} and empirical performance \cite{johndrow2020scalable}. 
More recently, the horseshoe prior~\cite{carvalho2010horseshoe}, which is one of the most popular global-local shrinkage priors, for sparse Bayesian modeling, has been used by~\cite{uribe2023horseshoe} for edge-preserving linear inverse problems. 

While the aforementioned priors provide useful regularization, they often lead to complex posterior distributions that are difficult to compute.
To sample these complex posterior effectively, many MCMC algorithms have been developed, including preconditioned Crank-Nicolson(pCN)~\cite{cotter2013mcmc}, and  Metropolis Hastings within Gibbs sampler~\cite{lucka2016fast}. Recently, a very interesting line of research is gradient based approximate MCMC approaches, which have been widely used in probabilistic machine learning~\cite{welling2011bayesian,chen2014stochastic,ma2015complete}. These approaches are scalable to high dimension data, and thus work for high resolution images. However, for the Bayesian inverse problems, the prior is often non-smooth. To remedy this issue, the proximal Langevin  dynamic has been proposed~\cite{durmus2018efficient,durmus2022proximal}. A central idea in this work is to replace the non-smooth prior with a carefully designed smooth approximation. The resulting approximated prior function is the Moureau-Yosida envelope (MYE) prior. Since this seminal work, numerous extensions and applications have been made which includes deriving priors schemes, in particular the plug and play prior~\cite{laumont2022bayesian},
and deriving theoretic analyses~\cite{brosse2017sampling}. 
However, a main limitation of MYE is that it is only well defined for priors with log-concave density, which restricts the application of this method.
 
In this paper, we focus on an alternative way to improve sampling efficiency, and proposed a Gibbs bouncy particle sampler (Gibbs-BPS) based on a piecewise deterministic Markov
process (PDMP). Davis proposed PDMP~\cite{davis1984piecewise} several decades ago and 
the MCMC sampler based on PDMP was first introduced in physics~\cite{peters2012rejection} and more recently extensively studied in statistics~\cite{bouchard2018bouncy,bierkens2019zig}. Examples of samplers based on PDMP include the bouncy particle sampler  (BPS)~\cite{bouchard2018bouncy} and the Zig-Zag sampler~\cite{bierkens2019zig}. 
We first introduce a non-Gaussian random field prior that belongs to the global-local shrinkage family in Bayesian sparse learning. More specifically, our basic building block is the $L_{1/2}$ prior~\cite{ke2021bayesian}, which is applied to each pixel of the image and its increment. We call this new prior fused $L_{1/2}$ prior. We show that the fused $L_{1/2}$ prior can be represented as a Gaussian mixture Markov random field prior and results in a simple Gibbs sampler in the linear inverse problem. We then propose the Gibbs bouncy particle sampler (Gibbs-BPS), allowing parameters to be updated in blocks, with a bouncy particle sampler~\cite{bouchard2018bouncy} applied to the pixels of the image, which are high dimensional multivariate Gaussian distribution in our case, and Gibbs style update applied to the global and local shrinkage parameters. Unlike most MCMC algorithms, such as the aforementioned Metropolis Hastings within Gibbs sampler, which are based on reversible discrete time Markov chains, the Gibbs-BPS is based non-reversible continuous time Markov chains. 
We show that this new sampler can converge to the target distribution without bias.

Samplers based on PDMP seem particularly well suited to Bayesian analysis in big data settings, as they allow access to only a small subset of data points at each iteration and are still guaranteed to target the true posterior distribution~\cite{fearnhead2018piecewise}. Although theoretically well justified, these samplers have not yet been widely used in Bayesian statistics. A major reason is the fact that sampling the event time between jumps from a non-homogeneous Poisson process is non-trivial for many of the applications.
However, we will show that the bouncy particle sampler is particularly fast for sampling the high dimensional Gaussian distribution in Bayesian linear inverse problems. In this case, sampling from the non-homogeneous Poisson process can be done by inverse the cumulative distribution function directly (the inverse transform sampling), with matrix multiplication as the only operation.  This nice property leads the new sampler to have the same computational complexity as the first order optimization approach.

The key novelty and contributions of this paper can be summarized as follows:
\begin{enumerate}[label=(\roman*)]
    \item We propose the fused $L_{\frac12}$ prior and formulate it as a Gaussian mixture Markov random field, which allows us to construct the Gibbs sampler with closed form expression for all the conditional posteriors involved.
    \item By integrating the Gibbs sampler and bouncy particle sampler (BPS), we demonstrate that the new Gibbs-BPS algorithm converges to the target distribution without bias.
    \item By circumventing the matrix inversion using BPS when sampling from the multivariate Gaussian distribution, we show that Gibbs-BPS has low computation complexity and achieves fast speed-up in dealing with high-resolution image.
    \item We verify the scalability of our approach on computed tomography (CT) imaging problems ranging from small size ($64\times64$) to very large size ($256 \times 256$), and show that our approach is competitive with existing popular alternative Bayesian methods and it is highly efficient.
\end{enumerate}
The paper is structured as follows. In Section \ref{sec:background}, we provide the background for Bayesian linear inverse problems. In Section \ref{sec:prior_setting}, we present the Gaussian mixture Markov random field representation for the fused $L_{1/2}$ prior. In Section \ref{sec:Gibbs-BPS}, we develop the Gibbs bouncy particle sampler (Gibbs-BPS) for efficient posterior sampling. In Section \ref{sec:numerical}, we illustrate our approach with computed tomography image problems. Finally, in Section \ref{sec:conclusions}, we conclude with a discussion.

\section{Bayesian linear inverse problem}
\label{sec:background}


Consider the linear observation model with independent Gaussian noise and known variance $\sigma_{\mathrm{obs}}^{2}$,
\begin{equation}\label{eq:linear}
\bm{y}=\bm{A}\bm{x}+\bm{e}, \quad \bm{e} \sim \mathcal{N}(0, \sigma_{\mathrm{obs}}^{2}I_{m}),
\end{equation}
where $\bm{A} \in \mathrm{R}^{m \times n}$ with $m<n$, is a known ill-conditioned matrix describing the forward model, $\bm{x} \in \mathrm{R}^n$ is the unknown (image) of interest and $\bm{y}$ is $m \times 1$ observation. For two dimensional problems, we have $d \times d$ matrix $\bm{X}$ with $\bm{x}=\operatorname{Vec}(\bm{X})$ and $n=d^{2}$.
Thus, the likelihood function of $\bm{y}$ given $\bm{x}$ takes the form
\begin{equation}\label{eq:likelihood}
\pi(\bm{y} \mid \bm{x}) \propto \exp \Big(-\frac{1}{2 \sigma_{\text {obs }}^2}\|\bm{y}-\bm{A} \bm{x}\|_2^2\Big).
\end{equation}
To tackle the
ill-posed problem, the Bayesian approach introduces the prior $\pi(\bm{x})$ to regularize the parameter space $\bm{x}$ and combines the prior and likelihood to form the posterior via Bayes rule, 
\begin{equation}\label{eq:posterior}
\pi(\bm{x} \mid \bm{y}) \propto \pi(\bm{y} \mid \bm{x})\pi(\bm{x}).
\end{equation}
This paper used the posterior mean as an estimator.

\section{Prior setting}
\label{sec:prior_setting}
If the true image $\bm{X}$ is sparse and has sharp edges, we can construct a prior by placing the $L_{1/2}$ prior~\cite{ke2021bayesian}, a subfamily of exponential power prior~\cite{zhu2009properties}, on both each pixel and its increment. We show that this family of prior has closed form Gaussian mixture representations, which is convenient for the development of MCMC schemes. Our discussion will start with a general class and then move to the special case, which is of our interest.

\subsection{Fused bridge prior}
We denote $x_{ij}$ as the pixel of image $\bm{X}$ at row $i$ column $j$ and define the horizontal and vertical increments as
$\Delta_{i,j}^{h}=x_{i,j}-x_{i,j-1}$ ($i=1,...,d$, $j=2,...,d$) and $\Delta_{i,j}^{v}=x_{i,j}-x_{i-1,j}$ ($i=2,...,d$, $j=1,...,d$), respectively. We consider the following non-Gaussian Markov random field prior,
\begin{equation}\label{eq:prior}
\resizebox{0.90\hsize}{!}{$\pi(\bm{x} \mid \lambda_{1},\lambda_{2},\lambda_{3})  \propto \exp\underbrace{\Big(-\lambda_{1}\sum_{i,j}|x_{ij}|^{\alpha_{1}}\Big.}_{\text{sparsity-promoting}}\underbrace{\Big.-\lambda_{2}\sum_{i,j}|\Delta_{i,j}^{h}|^{\alpha_{2}}-\lambda_{3}\sum_{i,j}|\Delta_{i,j}^{v}|^{\alpha_{2}}\Big)}_{\text{edge-preserving}}$},
\end{equation}
where $0< \alpha_{1} \leq 1$ and $0< \alpha_{2} \leq 1$. This construction is motivated by the bridge prior~\cite{polson2014bayesian,mallick2018bayesian} for sparse regression. When $\alpha_{1}=\alpha_{2}=1$, this prior is called fused LASSO prior in the statistical literature~\cite{casella2010penalized}, with the first term analogous to the LASSO prior for sparsity promotion and the last two terms analogous to the total variation prior for edge preservation. We call this family of priors the fused bridge prior.

\subsection{etermining $\boldsymbol{\lambda}$: full Bayesian vs empirical Bayes}
For the hyper parameter $\bm{\lambda}=(\lambda_{1},\lambda_{2},\lambda_{3})$, we can use full Bayesian approach by assigning the hyper-prior to $\boldsymbol{\lambda}$:
$$
\lambda_{1} \sim \mathrm{Gamma}(a_{1},b_{1}), \quad \lambda_{2} \sim \mathrm{Gamma}(a_{2},b_{2}), \quad \lambda_{3} \sim \mathrm{Gamma}(a_{3},b_{3}).
$$
It is also possible to combine $\lambda_{2}$ and $\lambda_{3}$ into one hyper parameter. In practice, we find no difference between these two settings empirically.

Apart from using full Bayesian approach by assigning the prior to $\boldsymbol{\lambda}$, another way to determine the hyperparameter is the empirical Bayes approach, which has recently been considered by \cite{de2020maximum, vidal2020maximum} in image recover problems. This technique is known as type II maximum likelihood, which maximizes the marginal likelihood with respect to the hyperparameter. They proposed to use the stochastic gradient descent algorithm to update the hyper-parameters with the intractable gradient of log marginal likelihood being replaced with Monte Carlo estimator.
We can also use this technique to determine the $\boldsymbol{\lambda}$ as an alternative, i.e.,
$$
\boldsymbol{\lambda}_{\star} \in \underset{\boldsymbol{\lambda}}{\arg \max } \log p(\boldsymbol{y} \mid \boldsymbol{\lambda}),
$$
where $p(\boldsymbol{y} \mid \boldsymbol{\lambda}) = \int p(\boldsymbol{y} \mid \boldsymbol{x})\pi( \boldsymbol{x} \mid \boldsymbol{\lambda})\mathrm{d}\boldsymbol{x}$. In addition, the gradient of  log marginal likelihood can be expressed as $\nabla_{\boldsymbol{\lambda}} \log p(\boldsymbol{y} \mid \boldsymbol{\lambda}) = \mathrm{E}_{\pi(\boldsymbol{x} \mid \boldsymbol{y},\boldsymbol{\lambda})}[\nabla_{\boldsymbol{\lambda}}\log \pi( \boldsymbol{x} \mid \boldsymbol{\lambda})]$. Since $\lambda_{i} \in \mathbb{R}^{+}$, we can use mirror descent algorithm with mirror map $\Phi(\lambda_{i}) = \lambda_{i} \log (\lambda_{i})$, which leads to the exponentiation gradient update:
$$
\lambda_{i,t+1} = \lambda_{i,t}\exp(\eta_{t}\nabla_{\lambda_{i} = \lambda_{i,t} } \log p(\boldsymbol{y} \mid \boldsymbol{\lambda})),
$$
with the gradient being replaced by Monte Carlo gradient sampled by MCMC.
\begin{itemize}
    \item Connection with Maximum a posterior estimation(MAP) of marginal posterior: If the parameters in the hyper-prior for $\boldsymbol{\lambda}$ are some constants that do not depend on the dimension of $\boldsymbol{y}$, 
the marginal posterior $p(\boldsymbol{\lambda} \mid \boldsymbol{y}) \propto p(\boldsymbol{y} \mid \boldsymbol{\lambda})\pi(\boldsymbol{\lambda})$ will be dominated by the marginal likelihood $p(\boldsymbol{y} \mid \boldsymbol{\lambda})$ in high-resolution image. In this case, the MAP of the marginal posterior will close to the maximum of  marginal likelihood.
    \item Connection with full Bayesian approach: To understand the connection with the full Bayesian approach, we see that $\pi(\boldsymbol{x} \mid \boldsymbol{y}) =\int \pi(\boldsymbol{x} \mid \boldsymbol{y},\boldsymbol{\lambda})\pi(\boldsymbol{\lambda} \mid \boldsymbol{y})d\boldsymbol{\lambda}$. Since  $p(\boldsymbol{\lambda} \mid \boldsymbol{y}) \propto p(\boldsymbol{y} \mid \boldsymbol{\lambda})\pi(\boldsymbol{\lambda})$ will be dominated by the marginal likelihood $p(\boldsymbol{y} \mid \boldsymbol{\lambda})$, most of the probability mass of $\pi(\boldsymbol{\lambda} \mid \boldsymbol{y})$ is around the neighborhood of empirical Bayes estimator. As a result, we expect that both the full Bayesian approach and the empirical approaches will deliver similar results. The numerical studies in Appendix E confirmed our conjecture.
\end{itemize}
In practice, we recommend the full Bayesian approach as the empirical Bayesian approach did not show superior performance. In addition, it is computation intensive as we need to run a short chain of MCMC to obtain the Monte Carlo gradient of log marginal likelihood at each iteration of exponential gradient update.

\subsection{Gaussian mixture Markov random fields}
We now show that the fused bridge prior has a Gaussian mixture representation. The scale mixture representation was first found by~\cite{west1987scale}, who showed that a function $f(h)$ is completely monotone if and only if it can be represented as a Laplace transform of some 
function $g(\cdot)$:
$$f(h)=\int_{0}^{\infty} \exp (-s h)g(s) \mathrm{d}s.$$
To represent the exponential power prior with $0<\alpha \leq 1$ as a Gaussian mixture, by setting $f(h)=\exp[-(2h)^{\frac{\alpha}{2}}]$,$h=\frac{ \lambda^{\frac{2}{\alpha}} t^{2}}{2}$ and $\tau^{2}=\frac{1}{s}$, we have
$$\exp \left(-\lambda |t|^{\alpha}\right)=\int_{0}^{\infty} \frac{\lambda^{\frac{1}{\alpha}}}{\sqrt{2 \pi\tau^{2}}}\exp \Big(- \frac{\lambda^{\frac{2}{\alpha}} t^{2}}  {2\tau^{2}}\Big) \frac{\sqrt{2\pi}}{\lambda^{\frac{1}{\alpha}}\tau^{3}} g\left(\frac{1}{\tau^{2}}\right) \mathrm{d} \tau^{2}, $$
where $\pi(\tau^{2}) \propto \frac{1}{\tau^{3}}g\left(\frac{1}{\tau^{2}}\right)$, $\tau$ is the local shrinakge parameters and $\lambda$ is the global shrinakge parameter, see~\cite{polson2014bayesian} for more details. For the fused bridge prior considered in (\ref{eq:prior}), we can apply the above argument to obtain the Gaussian mixture representation:

\begin{lemma}\label{lem:bridge_decomposition}
For the fused bridge prior in (\ref{eq:prior}), we have the Gaussian mixture representation:
\begin{equation}\label{eq:scale_mixture_normal}
\resizebox{0.90\hsize}{!}{$\pi(\bm{x} \mid \bm{\lambda},\bm{\tau},\bm{\tau}^{h},\bm{\tau}^{v})  \propto \exp\bigg(-\frac{\lambda_{1}^{\frac{2}{\alpha_{1}}}}{2}\sum \limits_{i,j}\left(\frac{x_{ij}}{\tau_{ij}}\right)^{2}-\frac{\lambda_{2}^{\frac{2}{\alpha_{2}}}}{2}\sum \limits_{i,j}\left(\frac{\Delta_{i,j}^{h}}{\tau_{ij}^{h}}\right)^{2}-\frac{\lambda_{3}^{\frac{2}{\alpha_{2}}}}{2}\sum \limits_{i,j}\left(\frac{\Delta_{i,j}^{v}}{\tau_{ij}^{v}}\right)^{2}\bigg)$},
\end{equation}
with 
$\pi(\bm{\tau},\bm{\tau}^{ h},\bm{\tau}^{v} )=\prod_{i,j}\pi(\tau_{i,j} )\prod_{i,j}\pi(\tau_{i,j}^{h})\prod_{i,j}\pi(\tau_{i,j}^{v})$ and the conditional posterior can be factorized as 

\begin{equation*}
\resizebox{.85\hsize}{!}{$
\begin{aligned}
& \,\, \pi(\bm{\tau},\bm{\tau}^{ h},\bm{\tau}^{v} | \bm{x},\bm{\lambda}) \\ 
= & \prod_{i,j}\pi( \tau_{ij}|x_{ij},\lambda_{1})\prod_{i,j}\pi(\tau_{ij}^{h} \mid \Delta_{ij}^{h},\lambda_{2})\prod_{i,j}\pi(\tau_{ij}^{v} \mid \Delta_{ij}^{v},\lambda_{3})\\
= &  \prod_{i,j}\frac{\exp \bigg[-\Big(\frac{\lambda_{1}^{\frac{1}{\alpha_{1}}}x_{ij}} {\sqrt{2}\tau_{ij}}\Big)^{2}\bigg] \pi\left(\tau_{j}^{2}\right)}{\exp(-\lambda |x_{ij}|^{\alpha_{1}})}
\prod_{i,j}\frac{\exp \bigg[-\Big(\frac{\lambda_{2}^{\frac{1}{\alpha_{2}}}\Delta_{ij}^{h}}{\sqrt{2}\tau_{ij}^{h}}\Big)^{2}\bigg] \pi\left((\tau_{j}^{h})^{2}\right)}{\exp(-\lambda |\Delta_{ij}^{h}|^{\alpha_{2}})}  \\
& \prod_{i,j}\frac{\exp \bigg[-\Big(\frac{\lambda_{3}^{\frac{1}{\alpha_{2}}}\Delta_{ij}^{v}}{\sqrt{2}\tau_{ij}^{v}}\Big)^{2}\bigg] \pi\left((\tau_{j}^{v})^{2}\right)}{\exp(-\lambda |\Delta_{ij}^{v}|^{\alpha_{2}})}
\end{aligned}$}
\end{equation*}
which are exponentially tilted stable distributions.
\end{lemma}

We see that the conditional prior distribution $\pi(\bm{x} \mid \bm{\lambda},\bm{\tau},\bm{\tau}^{h},\bm{\tau}^{v})$ is a Gaussian Markov random field, which can be rewritten in the compact form:
\begin{equation}\label{eq:GMRF}
\pi(\bm{x} \mid \bm{\lambda},\bm{\tau},\bm{\tau}^{ h},\bm{\tau}^{v}) \propto \exp \left(-\frac{1}{2} \bm{x}^T\left(\bm{\Lambda}+\bm{D}_{h}^{\top} \bm{\Lambda}_h \bm{D}_{h}+\bm{D}_{v}^{\top} \bm{\Lambda}_{v} \bm{D}_{v}\right) \bm{x}\right),
\end{equation}
where $\bm{D}_{h}=\bm{D} \,\, \otimes \,\,\bm{I}_{d}$, $\bm{D}_{v}=\bm{I}_{d} \,\,\otimes \,\, \bm{D}$ with $\otimes$ denotes the Kronecker product, $\bm{I}_{d}$ is $d \times d$ identity matrix and $\bm{D}$ is a $d \times (d-1)$ difference matrix, which is slightly different for different boundary conditions. A zero boundary condition  $X_{0, j}=X_{d+1, j}=X_{i, 0}=X_{i, d+1}=0$ is assumed here, which gives us
$$
\bm{D}=\left[\begin{array}{rrrrr}
-1 & 1 & 0 & \cdots & 0 \\
0 & -1 & 1 & \ddots & \vdots \\
\vdots & \ddots & \ddots & \ddots & 0 \\
0 & \cdots & 0 & -1 & 1 
\end{array}\right]_{(d-1) \times d .}
$$
This can be easily modified for the aperiodic or Neumann boundary condition. In addition, we have $\bm{\Lambda}^{\frac{1}{2}}=\operatorname{diag}\left(\operatorname{vec}\left(\lambda_{1}^{\frac{1}{\alpha_{1}}}/\tau_{i,j}\right)\right)$, $\bm{\Lambda}_{h}^{\frac{1}{2}}=\operatorname{diag}\left(\operatorname{vec}\left(\lambda_{2}^{\frac{1}{\alpha_{2}}}/\tau_{i,j}^{h}\right)\right)$ and $\bm{\Lambda}_{v}^{\frac{1}{2}}=\operatorname{diag}\left(\operatorname{vec}\left(\lambda_{3}^{\frac{1}{\alpha_{2}}}/\tau_{i,j}^{v}\right)\right)$.  
 For details of Gaussian Markov random field representation of $\pi(\bm{x} \mid \bm{\lambda},\bm{\tau},\bm{\tau}^{h},\bm{\tau}^{v})$ from equation (\ref{eq:scale_mixture_normal}) to equation (\ref{eq:GMRF}), please refer to Chapter 4 of~\cite{bardsley2018computational}. Finally, the posterior distribution in (\ref{eq:posterior}) can be expanded as the the joint posterior,
\begin{equation}\label{eq:joint_posterior}
\pi(\bm{x},\bm{\tau},\bm{\tau}^{h},\bm{\tau}^{v},\bm{\lambda} \mid \bm{y}) \propto \pi(\bm{y} \mid \bm {x})\pi(\bm{x} \mid \bm{\tau},\bm{\tau}^{h},\bm{\tau}^{v},\bm{\lambda})\pi(\bm{\tau})\pi(\bm{\tau}^{h})\pi(\bm{\tau}^{v})\pi(\bm{\lambda}).
\end{equation}

\subsection{Fused $L_{1/2}$ prior}

Unfortunately, for $0<\alpha < 1$, there is no closed form expression for $\pi(\bm{\tau}^{2})$. Therefore, sampling the conditional posterior of the local shrinkage parameter $\tau$ shown in Lemma \ref{lem:bridge_decomposition} is hard.~\cite{polson2012local} suggested using the double rejection sampling algorithm from~\cite{devroye2009random,hofert2011sampling} to sample these exponentially tilted stable distributions. However, this approach is complicated and cannot be easily scaled to high dimensions. For detailed implementation, see Algorithm 3.2 in~\cite{hofert2011sampling}. Recently, it was shown by~\cite{ke2021bayesian} that the exponential power prior with $\alpha=\frac{1}{2^{\gamma}},\gamma=\{0,1,2,\ldots\}$ has a Laplace mixture representation, which can be further decomposed as a scale mixture of Gaussian distributions. This representation introduces extra latent variables, allowing us to circumvent sampling the difficult conditional posterior distribution of the local shrinkage parameter $\tau$. Thus, this paper will focus on the cases that $\alpha_{1}=\frac{1}{2^{\gamma_{1}}}$ and $\alpha_{2}=\frac{1}{2^{\gamma_{2}}}$ with $\gamma_{1},\gamma_{2}\in \left\{0,1,2,3,...,\right\}$. For simplicity, we call this fused $L_{1/2}$ prior, which is a special case of the fused bridge prior:
\begin{equation}\label{eq:fused_prior}
\resizebox{0.90\hsize}{!}{$
\pi(\bm{x} \mid \bm{\lambda})  \propto \exp\Big(-\lambda_{1}\sum \limits_{i,j}|x_{ij}|^{\frac{1}{2^{\gamma_{1}}}}-\lambda_{2}\sum \limits_{i,j}|\Delta_{i,j}^{h}|^{\frac{1}{2^{\gamma_{2}}}}-\lambda_{3}\sum\limits_{i,j}|\Delta_{i,j}^{v}|^{\frac{1}{2^{\gamma_{2}}}}\Big)$}.
\end{equation}
Using similar argument as in~\cite{ke2021bayesian}, the fused $L_{1/2}$ prior has the following decomposition:

\begin{lemma}
For the fused $L_{1/2}$ prior with $\alpha_{1}=\frac{1}{2^{\gamma_{1}}}$, $\alpha_{2}=\frac{1}{2^{\gamma_{2}}}$ and $\gamma_{1},\gamma_{2}\in \left\{0,1,2,3,...,\right\}$, we have the Gaussian mixture representation as shown in equation (\ref{eq:GMRF}) with the local shrinkage parameters $\tau$ having the following latent variable representation:
\begin{equation}\label{eq:gls}
\begin{aligned}
& \tau_{ij}^{2}|v_{ij}^{1} \sim  \mathrm{Exp}\Big(\frac{1}{2(v_{ij}^{1})^{2}}\Big), \\ 
& v_{ij}^{l}|v_{ij}^{l+1} \sim \mathrm{Gamma}\Big(\frac{2^{l}+1}{2},\frac{1}{4(v_{ij}^{l+1})^{2}}\Big),\\
& v_{ij}^{\gamma_{1}} \sim \mathrm{Gamma}\Big(\frac{2^{\gamma_{1}}+1}{2},\frac{1}{4}\Big),
\end{aligned}
\end{equation}
for $l = 1,..,\gamma_{1}-1$. In addition, when $\gamma_{1}=1$, the terms $v_{ij}^{l}|v_{ij}^{l+1}$ vanish. For $\gamma_{1}=0$, we only have $\tau_{ij}^{2}\sim  \mathrm{Exp}\left(1/2\right)$.
\end{lemma}
To save space, we only write down the latent variable representation for $\pi(\tau_{ij}^{2})$. The same decomposition also holds for both $\pi((\tau_{ij}^{h})^{2})$ and $\pi((\tau_{ij}^{v})^{2})$. 

\section{MCMC method}
\label{sec:Gibbs-BPS}
\begin{figure}[!htb]
    \centering
    \includegraphics[width=0.93\linewidth]{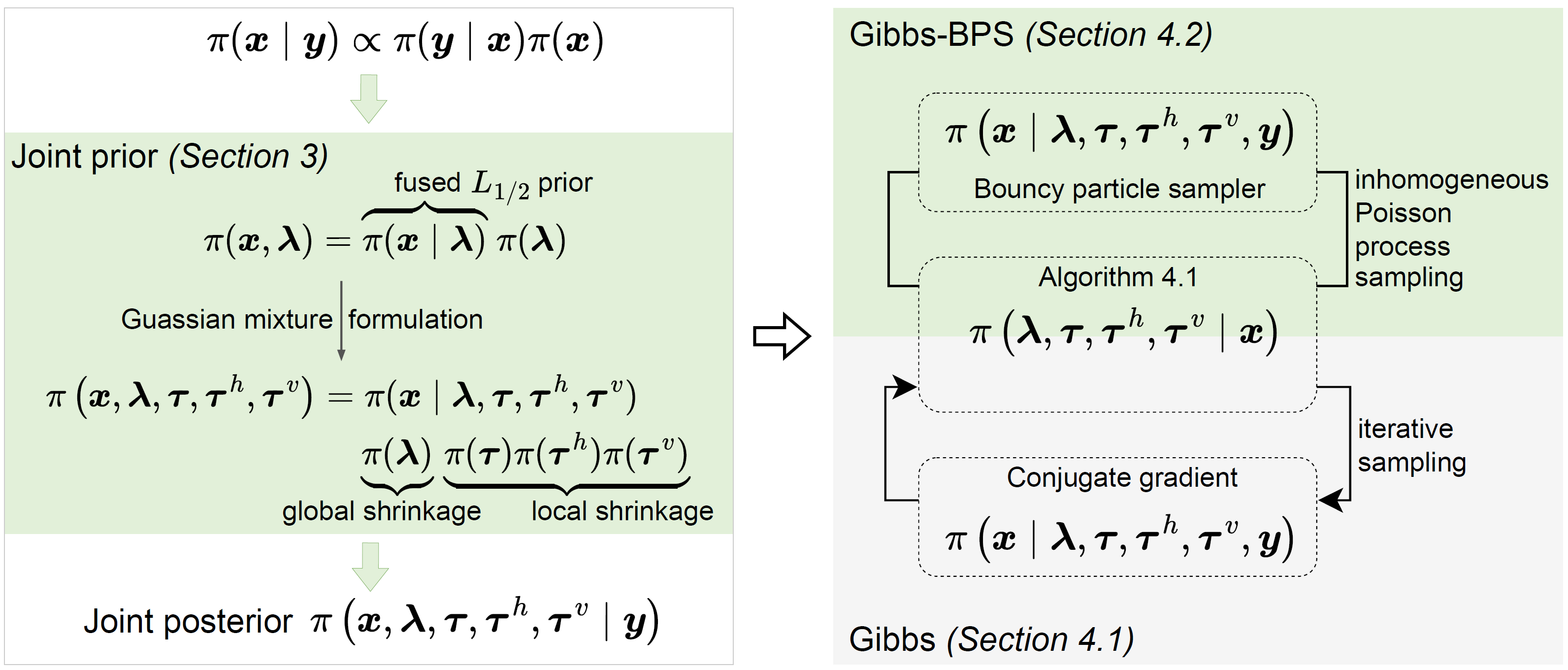}
    \caption{This schematic outlines the proposed method. The joint prior, represented as a Gaussian mixture representation of the fused $L_{1/2}$ prior, leads to the derivation of the joint posterior distribution, which is sampled using Gibbs-BPS. Unlike standard Gibbs sampler, which iteratively samples from conditional distributions, Gibbs-BPS leverages an inhomogeneous Poisson process to determine which conditional posterior to update.
    }
    \label{fig:Gibbs-BPS}
\end{figure}

In this section, we begin by building a Gibbs sampler to sample the joint conditional posterior $\pi(\bm{x},\bm{\lambda}, \bm{\tau},\bm{\tau}^{h},\bm{\tau}^{v} \mid \bm{y})$ in (\ref{eq:joint_posterior}). This is a Gaussian mixture Markov field given in~\eqref{eq:GMRF} and~\eqref{eq:gls},  based on the fused $L_{1/2}$ prior in~\eqref{eq:fused_prior}. 
This is done by iteratively sampling $\pi(\bm{x} \mid \bm{\lambda},\bm{\tau},\bm{\tau}^{ h},\bm{\tau}^{v}, \bm{y})$ in (\ref{eq:gaussian})  and $\pi(\bm{\lambda},\bm{\tau},\bm{\tau}^{ h},\bm{\tau}^{v} \mid \bm{x})$ in (\ref{eq:latent}). In order to reduce the computational burden in sampling $\pi(\bm{x} \mid \bm{\lambda},\bm{\tau},\bm{\tau}^{ h},\bm{\tau}^{v}, \bm{y})$, which is an ultra high dimensional Gaussian, we introduce the Gibbs bouncy particle sampler (Gibbs-BPS). This method combines elements of Gibbs sampler with the bouncy particle sampler, which is particularly well suited to sample the conditional posterior of the pixels $\bm{x}$ as it avoids any matrix inverse and matrix factorization. We show that, at each iteration, the computational complexity of Gibbs-BPS is an order of magnitude smaller than that of the standard Gibbs sampler. 
In Figure~\ref{fig:Gibbs-BPS}, we provide an overview of the Gibbs-BPS algorithm.

\subsection{Gibbs sampler}
With the Gaussian mixture Markov random fields representation of the fused $L_{1/2}$ prior, the  Gibbs sampler with closed form conditional posteriors can be constructed by slightly modifying the work from~\cite{ke2021bayesian}. Given the Gaussian likelihood of (\ref{eq:likelihood}) and Gaussian conditional prior of (\ref{eq:GMRF}), the conditional posterior of $\bm{x}$ is 
$$
\pi(\bm{x} \mid \bm{\lambda},\bm{\tau},\bm{\tau}^{ h},\bm{\tau}^{v}, \bm{y})\propto \pi(\bm{y} \mid \bm{x})\pi(\bm{x} \mid \bm{\lambda},\bm{\tau},\bm{\tau}^{ h},\bm{\tau}^{v}), 
$$
which is also Gaussian with precision matrix and mean vector given by
\begin{equation}\label{eq:gaussian}
\widetilde{\bm{\Lambda}}=\frac{1}{\sigma_{\mathrm{obs}}^2} \bm{A}^{\top} \bm{A}+\bm{\Lambda}+\bm{D}_{h}^{\top}\bm{\Lambda}_{h} \bm{D}_{h}+\bm{D}_{v}^{\top} \bm{\Lambda}_{v} \bm{D}_{v}, \quad \widetilde{\bm{\mu}}=\widetilde{\bm{\Lambda}}^{-1}\Big(\frac{1}{\sigma_{\mathrm{obs}}^{2}} \bm{A}^{\top} \bm{y}\Big).
\end{equation}
By using the latent variable representation of the local shrinkage parameters in Lemma 2, we can sample the conditional posterior of the global and local shrinkage parameters.

\begin{proposition}
The conditional posterior distribution of the global and local shrinkage parameters, $\lambda$ and $\tau$ respectively, can be factorized as 
\begin{equation}\label{eq:latent}
\begin{aligned}
& \pi(\bm{\lambda},\bm{\tau},\bm{\tau}^{ h},\bm{\tau}^{v} \mid \bm{x}) \\
= & \Big(\pi(\lambda_{1} \mid \bm{x})\prod_{i,j}\pi(\tau_{ij} \mid x_{ij},\lambda_{1})\Big) \Big(\pi(\lambda_{2} \mid \bm{\Delta}^{h})\prod_{i,j}\pi(\tau_{ij}^{h} \mid \Delta_{ij}^{h},\lambda_{2})\Big)\\
&  \Big(\pi(\lambda_{3} \mid \bm{\Delta}^{v})\prod_{i,j}\pi(\tau_{ij}^{v} \mid \Delta_{ij}^{v},\lambda_{3})\Big),
\end{aligned}
\end{equation}
which can be sampled using algorithm \ref{alg:ancestral_sampling}.
\end{proposition}

\begin{algorithm}
\begin{algorithmic}[1]
    \item[] \textbf{Input:} $\gamma_{1} \in \mathrm{N}_{0}$; $a_{1},b_{1} \in \mathrm{R}^{+}$
    \item[] \textbf{Output:} $\lambda_{1},\bm{\tau^{2}}$
          \vspace{4pt}
    \STATE {\bf Sample} $\lambda_{1} \mid \bm{x} \sim \mathrm{Gamma}\left(2^{\gamma_{1}}d^{2}+a_{1},\sum_{i=1}^{d}\sum_{j=1}^{d}\left|x_{ij}\right|^{\frac{1}{2^{\gamma_{1}}}}+b_{1}\right)$
    \STATE {\bf Sample} $\bm{\tau^{2}}$ {from} $\prod_{i,j}\pi(\tau_{ij} \mid x_{ij},\lambda_{1})$ {independently via:}
    \IF{$\gamma_{1}=0$}
        \STATE \mbox{Sample} $\frac{1}{\tau_{ij}^{2}} \mid x_{ij},\lambda_{1}  \sim \mathrm{IG}\left (\frac{1}{{\lambda_{1}}|x_{ij}|},1\right )$
    \ELSE
        \STATE \mbox{Sample} $\frac{1}{v_{ij}^{\gamma_{1}}}  \mid x_{ij},\lambda_{1}  \sim \mathrm{IG}\Big (\frac{1}{2\lambda_{1}|x_{ij}|^{\frac{1}{2^{\gamma_{1}}}}}, 1/2\Big)$
        \IF{$\gamma_{1} \geq 2$}
            \FOR{$l \leftarrow \gamma_{1}-1$ \TO $1$}
                \STATE \mbox{Sample} $\frac{1}{v_{ij}^{l}} \mid x_{ij}, \lambda_{1}, v_{ij}^{l+1} \sim \mathrm{IG}\Big (\frac{1}{2v_{ij}^{l+1}\lambda_{1}|x_{ij}|^{\frac{1}{2^{l}}}}, \big(\frac{1}{\sqrt{2}{v}_{ij}^{l+1}}\big)^{2}\Big)$
            \ENDFOR
        \ENDIF
        \STATE \mbox{Sample} $\frac{1}{\tau_{ij}^{2}} \mid x_{ij},\lambda_{1},v_{ij}^{1}  \sim \mathrm{IG}\Big (\frac{1}{\lambda_{1}^{2^{\gamma_{1}}}{v}_{ij}^{1}|x_{ij}|}, \big(\frac{1}{v_{ij}^{1}}\big)^{2}\Big)$ 
    \ENDIF
\end{algorithmic}
\caption{Sampling global and local shrinkage parameters}
\label{alg:ancestral_sampling}
\end{algorithm}

Again due to space constraints,  we will only write down the algorithm for sampling $\pi(\lambda_{1} \mid \bm{x})\prod_{i,j}\pi(\tau_{ij} \mid x_{ij},\lambda_{1})$. The sampling of $\pi(\lambda_{2} \mid \bm{\Delta}^{h})\prod_{i,j}\pi(\tau_{ij}^{h} \mid \Delta_{ij}^{h},\lambda_{2})$ and $\pi(\lambda_{3} \mid \bm{\Delta}^{v})\prod_{i,j}\pi(\tau_{ij}^{v} \mid \Delta_{ij}^{v},\lambda_{3})$ is exactly the same. Based on Algorithm \ref{alg:ancestral_sampling}, we construct a two-block Gibbs sampler as shown in Algorithm \ref{alg: Gibbs}.  

\begin{algorithm}
\begin{algorithmic}[1]
\item[] \textbf{Input:} $\gamma_{1},\gamma_{2} \in \mathrm{N}_{0}$; $a_{1}, b_{1},a_{2},b_{2},a_{3},b_{3} \in \mathrm{R}^{+}$; $T${: Num of iterations}
\item[] \textbf{Output:} {All the $T$ samples of $\bm{x}$}
\vspace{4pt}
\FOR{$t \leftarrow 1$ \TO T}
\STATE {\bf Block 1:} sample  $\bm{x} \mid \bm{\lambda},\bm{\tau},\bm{\tau}^{ h},\bm{\tau}^{v}, \bm{y}
\sim \mathcal{N}\Big(\widetilde{\bm{\Lambda}}^{-1}\big(\frac{1}{\sigma_{\mathrm{obs}}^2} \bm{A}^{\top} \bm{y}\big),\widetilde{\bm{\Lambda}}^{-1}\Big)$
\STATE {\bf Block 2:} sample $\bm{\lambda},\bm{\tau},\bm{\tau}^{ h},\bm{\tau}^{v} \mid \bm{x}$ via Algorithm \ref{alg:ancestral_sampling}
\ENDFOR
\end{algorithmic}
\caption{Two-Block Gibbs sampler}
\label{alg: Gibbs}
\end{algorithm}

\subsection{Bouncy particle sampler}
We first introduce Pecewise deterministic Markov process (PDMP)-based samplers~\cite{bouchard2018bouncy,bierkens2019zig}, then exemplify the approach using the bouncy particle sampler (BPS)~\cite{bouchard2018bouncy}, which forms the cornerstone for our construction of the Gibbs bouncy particle sampler in Section \ref{sec:Gibbs-BPS}. 

\subsubsection{PDMP} 
Intuitively speaking, the PDMP dynamic involves random events, with deterministic dynamics between events and possibly random transitions. The random events are distributed according to a non-homogeneous Poisson process. 

Suppose $\pi(\bm{x}) \propto \exp (-U(\bm{x}))$ is the target distribution with $U(\bm{x})$ as its potential. In the PDMP framework, an auxiliary variable, $\bm{v} \in \mathcal{V}$, which can be understood as the velocity of the particle $\bm{x}$, is introduced. PDMP based sampler explores the augmented state space $\mathrm{R}^{n} \times \mathcal{V}$, targeting a distribution $\pi(d \bm{x}, d \bm{v})$, with variable $\bm{z} = (\bm{x}, \bm{v})$ over $\mathrm{R}^{n} \times \mathcal{V}$, as its invariant distribution. By construction, the distribution ${\textcolor{blue}{\pi}}$ will enjoy independence between $\bm{x}$ and $\bm{v}$, so that $\pi(\bm{v},\bm{x})=\pi(\bm{v})\pi(\bm{x} )$. In the bouncy particle sampler, $\mathcal{V}$ is chosen to be the Euclidean space $\mathrm{R}^{n}$ and $\pi(\bm{v})$ are independent standard Gaussian distributions. A piecewise deterministic Markov process $\bm{z}_{t} = (\bm{x}_{t}, \bm{v}_{t})$ consists of three distinct components: 
\begin{itemize}
    \item A deterministic dynamic between the jumps according to some ordinary differential equation $\frac{d \mathbf{Z}_t}{d t}=\Psi\left(\mathbf{Z}_t\right)$.
    \item A jump event occurrence rate $\lambda\left(\mathbf{Z}_t\right)$. Here, an event refers to an occurrence of a time inhomogeneous Poisson process.
    \item Transition immediately after the event, $Q\left(\mathbf{z}_s \mid \mathbf{z}_{s-}\right)$.    
\end{itemize}
Davis~\cite{davis1984piecewise,davis2018markov} gives the generator for a piecewise deterministic process:
\begin{equation}\label{eq:generator}
\mathcal{L} f(\mathbf{z})=\nabla f(\mathbf{z}) \cdot \Psi(\mathbf{z})+\lambda(\mathbf{z}) \int_{\mathbf{z}^{\prime}}\left(f\left(\mathbf{z}^{\prime}\right)-f(\mathbf{z})\right) Q\left(d \mathbf{z}^{\prime} \mid \mathbf{z}\right).
\end{equation}
To guarantee the invariant distribution of this process is  $\rho(d \mathbf{z})$, the generator needs to satisfy $\int \mathcal{L} f(\mathbf{z}) \rho(d \mathbf{z})=0$ for all function $f$ in the domain of the generator $\mathcal{L}$. For details, see Proposition 34.7 from~\cite{davis2018markov}. 

\subsubsection{BPS} 
We now use the bouncy particle sampler~\cite{bouchard2018bouncy} as a concrete example:
\begin{itemize}
    \item The corresponding deterministic dynamic is
    \begin{equation}\label{eq:straight_line}
    \frac{d \bm{x}_t}{d t}=\bm{v}_t, \,\, \frac{d \bm{v}_t}{d t}=\mathbf{0}.
    \end{equation}
    \item The event rate satisfies
    \begin{equation}\label{eq:event_rate}
    \lambda(\mathbf{z}_{t})=\lambda(\mathbf{x}_{t}, \mathbf{v}_{t})=\langle\mathbf{v}_{t}, \nabla U(\mathbf{x}_{t})\rangle_{+}+\lambda^{\mathrm{ref}}.
    \end{equation}
    \item The transition kernel
    \begin{equation}\label{eq:transition}
    \begin{aligned}
    & Q\left(\left(d \mathbf{x}^{\prime}, d \mathbf{v}^{\prime}\right) \mid(\mathbf{x}, \mathbf{v})\right)\\
    = &\frac{\langle\mathbf{v}, \nabla U(\mathbf{x})\rangle_{+}}{\lambda(\mathbf{x}, \mathbf{v})} \delta_{\mathbf{x}}\left(d \mathbf{x}^{\prime}\right) \delta_{R_{\nabla U(\mathbf{x})}}\left(d \mathbf{v}^{\prime}\right) +\frac{\lambda^{\mathrm{ref}}}{\lambda(\mathbf{x}, \mathbf{v})} \delta_{\mathbf{x}}\left(d \mathbf{x}^{\prime}\right) \varphi\left(d \mathbf{v}^{\prime}\right),
    \end{aligned}
    \end{equation}
\end{itemize}
where $\langle \cdot, \cdot \rangle_{+}$ is the operator taking the positive part of the inner product of two vectors, $\lambda^{\mathrm{ref}}$ is a user chosen positive constant and the velocity after bouncing is given by
\begin{equation}
R_{\nabla U(\bm{x})}(\bm{v})=\bm{v}-2 \frac{ \bm{v}^{T} \nabla U(\bm{x})}{\|\nabla U(\bm{x})\|^2} \nabla U(\bm{x}),
\end{equation}
and $\varphi(d \mathbf{v})=\mathcal{N}\left(d \mathbf{v} \mid 0_{n}, I_n \right)$. Plugging equations (\ref{eq:straight_line}),(\ref{eq:event_rate}) and (\ref{eq:transition}) into equation (\ref{eq:generator}), one can verify that the bouncy particle sampler is invariant to the target distribution. The basic version of the BPS algorithm proceeds is described in Algorithm~\ref{algo:BPS}.
\begin{algorithm}[!ht]
    \caption{BPS algorithm}
    \label{algo:BPS}
    \begin{algorithmic}[1]
       \item[] \textbf{Input:} $\lambda_{\mathrm{ref}} \in \mathrm{R}^{+}$, $T$: length of the trajectory.
      \item[] \textbf{Output:} $\left\{(\bm{x}^{(k)},\bm{v}^{(k)},s^{(k)})\right\}_{k=1}^{i}$ and $t^{(i)}$
      \vspace{4pt}
        \STATE {\bf Initialize:} $\left(\bm{x}^{\left(0\right)},\bm{v}^{\left(0\right)}\right)$ arbitrarily on $\mathrm{R}^{n}\times\mathrm{R}^{n}$, $t^{(0)}=0$, $i=0$.
        \WHILE{$t^{(i)} \leq T$}
            \STATE $i \leftarrow i+1$
            \STATE Simulate the first arrival time $s_{\mathrm{bounce}}$: 
            \begin{equation}\label{eq:first_arrival_time}
            \begin{aligned}
            \int_{0}^{s} \lambda(\bm{x}^{\left(i-1\right)}+\bm{v}^{\left(i-1\right)}t,\bm{v}^{\left(i-1\right)}) \, dt &= \int_{0}^{s} \langle \bm{v}^{i-1}, \nabla U(\bm{x}^{\left(i-1\right)}+\bm{v}^{i-1}t)\rangle_{+} dt \\
            &= -\log (u),~~u \sim U(0,1).
            \end{aligned}
            \end{equation}
            \STATE Simulate $s_{\mathrm{ref}}\sim\mathrm{Exp\left(\lambda^{\mathrm{ref}}\right)}$.
            \STATE Set $s^{(i)}\leftarrow\mathrm{min}\left(s_{\mathrm{bounce}},s_{\mathrm{ref}}\right)$ and compute the next position using
            \begin{align*}
            \bm{x}^{\left(i\right)}\leftarrow \bm{x}^{\left(i-1\right)}+\bm{v}^{\left(i-1\right)}s^{(i)} 
            \end{align*}
            \vspace{-1.5em}
            \IF{$s^{(i)}=s_{\mathrm{ref}}$}
                \STATE Sample the next velocity  $\bm{v}^{\left(i\right)}\sim\mathcal{N}\left(0_{n},I_{n}\right)$.
            \ELSE
                \STATE Compute the next velocity using 
                $\bm{v}^{\left(i\right)}\leftarrow R_{\nabla_{\bm{x}} U(\bm{x^{(i)}})}(\bm{v}^{\left(i-1\right)})$.
            \ENDIF
            \STATE $t^{(i)} \leftarrow t^{(i-1)}+s^{(i)}$
        \ENDWHILE
    \end{algorithmic}
\end{algorithm}

In practice, the main difﬁculty in implementing the BPS sampler is the generation of the occurrence times of the time inhomogeneous Poisson process with event rate $\lambda(\mathbf{z}_{t})$. We can apply the superposition theorem~\cite{kingman1992poisson}, which allows us to simulate two arrival times from two Poisson processes with rate $\langle\mathbf{v}_{t}, \nabla U(\mathbf{x}_{t})\rangle_{+}$ and $\lambda^{\mathrm{ref}}$, respectively and take their minimum. See Algorithm \ref{algo:BPS}, lines 5-7. Since it is generally impossible to simulate Poisson processes with the rate function $\langle\mathbf{v}_{t}, \nabla U(\mathbf{x}_{t})\rangle_{+}$ using inverse transform sampling as shown in equation (\ref{eq:first_arrival_time}), we need to find its tight upper bound and use the Poisson thinning~\cite{lewis1979simulation} to simulate the arrival times. If the upper bound is not tight, the bouncy particle sampler is not efficient.

\subsection{Gibbs-BPS sampler}

Sampling the conditional posterior of global and local shrinkage parameters is very fast due to the conditional independent structure. The main computation bottleneck of the Gibbs sampler is the need to sample from multivariate Gaussian distribution of the form (\ref{eq:gaussian}) repeatedly, which requires solving the linear system or doing matrix factorization and, thus, has computational complexity $O(\min(n^{3}, mn^{2}))$~\cite{rue2001fast,bhattacharya2016fast}. A significant speed up can be achieved by using the conjugate gradient method with a preconditioner on the prior precision matrix~\cite{parker2012sampling,nishimura2022prior}. Recently, an approximate Gibbs sampler algorithm has been proposed for the horseshoe prior in a linear model with Gaussian likelihood~\cite{johndrow2020scalable}. It reduced the task of sampling a multivariate Gaussian distribution from solving a $n \times n$ linear system to $s \times s$ linear system with $s \ll n$ in a linear regression setting. The computational complexity of their approach is 
$O(\max(s^{2}n,mn))$ with $s$ depends on the user defined threshold and the sparsity level of the true $\bm{x}$. However, their approach only works well when the true signal is extremely sparse, and it does not work for the image problem.

We now propose the Gibbs bouncy particle sampler (Gibbs-BPS), whose computational complexity is an
order of magnitude smaller than the Gibbs sampler in linear inverse problem. Let $\bm{\phi}=(\bm{\lambda},\bm{\tau},\bm{\tau}^{ h},\bm{\tau}^{v})$. The idea of the Gibbs bouncy particle sampler(Gibbs-BPS) is to combine updates of the component $\bm{x}$ given $\bm{\phi}$ via the bouncy particle sampler and update $\bm{\phi}$ given $\bm{x}$ fixed via some Markov kernels, which are invariant to $\pi(\bm{\phi} \mid \bm{x})$. It should be pointed out that these two updates are not combined in the same way as Metropolis Hastings within the Gibbs algorithm framework as shown in the work~\cite{zhao2021analysis,zhang2021large}, but combined in a way that still keeps the whole algorithm as PDMP, similar to the Gibbs-ZigZag sampler~\cite{sachs2023posterior}.

More precisely, let $\mathcal{L}_{\mathrm{BPS}}$ denote the generator of the process which leaves the $\bm{\phi}$ fix and evolves $\bm{x}$  according to a bouncy particle sampler with the event occurrence rate
$$\lambda(\bm{x}, \bm{v},\bm{\phi})=\langle\bm{v}, \nabla_{\bm{x}} U(\bm{x},\bm{\phi})\rangle_{+}+\lambda^{\mathrm{ref}},$$
where $\pi(\bm{x} \mid \bm{\phi},\bm{y})\propto \exp(-U(\bm{x}, \bm{\phi}))$. Then the generator $\mathcal{L}_{\mathrm{BPS}}$ for updating $\bm{x}$ takes the form:
\begin{equation}\label{eq:BPS_generator}
\begin{aligned}
\left(\mathcal{L}_{\mathrm{BPS}} f\right)(\bm{x}, \bm{v},\bm{\phi}) & = \bm{v}^{T} \nabla_{\bm{x}} f(\bm{x}, \bm{v},\bm{\phi})\\
+ & \lambda^{\mathrm{ref}}  \int  [f(\bm{x}, \bm{v}^{\prime}, \bm{\phi})-f(\bm{x},\bm{v}, \bm{\phi})]\pi(\bm{v}^{\prime})d\bm{v}^{\prime}\\
+  \langle & \bm{v},   \nabla_{\bm{x}} U(\bm{x},\bm{\phi})\rangle_{+}
[f(\bm{x},R_{\nabla_{\bm{x}} U(\bm{x},\bm{\phi})}(\bm{v}), \bm{\phi})-f(\bm{x},\bm{v}, \bm{\phi})].
\end{aligned}
\end{equation}
Let $\mathcal{Q}$ be a Markov kernel for $\bm{\phi}$, which is invariant with respect to $\pi(\bm{\phi} \mid \bm{x})$. Then the generator of Gibbs-type update for $\bm{\phi}$ takes the form:
\begin{equation}\label{eq:Gibbs_generator}
\left(\mathcal{L}_{\mathrm{Gibbs}} f\right)(\bm{x}, \bm{v},\bm{\phi})=\int \left\{f(\bm{x}, \bm{v},\bm{\phi}^{\prime})-f(\bm{x}, \bm{v},\bm{\phi})\right\} \mathcal{Q}(\bm{\phi}, \mathrm{d} \bm{\phi}^{\prime}).
\end{equation}
We obtain the Gibbs-BPS by combining the two processes described above, whose generator can be written as  
\begin{equation}\label{eq:Gibbs_BPS}
\mathcal{L}_{\mathrm{Gibbs-BPS}}=\mathcal{L}_{\mathrm{BPS}}+\eta \mathcal{L}_{\mathrm{Gibbs}},
\end{equation}
where $\eta$ is a user-chosen positive constant. 

For the Gibbs-BPS, the $\bm{\phi}$ is constant between the jump events. Given the 'jump' event happens, with probability $\frac{\eta}{\langle\bm{v}_{t}, \nabla_{x} U(\bm{x}_{t},\bm{\phi}_{t})\rangle_{+}+\lambda^{\mathrm{ref}}+\eta}$, $\bm{\phi}$ will be updated with Markov kernel  $\mathcal{Q}(\bm{\phi}, \mathrm{d} \bm{\phi}^{\prime})$. In practice, we use the superimposition technique to simulate the PDMP process described by (\ref{eq:Gibbs_BPS}). By verifying $\int \mathcal{L}_{\mathrm{GBPS}} f(\bm{\theta}) \rho(d \bm{\theta})=0$ with $\bm{\theta}=(\bm{x},\bm{v}, \bm{\phi})$, we obtain the next theorem.
\begin{theorem}
The Gibbs-BPS with the generator (\ref{eq:Gibbs_BPS}) is invariant with respect to $\pi(\bm{x},\bm{\phi} \mid \bm{y})\pi(\bm{v})$.
\end{theorem}
The proof of Theorem 4.2 is given in the Appendix. In the Gibbs-BPS algorithm, the conditional posterior $\pi(\bm{x} \mid \bm{\lambda},\bm{\tau},\bm{\tau}^{ h},\bm{\tau}^{v}, \bm{y})$ is updated by bouncy particle sampler. The following lemma demonstrates the efficiency of the bouncy particle sampler for sampling the high dimensional Gaussian distribution in (\ref{eq:gaussian}). 

\begin{lemma}
For the Gaussian distribution $\bm{x} \sim \mathcal{N}(\bm{\mu}, \bm{\Sigma})$, we can use the inverse cumulative distribution function technique to simulate the arrival time $s$ with rate $\langle\mathbf{v}_{t}, \nabla U(\mathbf{x}_{t})\rangle_{+}$ in equation (\ref{eq:first_arrival_time}). Specifically, we have
\begin{equation}\label{eq:inverse_CDF}
\resizebox{0.90\hsize}{!}{$
s=\left(\bm{v}^T \bm{\Sigma}^{-1} \bm{v}\right)^{-1}\left[-(\bm{x}-\bm{\mu})^T \bm{\Sigma}^{-1} \bm{v}+\sqrt{\left(\left((\bm{x}-\bm{\mu})^T \bm{\Sigma}^{-1} \bm{v}\right)_{+}\right)^2-2 \bm{v}^T \bm{\Sigma}^{-1} \bm{v} \log u}\right]$,}
\end{equation}
where $u \sim U(0,1)$.
\end{lemma}
Plug equation (\ref{eq:gaussian}) into equation (\ref{eq:inverse_CDF}), we have 
\begin{equation}
s=(-c_{1}+\sqrt{((c_{1})_{+})^{2}-2c_{2}\log u})/c_{2},
\end{equation}
where $c_{1}=\bm{x}^{\top}\widetilde{\bm{\Lambda}}\bm{v}-\frac{\bm{y}^{\top}\bm{A}\bm{v}}{\sigma_{obs}^{2}}$, $c_{2}=\bm{v}^{\top}\widetilde{\bm{\Lambda}}\bm{v}$ and $\widetilde{\bm{\Lambda}}=\frac{1}{\sigma_{\mathrm{obs}}^2} \bm{A}^{\top} \bm{A}+\bm{\Lambda}+\bm{D}_{h}^{\top} \bm{\Lambda}_{h} \bm{D}_{h}+\bm{D}_{v}^{\top} \bm{\Lambda}_{v} \bm{D}_{v}$. In addition, we have 
\begin{equation}
\begin{aligned}
\nabla_{\bm{x}} U(\bm{x},\bm{\phi}) & =-\frac{\bm{A}^{\top}\bm{y}}{\sigma_{obs}^{2}}+\frac{\bm{A}^{\top}\bm{A}\bm{x}}{\sigma_{obs}^{2}}+\bm{\Lambda}\bm{x}+\bm{D}_{h}^{\top} \bm{\Lambda}_{h} \bm{D}_{h}\bm{x}+\bm{D}_{v}^{\top} \bm{\Lambda}_{v} \bm{D}_{v}\bm{x},\\
\bm{v}^{\top}\nabla_{\bm{x}} U(\bm{x},\bm{\phi}) & =c_{1}.
\end{aligned}
\end{equation}
We see that $\bm{A}^{\top}\bm{y}$ and $\bm{A}^{\top}\bm{A}$ can be precomputed. In addition, due to the sparsity structure of the Markov property of the conditional Gaussian prior, the computational complexity of calculating $\bm{\Lambda}\bm{x}+\bm{D}_h^{\top} \bm{\Lambda}_h \bm{D}_h\bm{x}+\bm{D}_v^{\top} \bm{\Lambda}_v \bm{D}_v\bm{x}$ is only $O(n)$. The main computational burden is calculating $\bm{A}^{\top}\bm{A}\bm{x}$, which has complexity $O(n^{2})$ per iteration. This is an order lower than the previous approaches. In Algorithm \ref{alg:Gibbs-BPS}, we provide detailed implementation of Gibbs-BPS.

\begin{algorithm}[!ht]
    \caption{Gibbs-BPS algorithm}
    \label{alg:Gibbs-BPS}
    \begin{algorithmic}[1]
     \item[] \textbf{Input:} $\gamma_{1},\gamma_{2} \in \mathrm{N}_{0}$; $a_{1}, b_{1},a_{2},b_{2},a_{3},b_{3},\lambda_{\mathrm{ref}},\eta \in \mathrm{R}^{+}$, $T$: length of the trajectory.   
     \item[] \textbf{output:} $\left\{(\bm{x}^{(k)},\bm{v}^{(k)},s^{(k)})\right\}_{k=1}^{i}$ and $t^{(i)}$  
           \vspace{4pt}
        \STATE {\bf Initialize:} $\left(\bm{x}^{\left(0\right)},\bm{v}^{\left(0\right)}\right)$ arbitrarily on $\mathrm{R}^{n}\times\mathrm{R}^{n}$, $t^{(0)}=0$, $i=0$.
        \WHILE{$t^{(i)} \leq T$}
            \STATE $i \leftarrow i+1$
            \STATE Simulate the first arrival time $s_{\mathrm{bounce}}$: 
            $$s_{\mathrm{bounce}} \leftarrow (-c_{1}+\sqrt{((c_{1})_{+})^{2}-2c_{2}\log u})/c_{2},~~u \sim U(0,1).$$
            \STATE Simulate $s_{\mathrm{ref}}\sim\mathrm{Exp\left(\lambda^{\mathrm{ref}}\right)}$ and $s_{\mathrm{Gibbs}} \sim \mathrm{Exp\left(\eta\right)}$.
            \STATE Set $s^{(i)}\leftarrow\mathrm{min}\left(s_{\mathrm{bounce}},s_{\mathrm{ref}},s_{\mathrm{Gibbs}}\right)$ and compute the next position using 
            \begin{align*}
            \bm{x}^{\left(i\right)}\leftarrow \bm{x}^{\left(i-1\right)}+\bm{v}^{\left(i-1\right)}s^{(i)}
            \end{align*}
            \vspace{-1.5em} 
            \IF{$s^{(i)}=s_{\mathrm{ref}}$}
                \STATE Sample the next velocity  $\bm{v}^{\left(i\right)}\sim\mathcal{N}\left(0_{n},I_{n}\right)$.
            \ELSIF{$s^{(i)}=s_{\mathrm{bounce}}$}
                \STATE Compute the next velocity using 
                $\bm{v}^{\left(i\right)}\leftarrow R_{\nabla_{\bm{x}} U(\bm{x^{(i)}})}(\bm{v}^{\left(i-1\right)})$. 
            \ELSE
                \STATE Sample $\bm{\phi}^{\left(i\right)} \sim \pi(\bm{\phi} \mid \bm{x}^{\left(i\right)}, \bm{y})$ using Algorithm \ref{alg:ancestral_sampling}
            \ENDIF
            \STATE $t^{(i)} \leftarrow t^{(i-1)}+s^{(i)}$
        \ENDWHILE
    \end{algorithmic}
\end{algorithm}

Given a realization of $\bm{x}(t)$ over the interval $[0, T]$, where $T$ is the total trajectory length, the expectation of a function $\varphi: \mathrm{R}^{n} \rightarrow \mathrm{R}$ with respect to $\pi(\bm{x} \mid \bm{y})$ can be estimated using
\begin{equation}\label{eq:MC_PDMP}
\begin{aligned}
\frac{1}{T} \int_0^T \varphi(x(t)) d t =& \frac{1}{T}\bigg(\sum_{k=1}^{i-1} \int_0^{s^{(k)}} \varphi\big(x^{(k-1)}+v^{(k-1)} t\big) d t \bigg.\\
& \,\, \bigg. +\int_0^{s^{(i)}-(t^{(i)}-T)} \varphi\big(x^{(i-1)}+v^{(i-1)} t\big) dt\bigg).
\end{aligned}
\end{equation}
When $\varphi(x)=x$, we have 
$$ 
\int_0^{s^{(k)}} \varphi\left(x^{(k-1)}+v^{(k-1)} t\right) dt=x^{(k-1)}s^{(k)}+\frac{1}{2}v^{(k-1)}(s^{(k-1)})^{2}.$$
When $\varphi(x)=x^{2}$, we have
$$
\begin{aligned}
\int_0^{s^{(k)}}\varphi\left(x^{(k-1)}+v^{(k-1)} t\right)dt & = (x^{(k-1)})^{2}s^{(k)}+x^{(k-1)}v^{(k-1)}(s^{(k-1)})^{2}\\
& \quad +\frac{1}{3}(v^{(k-1)})^{2}(s^{(k-1)})^{3}.
\end{aligned}
$$
The above formulas allow us to compute both the posterior mean and the posterior standard deviation.

\section{Numerical experiment}
\label{sec:numerical}
In this section, we demonstrate the performance of the proposed algorithm by applying it to an X-ray CT image reconstruction problem using both synthetic and real world data. 
The CT inverse problem can be formulated as the linear system described in \eqref{eq:linear}, where $\bm{x} \in \mathrm{R}^{n}$ represents the vectorized image to be reconstructed and $\bm{y} \in \mathrm{R}^{m}$ denotes the measurement projection data. The system matrix $\bm{A}\in \mathbb{R}^{m \times n}$ represents the discretized Radon transform, with $m$ equals to the product of the number of detector elements and the number of projections.
All experiments are implemented in \texttt{Pytorch} with RTX4090 GPU. The codes are available at \url{https://github.com/kexiongwen/Bayesian_Linear_inverse.git}.

\subsection{Comparison}
We compare our method against three state of the art methods for Bayesian CT restoration~\cite{iordache2012total,yao2016tv,banerjee2022horseshoe}, 
which are briefly described below.
\begin{enumerate}
\item \textbf{Total variation Gaussian prior}~\cite{yao2016tv} with Bayesian inference implemented with the preconditioned Crank-Nicolson MCMC sampler (\textbf{pCN})~\cite{cotter2013mcmc}. This is a popular edge-preserving prior, enabling model the sharp jumps in the unknown, which often occur in medical images.  
\item \textbf{Fused horseshoe prior}. We notice that, in recent papers for linear inverse problem,~\cite{uribe2023horseshoe} put horseshoe prior to the increment of each pixel for edge-preserving property. They used Gibbs sampler to evaluate posterior mean as estimator. Concurrently,~\cite{dong2023inducing} put horseshoe prior to each pixel for sparsity promotion. They proposed the block coordinate descent algorithm to find the posterior mode as estimator.  Motivated by the construction of the fused bridge prior in equation (\ref{eq:fused_prior}), we put the horseshoe shrinkage to both pixel and its increment. Such prior is called fused horseshoe prior has been used in graph denoise in statistics literature~\cite{banerjee2022horseshoe} and has excellent performance in our numerical studies. Both this prior and our Fused $L_{1/2}$ prior belongs to the global-local shrinkage family~\cite{polson2014bayesian}. The corresponding posterior sampling only requires a minor modification of the Gibbs sampler(\textbf{Gibbs}) discussed in ~\cite{uribe2023horseshoe}. See Section A of the supplementary for details.
\item \textbf{Fused LASSO prior}~\cite{iordache2012total} with the corresponding posterior sampled by proximal Langevin  dynamic (\textbf{PLD})~\cite{durmus2018efficient,durmus2022proximal}. This prior is the combination of total variation prior and LASSO prior (i.e. $\gamma_{1}=\gamma_{2}=0$ in equation \ref{eq:fused_prior}). For the fused LASSO prior, rather than assigning the hyper prior, we tune the hyper parameter $\lambda_{1}$,$\lambda_{2}$ and $\lambda_{3}$ manually. When $\lambda_{3} = 0$, this prior is just total variation prior. Fused LASSO prior has log-concave density, which is required by PLD algorithm. In addition, we use the ADMM algorithm~\cite{boyd2011distributed} to solve the proximal operator involved. See Appendix D for details.
\end{enumerate}
In the later section, we refer different methods by the abbreviation of their posterior sampling approach. To ensure a fair comparison, all hyper parameters involved in the competing methods are either 
manually tuned optimally or automatically chosen as described in the reference papers.In addition, all the methods are started with same initialization.
It is important to determine the hyper parameters for both the fused $L_{1/2}$ prior and the sampling algorithm Gibbs-BPS; therefore, details for tuning these hyper parameters will be thoroughly discussed in section~\ref{sec:hyper-setting}.

Throughout all the experiments, the Gibbs sampler will be run for 5,000 iterations, the PLD will be run for 10,000 iterations, the pCN will be run for 4,000,000 iterations and the Gibbs-BPS will be run for 600,000 iterations. Unlike the traditional discrete time MCMC, the iteration of Gibbs-BPS algorithm from $i$ to $i+1$ produces the continuous time trajectory of  parameters $\bm{x}$ between the $i$th jump event at time $t_{i}$ and the $(i+1)$th jump event at time $t_{i+1}$. 
 
\subsection{Results and discussion}\label{sec:results}
\subsubsection{Case S}
\begin{figure}[htbp]
\centering
\subfloat[Shepp–Logan Phantom]{\includegraphics[width=0.4\textwidth]{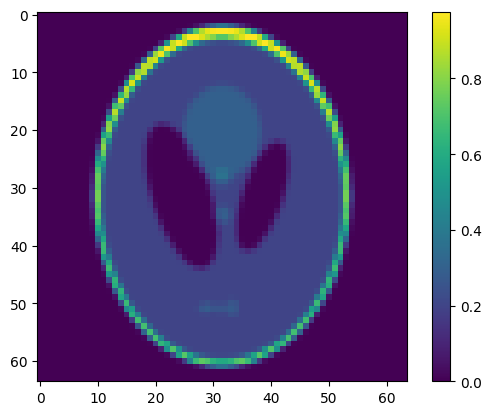}}  \hspace{10pt}
\subfloat[Grains Phantom]{\includegraphics[width=0.4\textwidth]{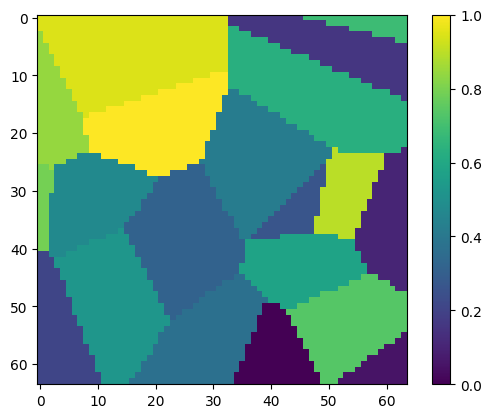}} 
\vspace{-8pt}
\caption{Two ground truth small-scale images: Shepp-Logan has many zero-valued pixels (sparse), while Grains has many non-zero pixels (dense).}
\label{fig:true_image1}
\end{figure}

First, we consider a small scale image setting with $64\times 64$ pixels used in many literature~\cite{uribe2023horseshoe,bardsley2018computational}. We tested the algorithms using the Shepp-Logan phantom image and the Grains phantom image  shown in Figure \ref{fig:true_image1}. In our simulation, we used $32$ projections equi-spatially sampled from $0$ to $\pi$. The noise are taken to be Gaussian with zero mean and $\sigma_{obs}=0.01 \times \|\bm{A}\bm{x}\|_{\infty}$ in the numerical experiments, which leads to 32.88db and 32.17db Signal-to-Noise Ratio(SNR) respectively. 

\begin{figure}[!htb]
\centering
\subfloat[Shepp–Logan Phantom]{\includegraphics[width=0.45\textwidth]{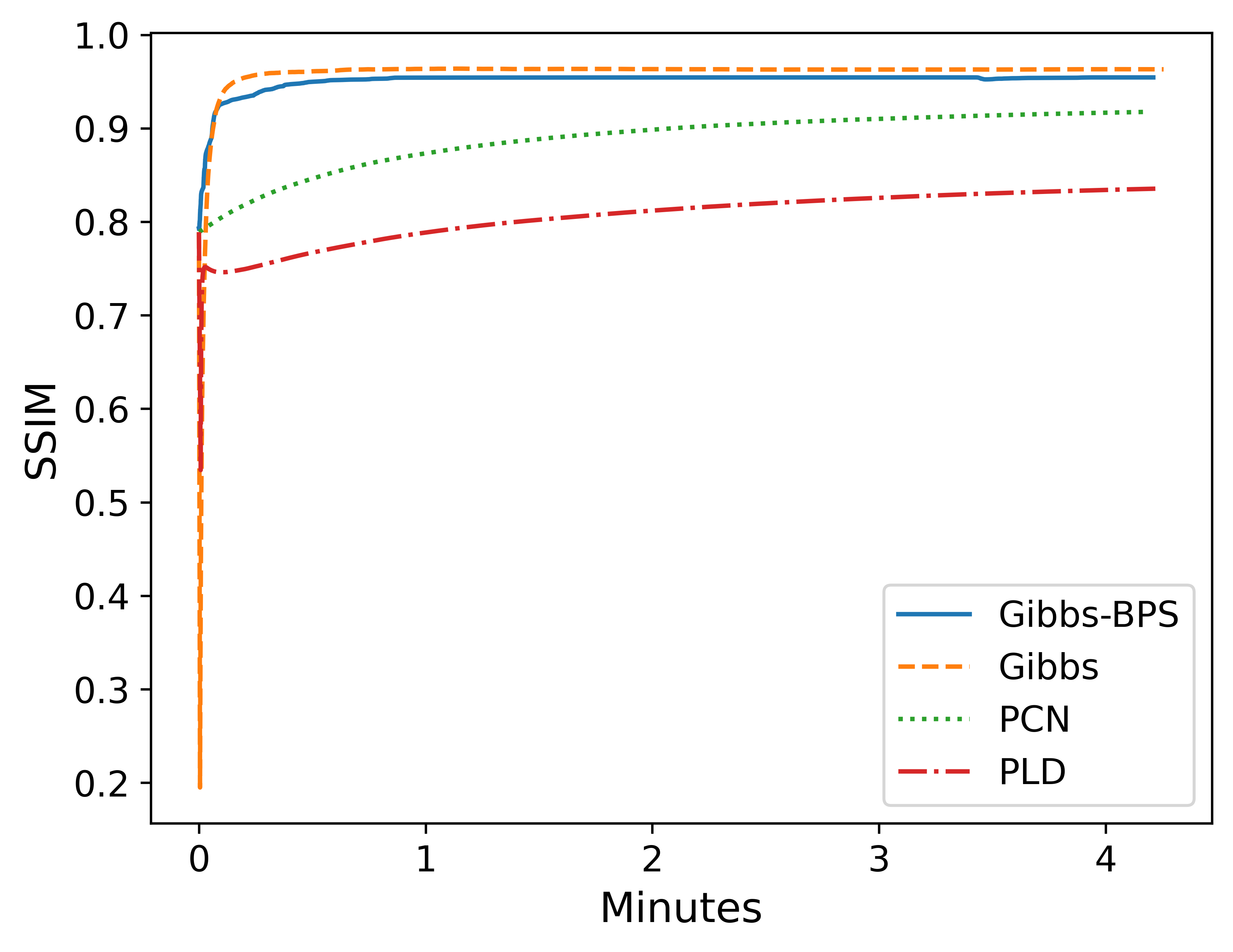}}   \hspace{10pt}
\subfloat[Grains Phantom]{\includegraphics[width=0.45\textwidth]{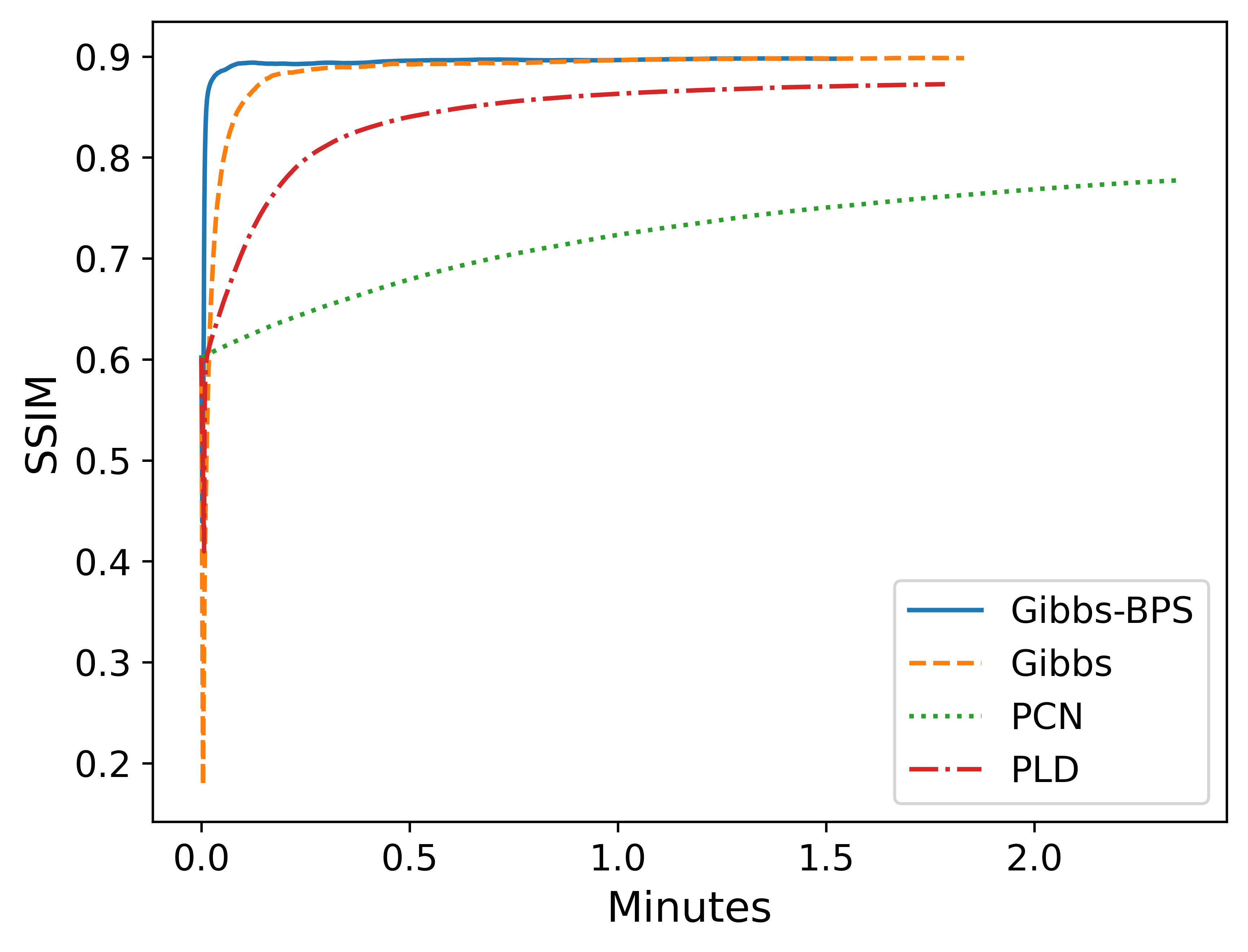}} 
\vspace{-8pt}
\caption{Comparison of convergence speed of posterior mean estimator from MCMC samplers for two small size image.}
\label{fig:speed_s}
\end{figure}

Table \ref{tab:result-case-S} reports the image recover quality of different algorithms in terms of peak signal-to-noise ratio (PSNR) and structural similarity index measure (SSIM). Their computation time per 10,000 iterations is also given in the table. It should point out that it is difficult to empirically characterize the convergence speed of a high-dimensional Markov chain. Standard MCMC diagnostics such as the effective sampler size, integrated autocorrelation time and Gelman–Rubin statistics are not suitable for approximated MCMC as they do not account for asymptotic bias. They are also not directly applicable to the Gibbs-BPS as they are calculated in the discrete setting. To evaluate the mixing speed of PDMP based sample, a common practice is to discretize $(\boldsymbol{x}(t),\boldsymbol{v}(t))$ in equation (\ref{eq:MC_PDMP}) at regular time intervals and calculate the effective sampler size per second\cite{bouchard2018bouncy,bierkens2019zig}. Since the effective sampler size does not work for
approximate MCMC, this comparison is only among Gibbs, pCN and Gibbs-BPS. In addition, we recalculate the posterior mean and record the computation time at each iteration. In Figure \ref{fig:speed_s}, we show the change of SSIM for posterior mean with respect to the accumulate computation time for different MCMC algorithms. 

From Table \ref{tab:result-case-S}, Table \ref{tab:ESS_small} and Figure \ref{fig:speed_s}, we see that, although the Gibbs sampler has the highest computation complexity at each iteration due to the require of sampling the high dimensional Gaussian distribution, it can converge with a very short chain. On the other hand, the pCN has the lowest computation complexity at each iteration, but its mixing is slow. We need to run it with a very long chain. For the PLD algorithm, its computation time at each iteration depends on the convergent speed of the ADMM solver involved. Even with the same pixel and same length of the chains, the computation time of PLD for Shepp-Logan is roughly the twice of Grains. In terms of convergence speed with respect to the posterior mean in real computation time, both Gibbs sampler and Gibbs-BPS are quite efficient for small size images. But for mixing speed of the chain, the Gibbs sampler dominates the Gibbs-BPS for small size images.

\begin{figure}[!htb]
     \centering
     \begin{subfigure}[b]{0.24\textwidth}
         \centering
         \includegraphics[width=\textwidth]{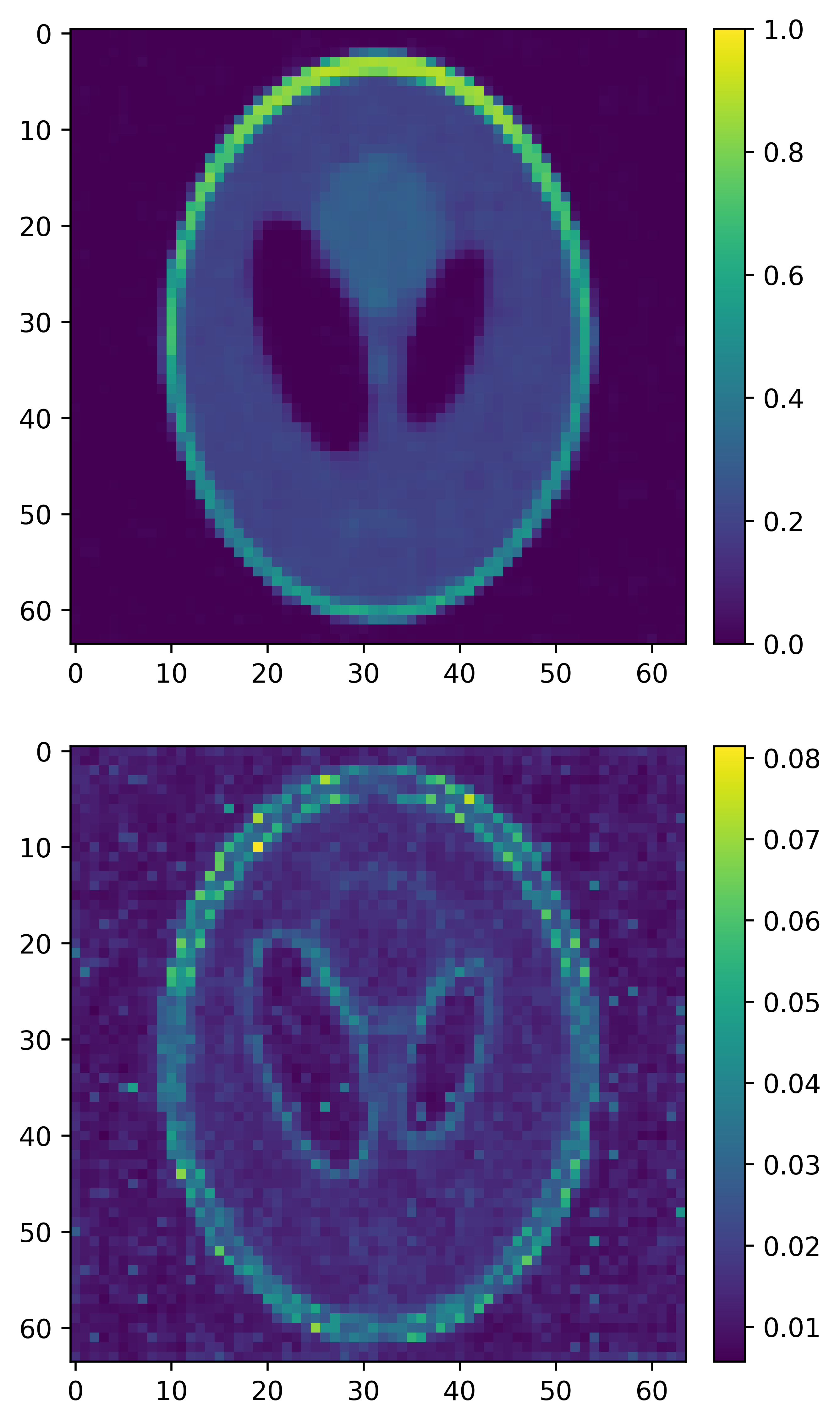}
         \subcaption{Gibbs-BPS}
     \end{subfigure}
     \hfill
     \begin{subfigure}[b]{0.24\textwidth}
         \centering
         \includegraphics[width=\textwidth]{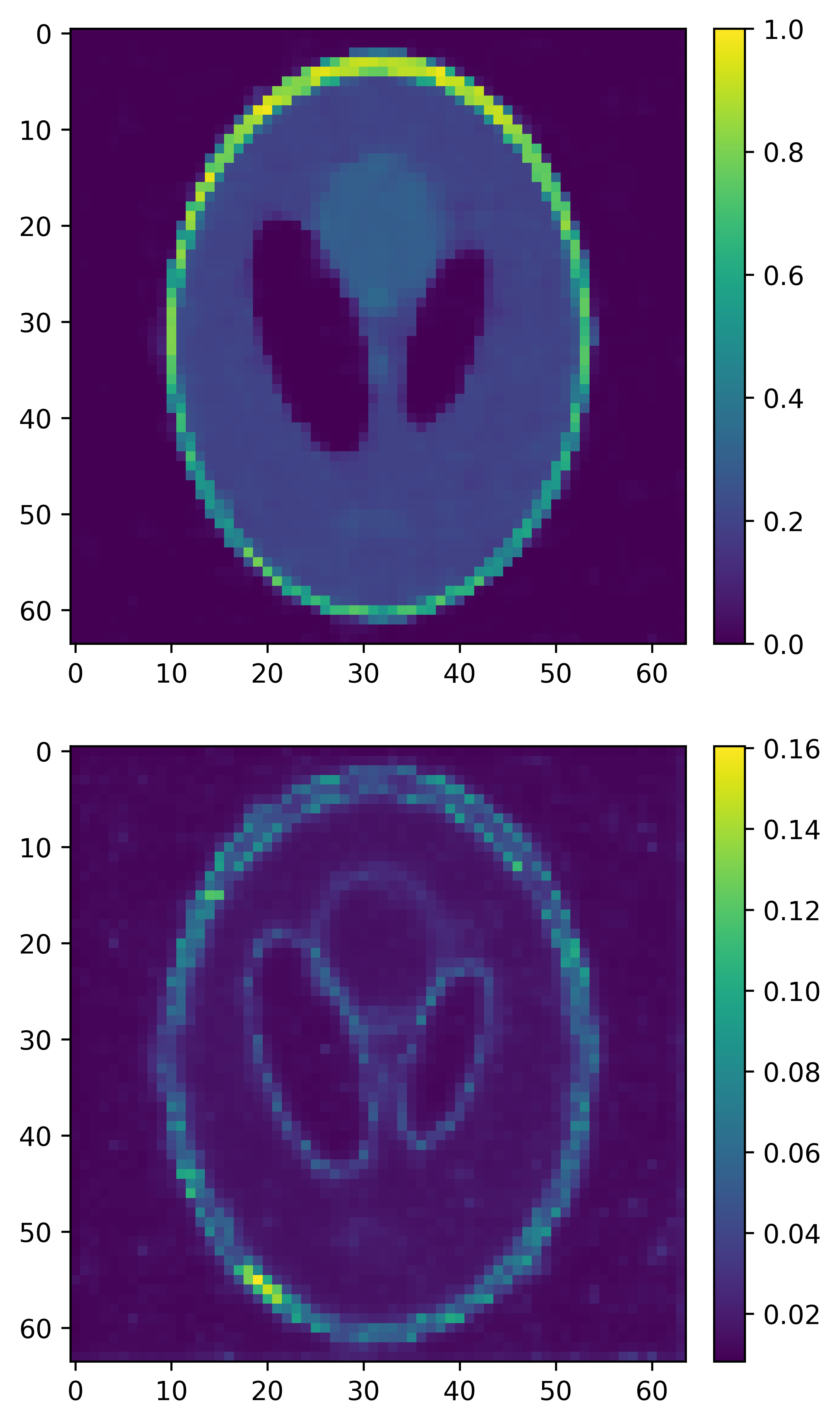}
         \caption{Gibbs}
     \end{subfigure}
     \hfill
     \begin{subfigure}[b]{0.24\textwidth}
         \centering
         \includegraphics[width=\textwidth]{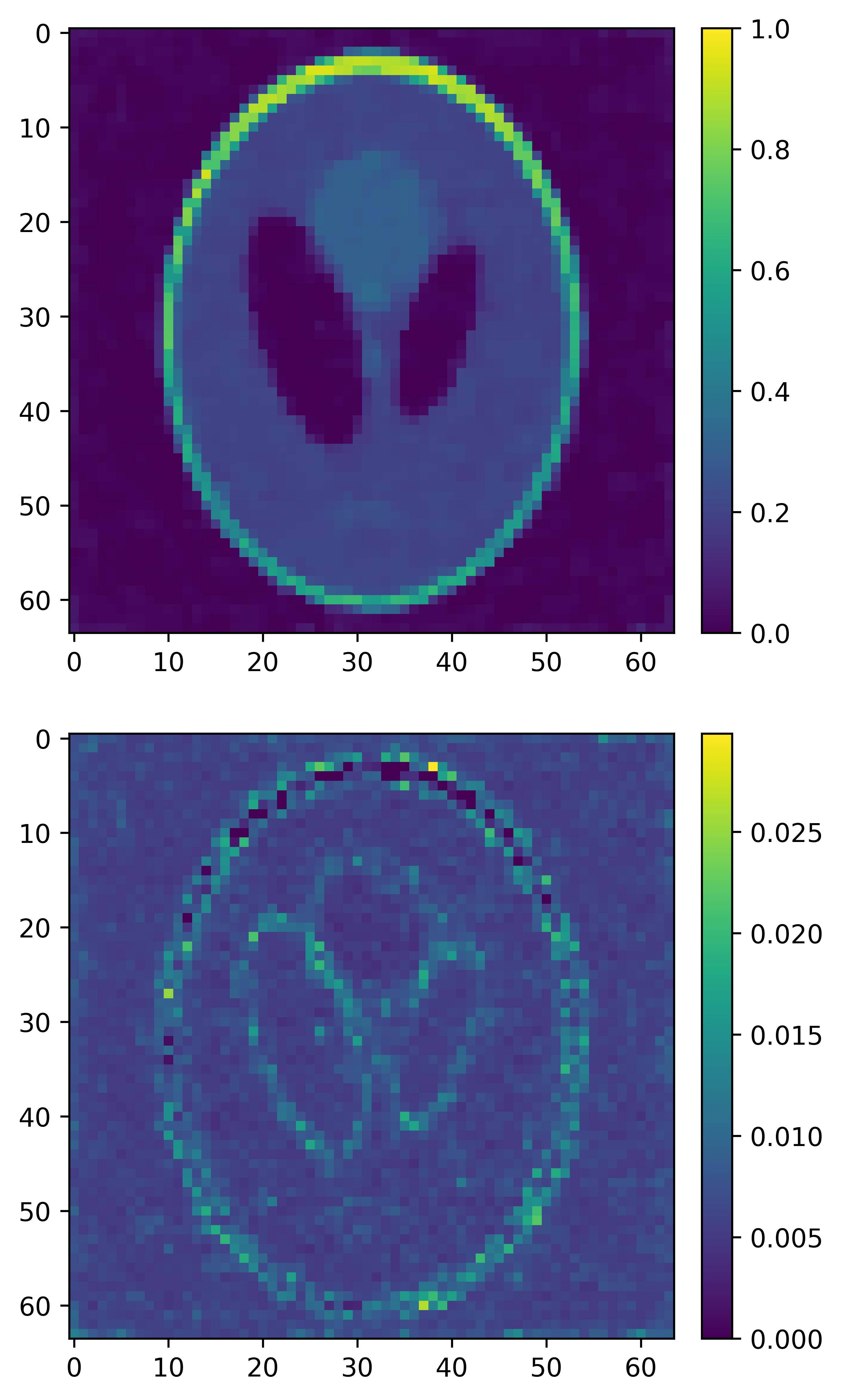}
         \caption{PLD}
     \end{subfigure}
     \hfill
     \begin{subfigure}[b]{0.24\textwidth}
         \centering
         \includegraphics[width=\textwidth]{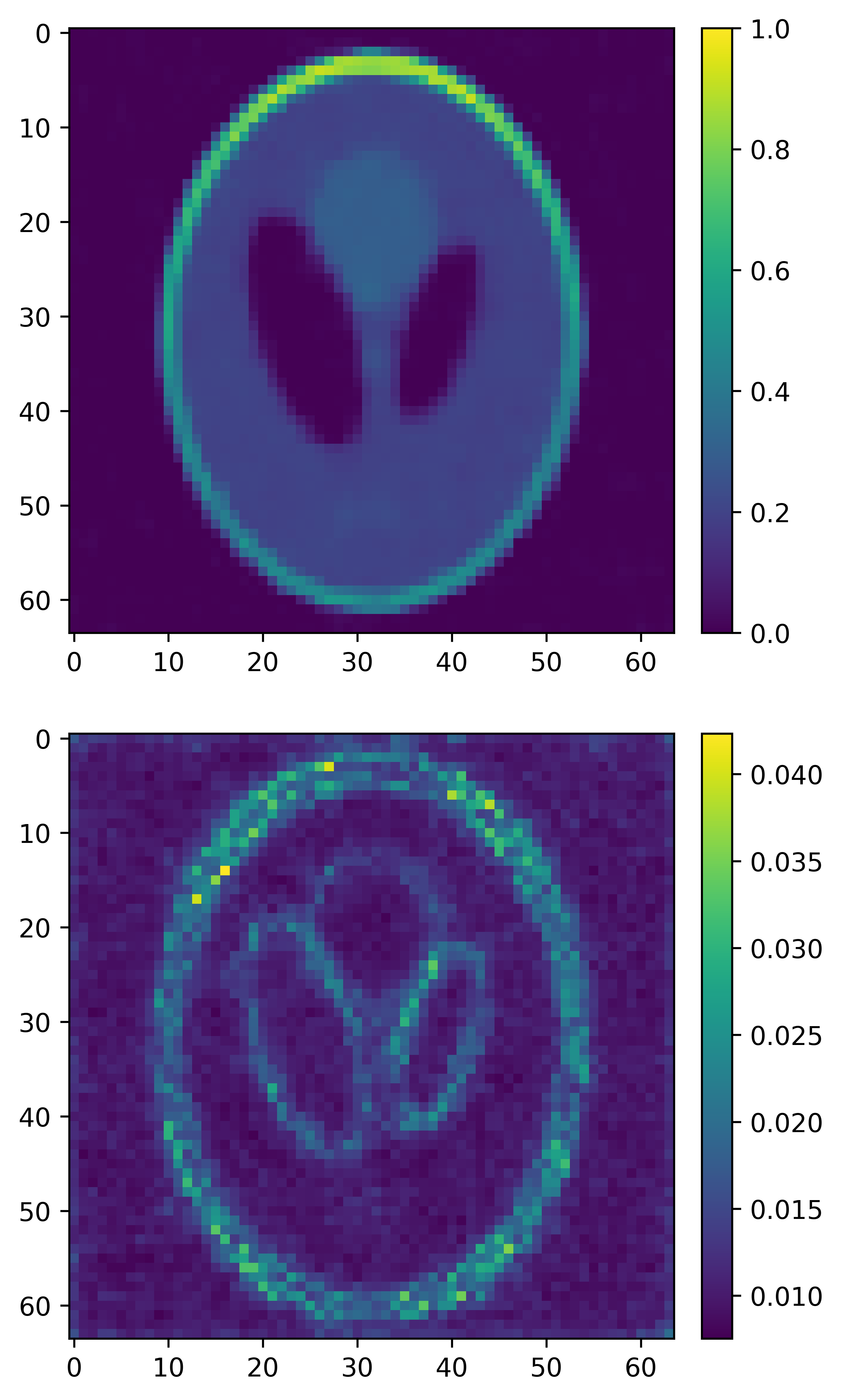}
         \caption{pCN}
     \end{subfigure}
     \vspace{-8pt}
        \caption{Comparison of CT reconstruction for Shepp–Logan phantom with different priors. The upper images are posterior mean. The bottom images are posterior standard deviations.} 
        \label{fig:shepplogan}
\end{figure}

\begin{figure}[!htb]
     \centering
     \begin{subfigure}[b]{0.24\textwidth}
         \centering
         \includegraphics[width=\textwidth]{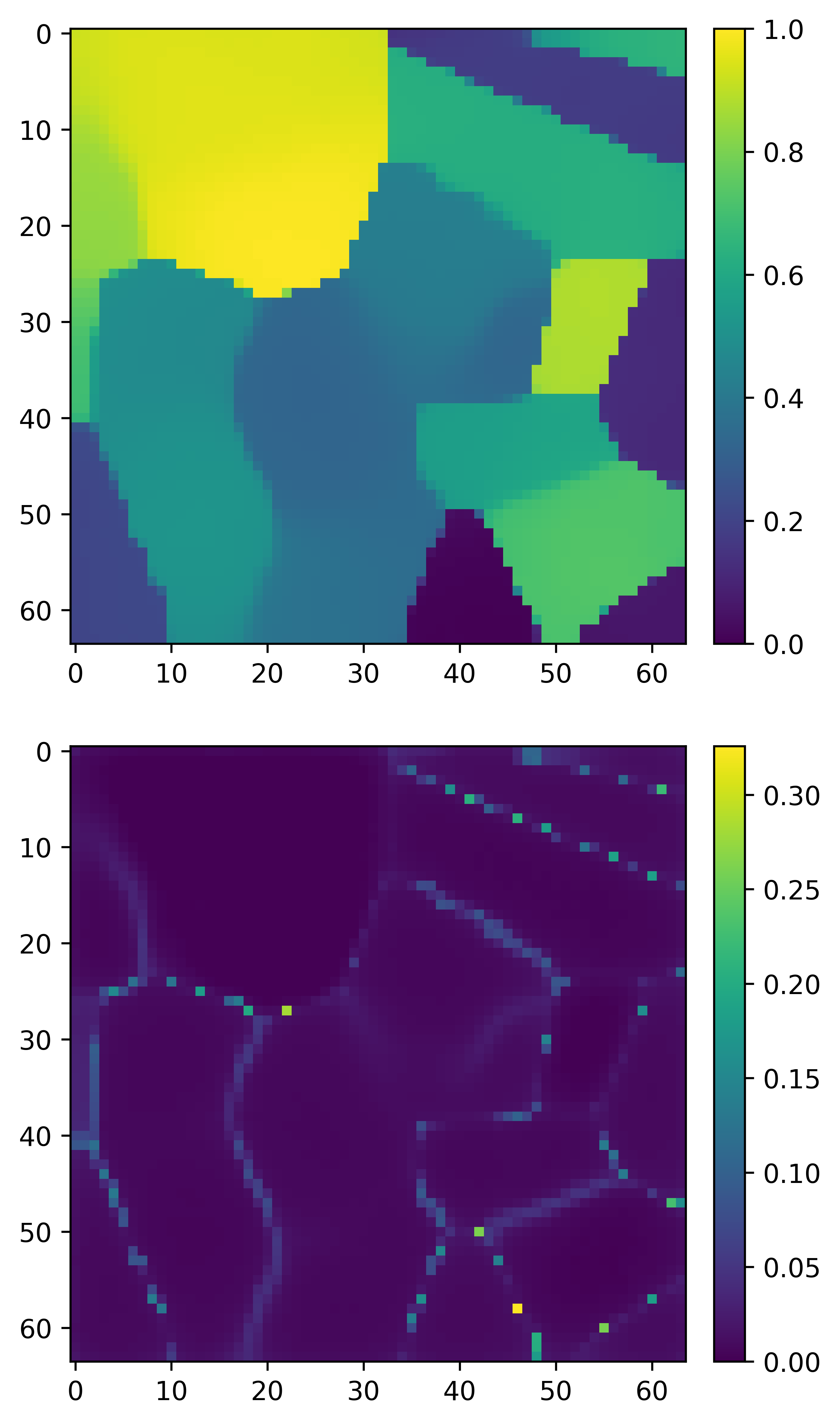}
         \subcaption{Gibbs-BPS}
     \end{subfigure}
     \hfill
     \begin{subfigure}[b]{0.24\textwidth}
         \centering
         \includegraphics[width=\textwidth]{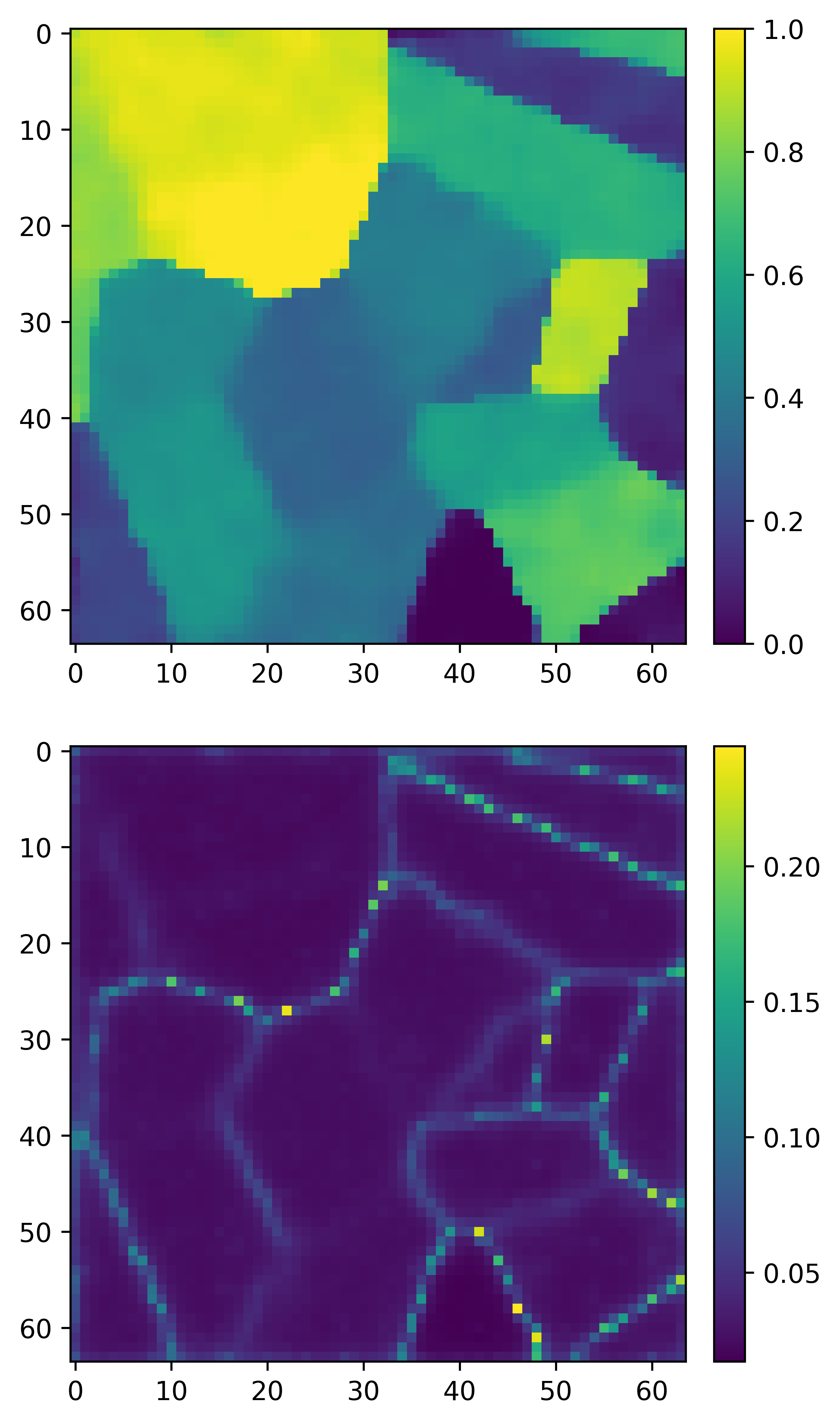}
         \caption{Gibbs}
     \end{subfigure}
     \hfill
     \begin{subfigure}[b]{0.24\textwidth}
         \centering
         \includegraphics[width=\textwidth]{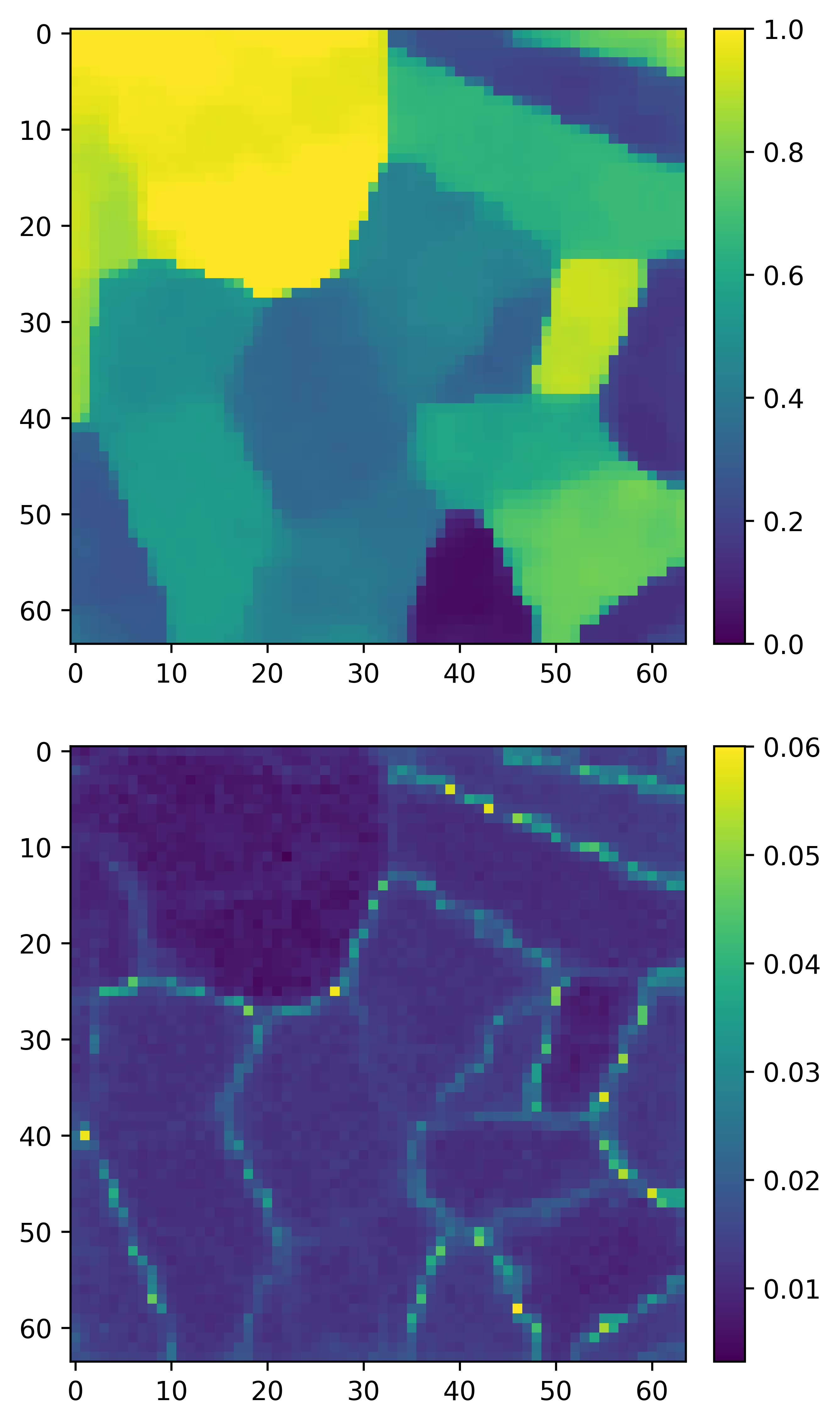}
         \caption{PLD}
     \end{subfigure}
     \hfill
     \begin{subfigure}[b]{0.24\textwidth}
         \centering
         \includegraphics[width=\textwidth]{grains_PCN_mean_std.png}
         \caption{pCN}
     \end{subfigure}
     \vspace{-8pt}
        \caption{Comparison of CT reconstruction for grains phantom with different priors. The upper images are posterior mean. The bottom images are posterior standard deviation.} 
        \label{fig:grains}
\end{figure}

The posterior mean and posterior standard deviation are shown in Figure \ref{fig:shepplogan} and Figure \ref{fig:grains}. Overall, for two small scale image problems, the fused $L_{1/2}$ prior and the fused horseshoe prior have comparable performances in terms of both PSNR and SSIM. The images recovered by Total variation Gaussian prior always has the worst quality. The fused LASSO prior works reasonably well in Shepp–logan phantom. The main performance difference among all the algorithms is in Grains phantom, both the fused LASSO prior and the TV-Gaussian prior significantly fall behind. The fused $L_{1/2}$ prior allows us to obtain a sharper reconstruction in Grains phantom, despite some of the grain features missing in the reconstruction and its PSNR is slightly lower than the fused horseshoe prior.

For uncertainty quantification, we observe that, all the posterior standard deviations are relatively large at the edge locations and almost zero in the rest of the image. Among all these priors, the posterior standard deviations based on the fused LASSO prior are particularly small in these two cases. 

\begin{table}[!htb]
\centering
\caption{ Quantitative results (PSNR and SSIM) and computation times for every 10 000 samples of different MCMC algorithms run in \texttt{Pytorch} with RTX4090 GPU.}
\label{tab:my-table}
\begin{tabular}{cccclccc}
\hline
\multirow{2}{*}{} & \multicolumn{3}{c}{Shepp-Logan} &  & \multicolumn{3}{c}{Grains} \\ \cline{2-4} \cline{6-8} 
                  & PSNR     & SSIM   & Time(min)   &  & PSNR   & SSIM  & Time(min) \\ \hline
Gibbs-BPS         & 31.20    & 0.96   & 0.16        &  & 27.11  & 0.90  & 0.17      \\
Gibbs~\cite{banerjee2022horseshoe}             & 31.52    & 0.96   & 18.35       &  & 27.95  & 0.90  & 18.40     \\
PLD~\cite{durmus2018efficient}               & 31.37    & 0.85   & 21.11       &  & 24.43  & 0.90  & 10.20     \\
pCN~\cite{yao2016tv}               & 28.13    & 0.92   & 0.06        &  & 23.72  & 0.83  & 0.06      \\ \hline
\end{tabular}
\label{tab:result-case-S}
\end{table}

\begin{table}[!htb]
\centering
\caption{The mean, median, maximum and minimum of ESS per second for Gibbs-BPS, Gibbs sampler and Preconditioned Crank–Nicolson cross all the pixels in the Shepp-Logan and Grains.}
\begin{tabular}{ccccc|cccc}
\hline
     & \multicolumn{4}{c|}{\begin{tabular}[c]{@{}c@{}}Shepp-Logan\\ ($64 \times 64$)\end{tabular}} & \multicolumn{4}{c}{\begin{tabular}[c]{@{}c@{}}Grains\\ ($64 \times 64$)\end{tabular}} \\ \cline{2-9} 
     & Mean                 & Median                 & Max                   & Min                 & Mean                & Median               & Max                 & Min                \\ \hline
Gibbs-BPS & 3.24                 & 3.08                   & 5.09                  & 0.63                & 4.57                & 4.98                 & 7.64                & 0.37               \\ \hline
Gibbs~\cite{banerjee2022horseshoe}   & 6.11                 & 6.18                   & 14.07                 & 0.04                & 6.12                & 6.25                 & 13.64               & 0.08               \\ \hline
pCN~\cite{yao2016tv} & 0.67                 & 0.59                   & 3.66                  & 0.11                & 0.45                & 0.41                 & 1.82                & 0.05               \\ \hline
\end{tabular}
\label{tab:ESS_small}
\end{table}

\subsubsection{Case L}
\begin{figure}[!htb]
     \centering
     \begin{subfigure}[b]{0.32\textwidth}
         \centering
         \includegraphics[width=\textwidth]{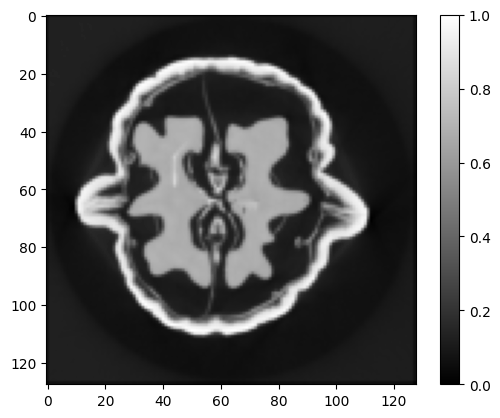}
         \subcaption{Walnut Phantom}
     \end{subfigure}
     \hfill
     \begin{subfigure}[b]{0.32\textwidth}
         \centering
         \includegraphics[width=\textwidth]{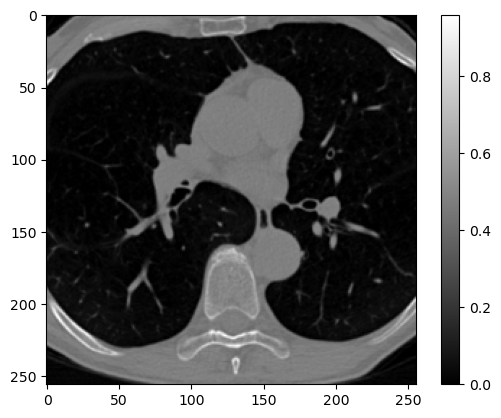}
         \caption{Lung 1}
     \end{subfigure}
     \hfill
     \begin{subfigure}[b]{0.32\textwidth}
         \centering
         \includegraphics[width=\textwidth]{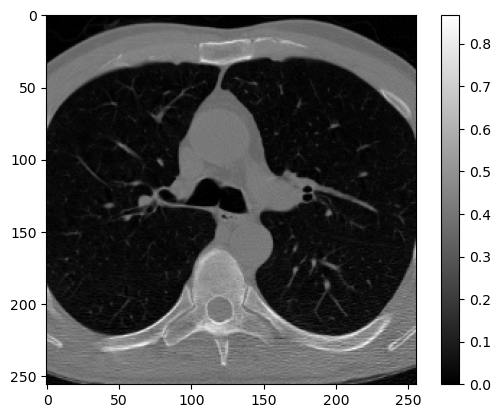}
         \caption{Lung 2}
     \end{subfigure}
     \vspace{-8pt}
        \caption{The three ground truth large scale images.}
        \label{fig:true_image2}
\end{figure}

We consider a large scale image setting with Walnut phantom image ~\cite{hamalainen2015tomographic} of size $128\times 128$ and two lung CT images of size $256 \times 256$ taken from the LoDoPaB-CT dataset~\cite{leuschner2021lodopab} to verify the scalability of the Gibbs-BPS sampler. The images are shown in Figure \ref{fig:true_image2}. In the simulation, we used $64$ projections equi-spatially sampled from $0$ to $\pi$ for Walnut Phantom and $128$ projections equi-spatially sampled from $0$ to $\pi$ for two lung CT images. The noise are taken to be Gaussian with zero mean and $\sigma_{obs}=0.01 \times \|\boldsymbol{A}\boldsymbol{x}\|_{\infty}$ for Walnut Phantom and $\sigma_{obs}=0.02 \times\frac{ \|\bm{A}\bm{x}_{true}\|_{2}}{\sqrt{m}}$ for two lung CT images. These setting leads to 32.77db SNR for Walnut Phantom and 34db for two lung CT images.

Table~\ref{tab:result-case-L} demonstrates that Gibbs-BPS consistently outperforms all other methods across datasets in both reconstruction quality and speed. While the smaller $128\times128$ Walnut dataset shows modest improvements of 0.01~dB, the more challenging $256\times256$ Lung~1 medical images achieve substantial gains of 1.42~dB. Notably, Gibbs-BPS requires only 0.18 to 2.18 minutes for reconstruction compared to PLD's 20 to 132 minutes, delivering a speedup of up to $100\times$ with no loss in quality.

\begin{table}[!htb]
\centering
\caption{Quantitative results (PSNR and SSIM) and computation times for every 10 000 samples of different MCMC algorithms run in \texttt{Pytorch} with RTX4090 GPU.  }
\resizebox{\textwidth}{!}{%
\begin{tabular}{lccclccclccc}
\toprule[0.8pt]
          & \multicolumn{3}{c}{Walnut} &  & \multicolumn{3}{c}{Lung 1} &  & \multicolumn{3}{c}{Lung 2} \\ \cline{2-4} \cline{6-8} \cline{10-12} 
          & PSNR   & SSIM  & Time(min) &  & PSNR   & SSIM  & Time(min) &  & PSNR   & SSIM  & Time(min) \\ \hline
Gibbs-BPS & 27.91  & 0.92  & 0.18      &  & 32.55  & 0.83  & 2.16      &  & 31.11  & 0.73  & 2.18      \\
Gibbs~\cite{banerjee2022horseshoe}    & 27.90  & 0.92  & 425.15    &  & NA     & NA    & NA        &  & NA     & NA    & NA        \\
PLD ~\cite{durmus2022proximal}  & 27.90  & 0.92  & 20.03     &  & 31.13  & 0.83  & 128.86    &  & 30.15  & 0.73  & 132.94    \\
pCN ~\cite{yao2016tv}  & 27.39  & 0.92  & 0.25      &  & NA     & NA    & NA        &  & NA     & NA    & NA        \\ 
\bottomrule[0.8pt]
\end{tabular}%
}
\label{tab:result-case-L}
\end{table}

\begin{figure}[!htb]
     \centering
     \begin{subfigure}[b]{0.32\textwidth}
         \centering
         \includegraphics[width=\textwidth]{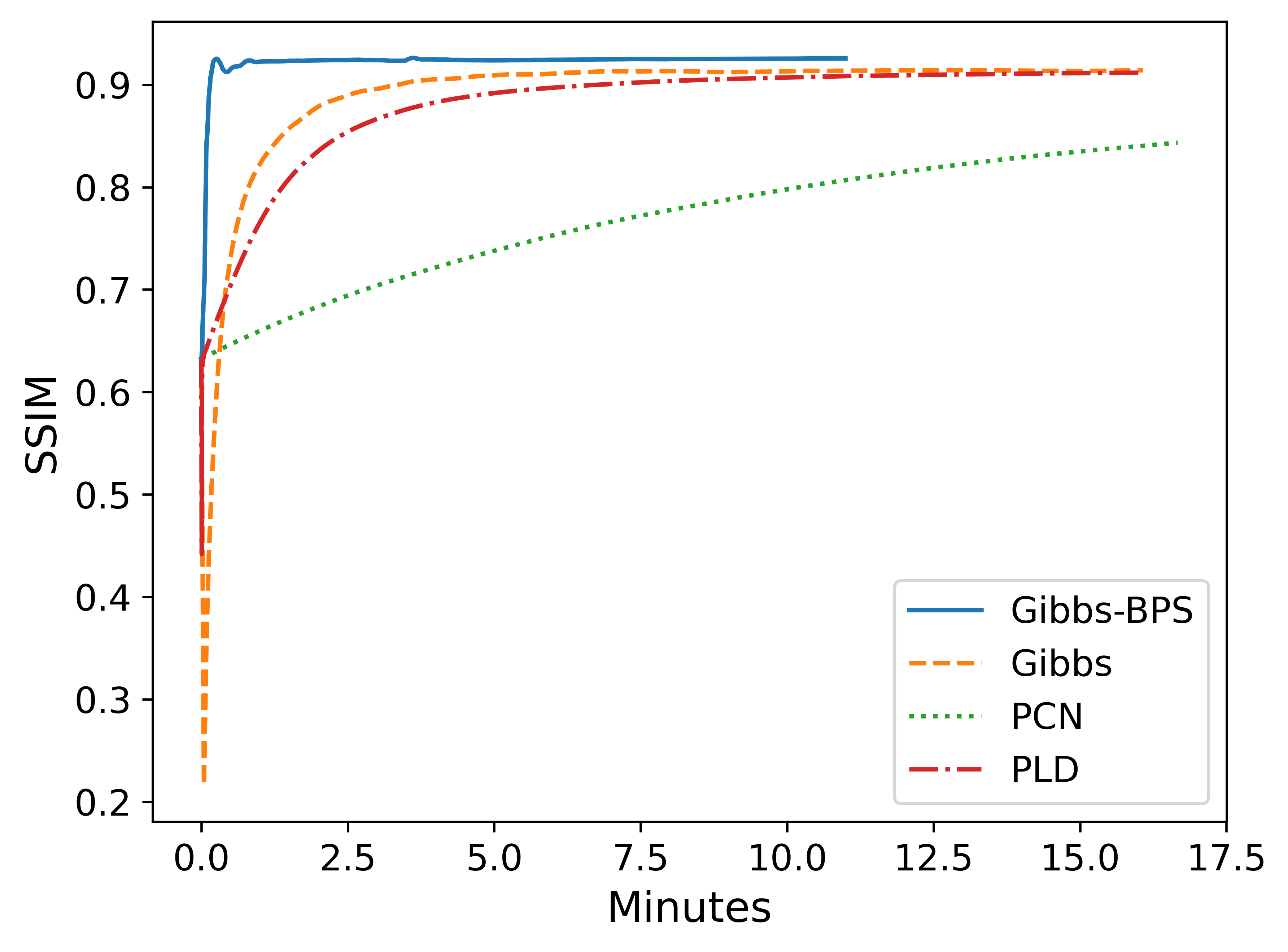}
         \subcaption{Walnut Phantom}
     \end{subfigure}
     \hfill
     \begin{subfigure}[b]{0.32\textwidth}
         \centering
         \includegraphics[width=\textwidth]{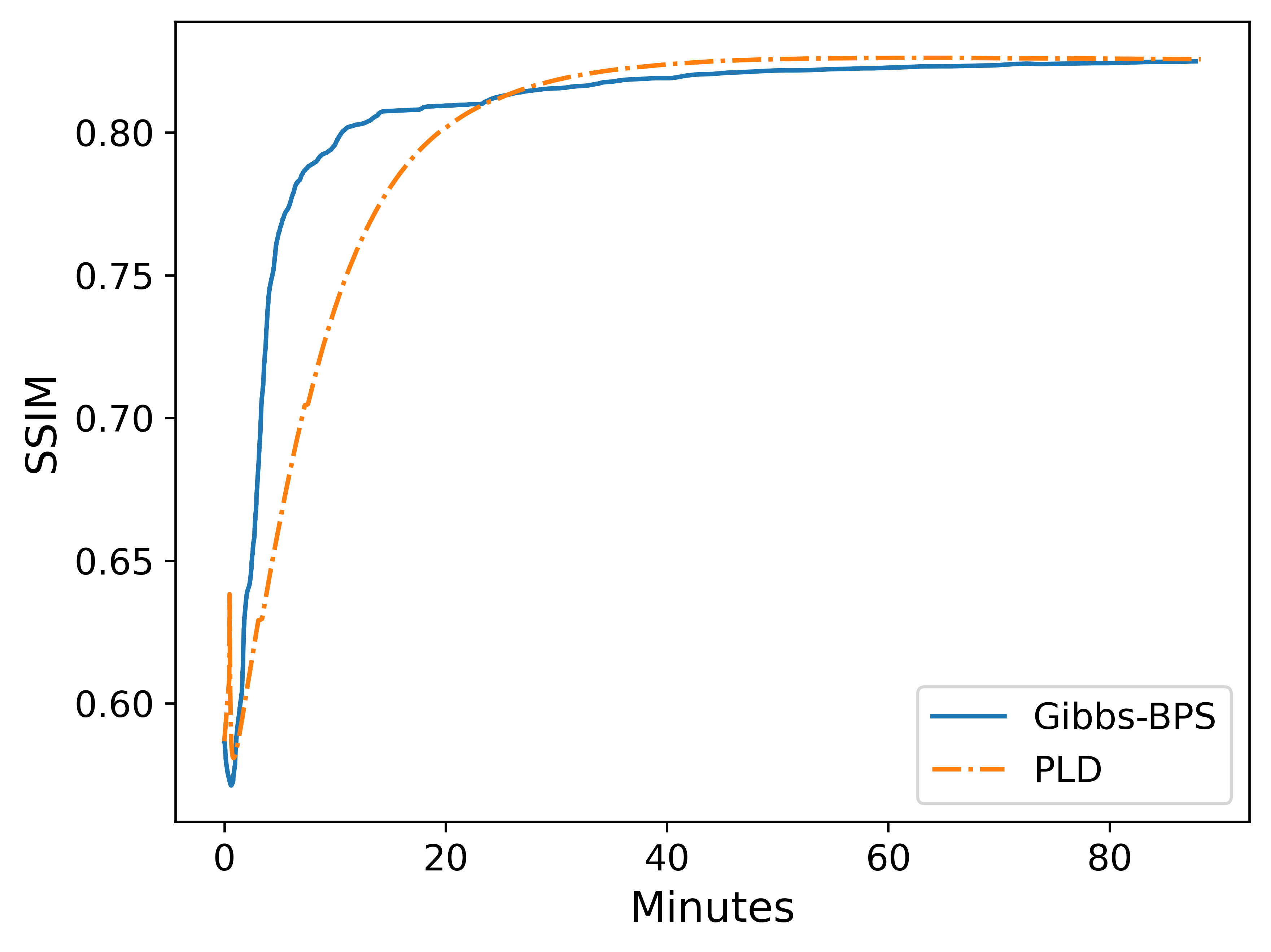}
         \subcaption{Lung 1}
     \end{subfigure}
     \hfill
     \begin{subfigure}[b]{0.32\textwidth}
          \centering
          \includegraphics[width=\textwidth]{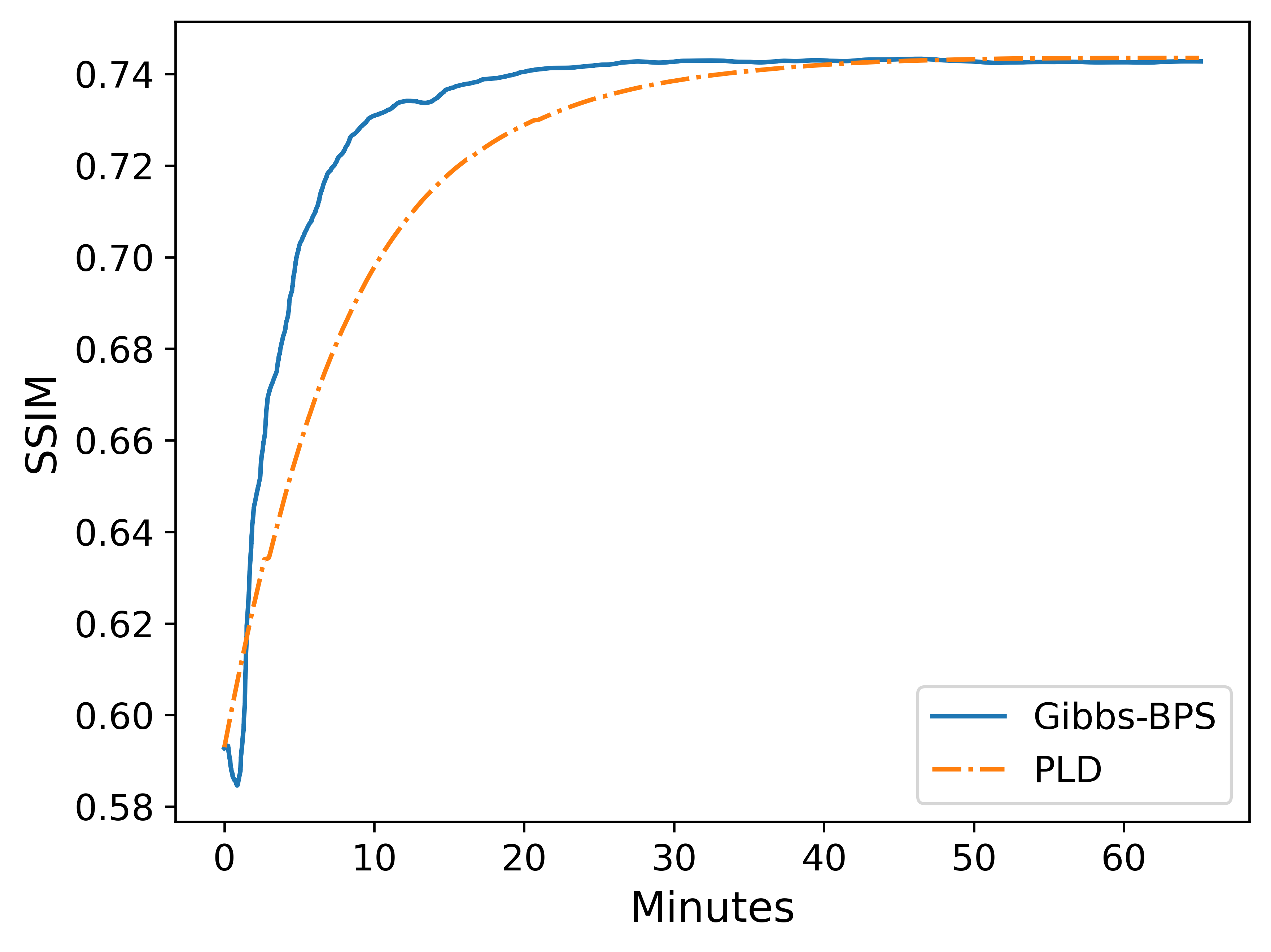}
          \caption{Lung 2}
      \end{subfigure}
      \vspace{-8pt}
        \caption{Comparison of convergence speed of MCMC algorithms for three large size images. For a clear vision, we only plot the first 2500 iterations of Gibbs sampler in Walnut Phantom image.}
        \label{fig:speed_L}
\end{figure}

\begin{table}[!htb]
\centering
\caption{The statistics of ESS per second cross all the pixels in the Wallnut.}
\begin{tabular}{ccccc}
\hline
\multicolumn{5}{c}{Wallnut($128 \times 128$)} \\ \hline
        & Mean   & Median   & Max    & Min    \\ \hline
Gibbs-BPS    & 2.52   & 2.83     & 3.81   & 0.22   \\ \hline
Gibbs~\cite{banerjee2022horseshoe}      & 0.42   & 0.39     & 1.41   & 0.06   \\ \hline
pCN~\cite{yao2016tv}     & 0.15   & 0.12     & 0.94   & 0.03   \\ \hline
\end{tabular}
\label{tab:ESS_large}
\end{table}

\begin{figure}[!htb]
     \centering
     \begin{subfigure}[b]{0.24\textwidth}
         \centering
         \includegraphics[width=\textwidth]{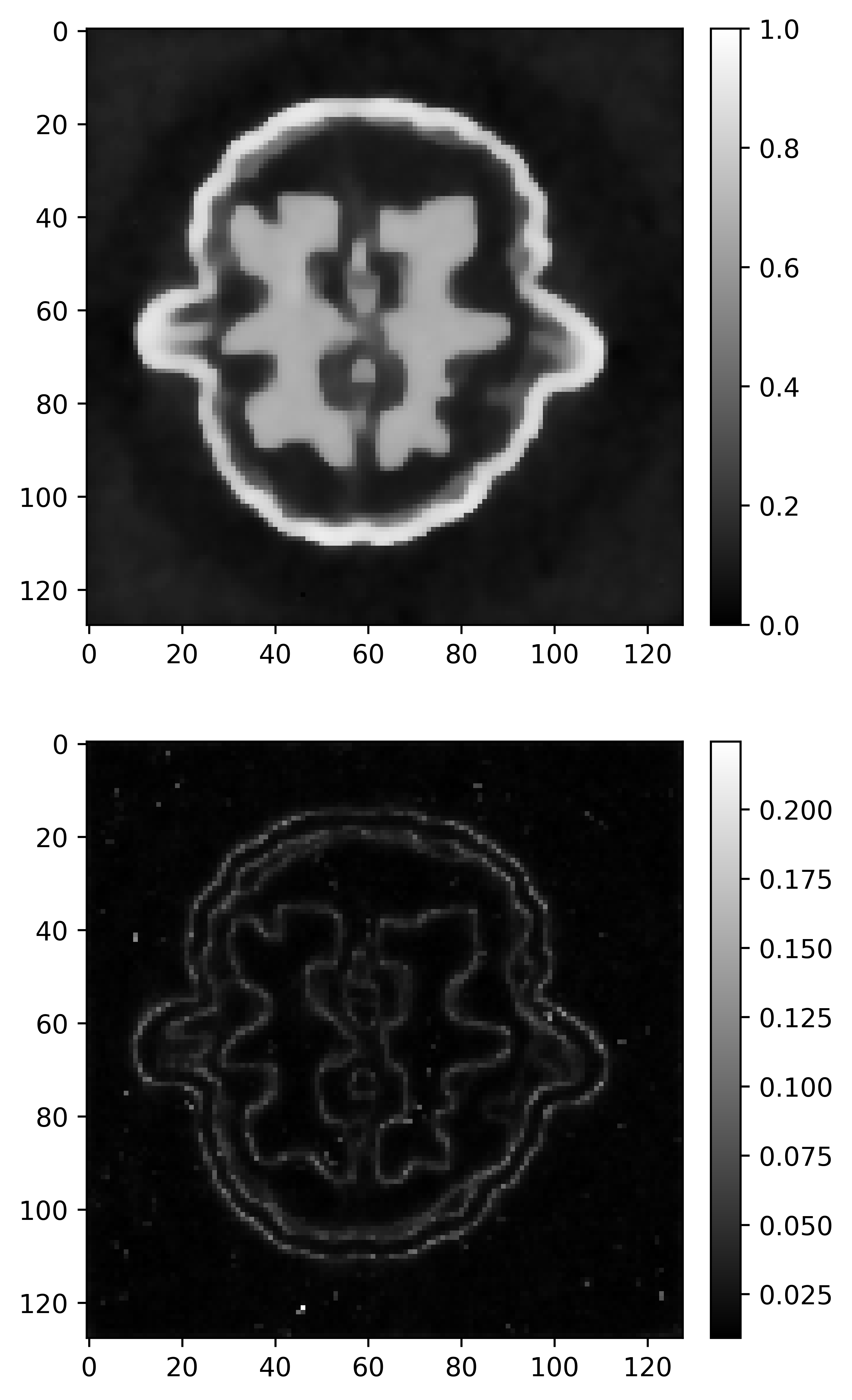}
         \subcaption{Gibbs-BPS}
     \end{subfigure}
     \hfill
     \begin{subfigure}[b]{0.24\textwidth}
         \centering
         \includegraphics[width=\textwidth]{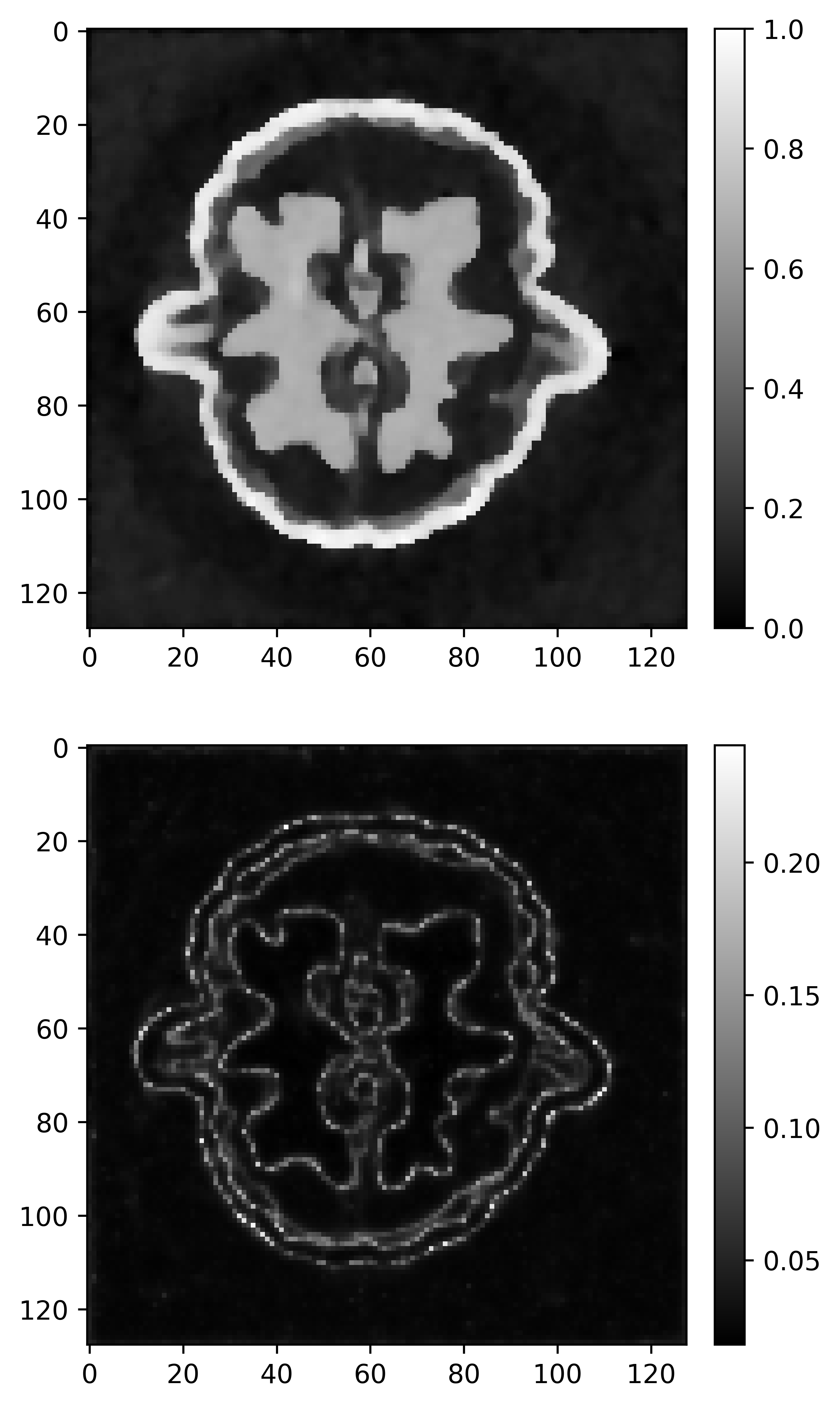}
         \caption{Gibbs}
     \end{subfigure}
     \hfill
     \begin{subfigure}[b]{0.24\textwidth}
         \centering
         \includegraphics[width=\textwidth]{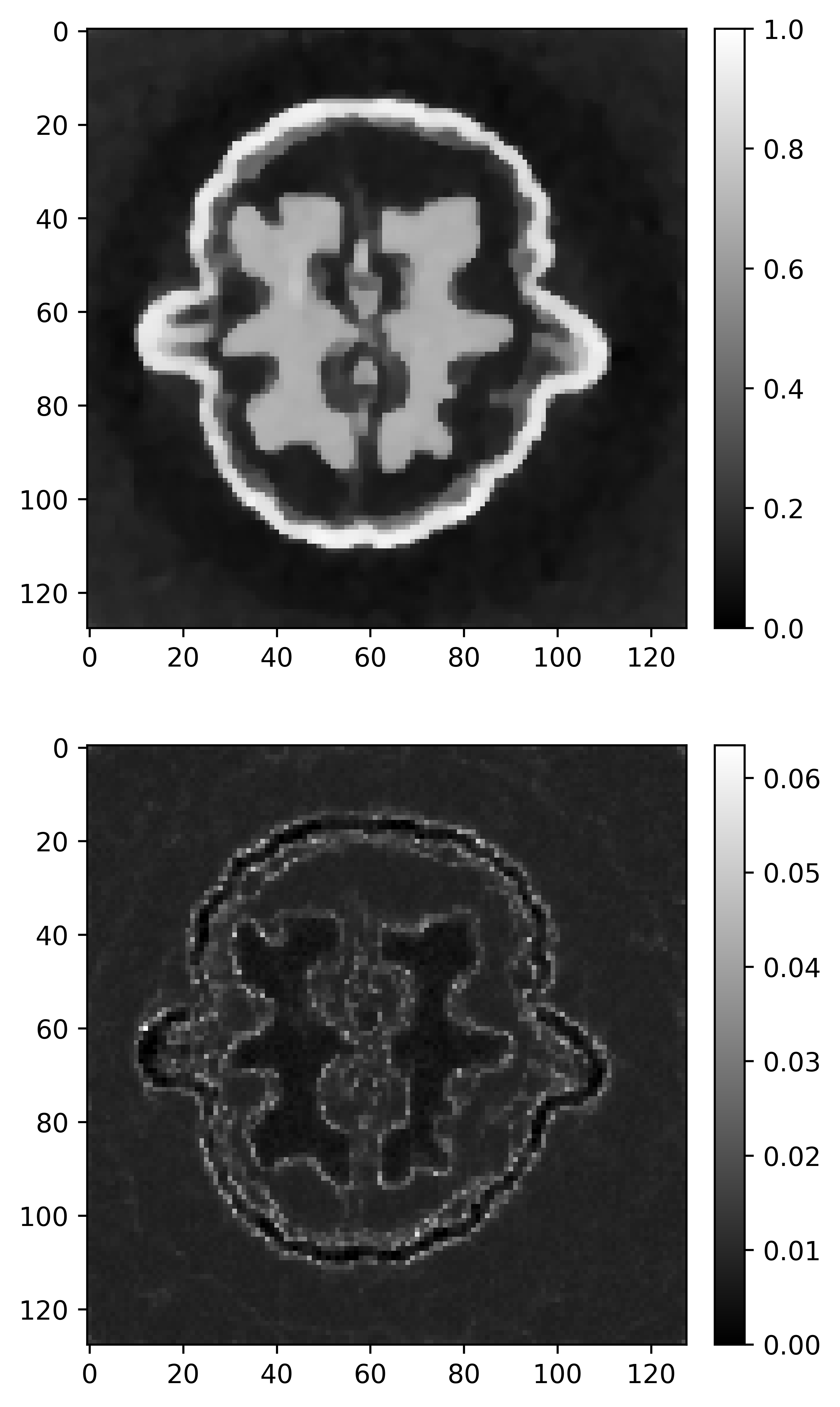}
         \caption{PLD}
     \end{subfigure}
     \hfill
     \begin{subfigure}[b]{0.24\textwidth}
         \centering
         \includegraphics[width=\textwidth]{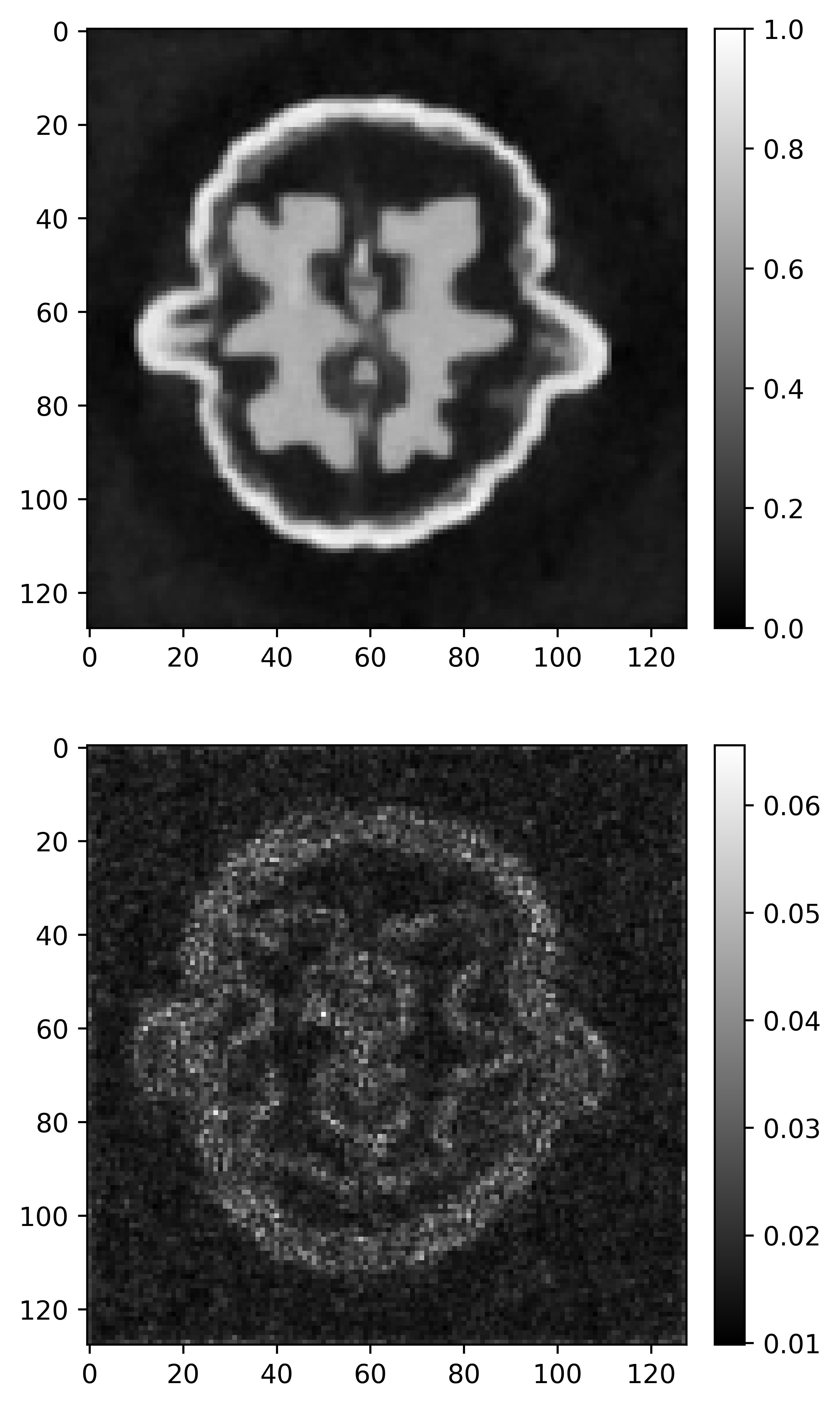}
         \caption{pCN}
     \end{subfigure}
     \vspace{-8pt}
        \caption{Comparison of CT reconstruction for walnut phantom with different priors. The upper images are posterior mean. The bottom images are posterior standard deviation.}
        \label{fig:walnut}
\end{figure}

\begin{figure}[!htb]
     \centering
     \begin{subfigure}[b]{0.4\textwidth}
         \centering
         \includegraphics[width=\textwidth]{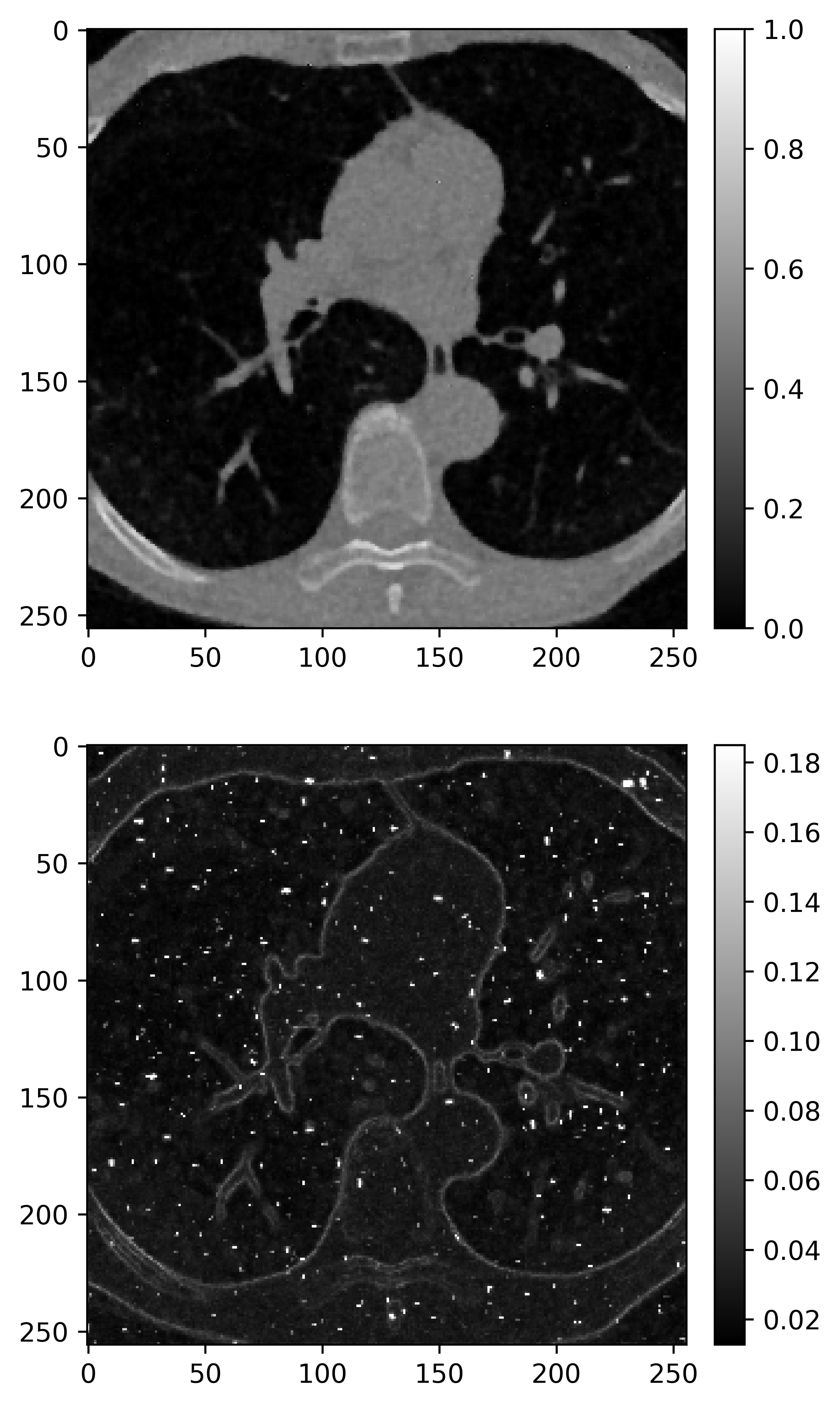}
         \subcaption{Gibbs-BPS}
     \end{subfigure}
     \begin{subfigure}[b]{0.41\textwidth}
         \centering
         \includegraphics[width=\textwidth]
         {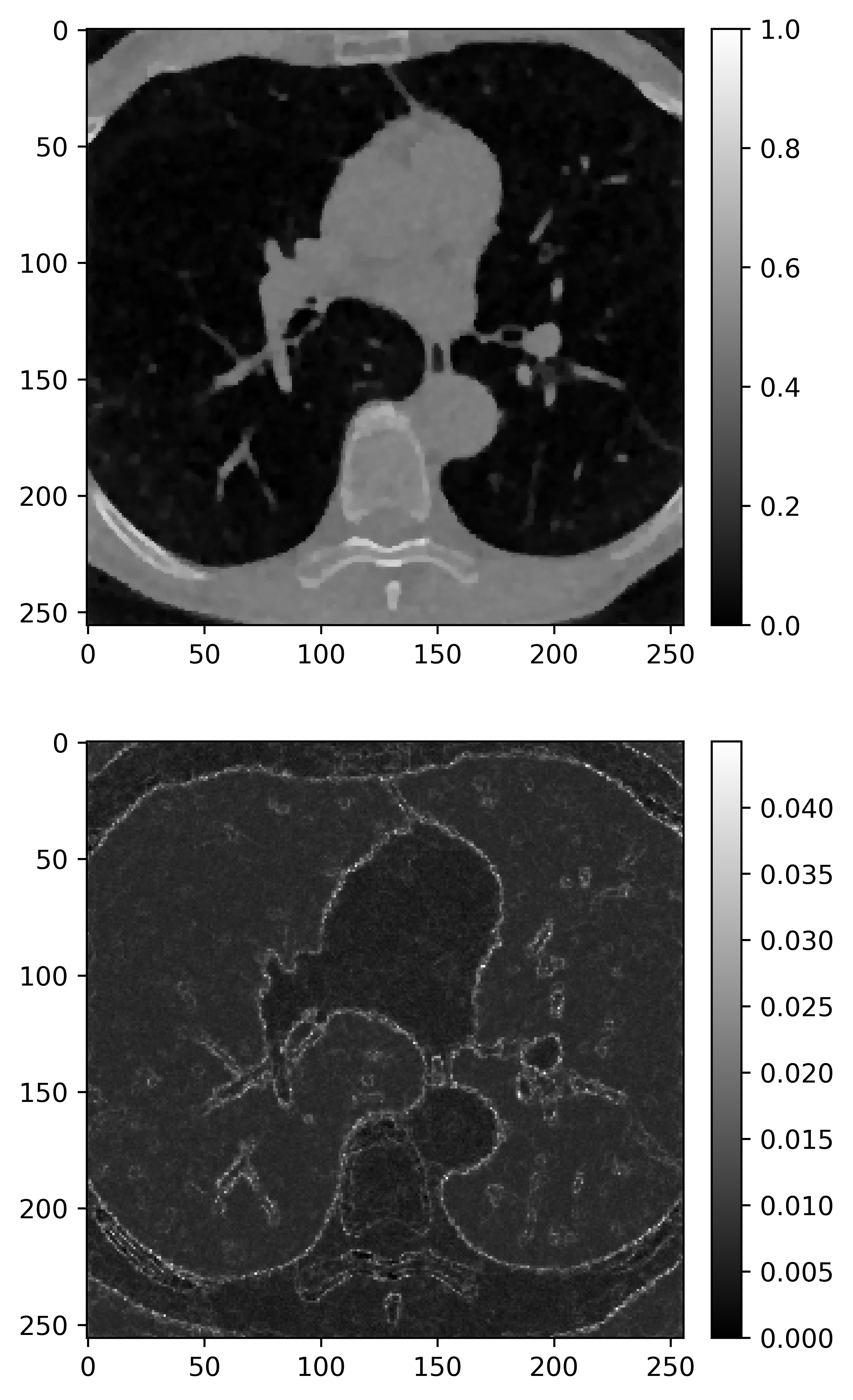}
         \caption{PLD}
     \end{subfigure}
\caption{Comparison of CT reconstruction for Lung CT Image 1 with fused $L_{1/2}$ prior and fused LASSO prior. The upper images are posterior mean. The bottom images are posterior standard deviations.}
\label{fig:imageA}
\end{figure}

\begin{figure}[!htb]
     \centering
     \begin{subfigure}[b]{0.4\textwidth}
         \centering
         \includegraphics[width=\textwidth]{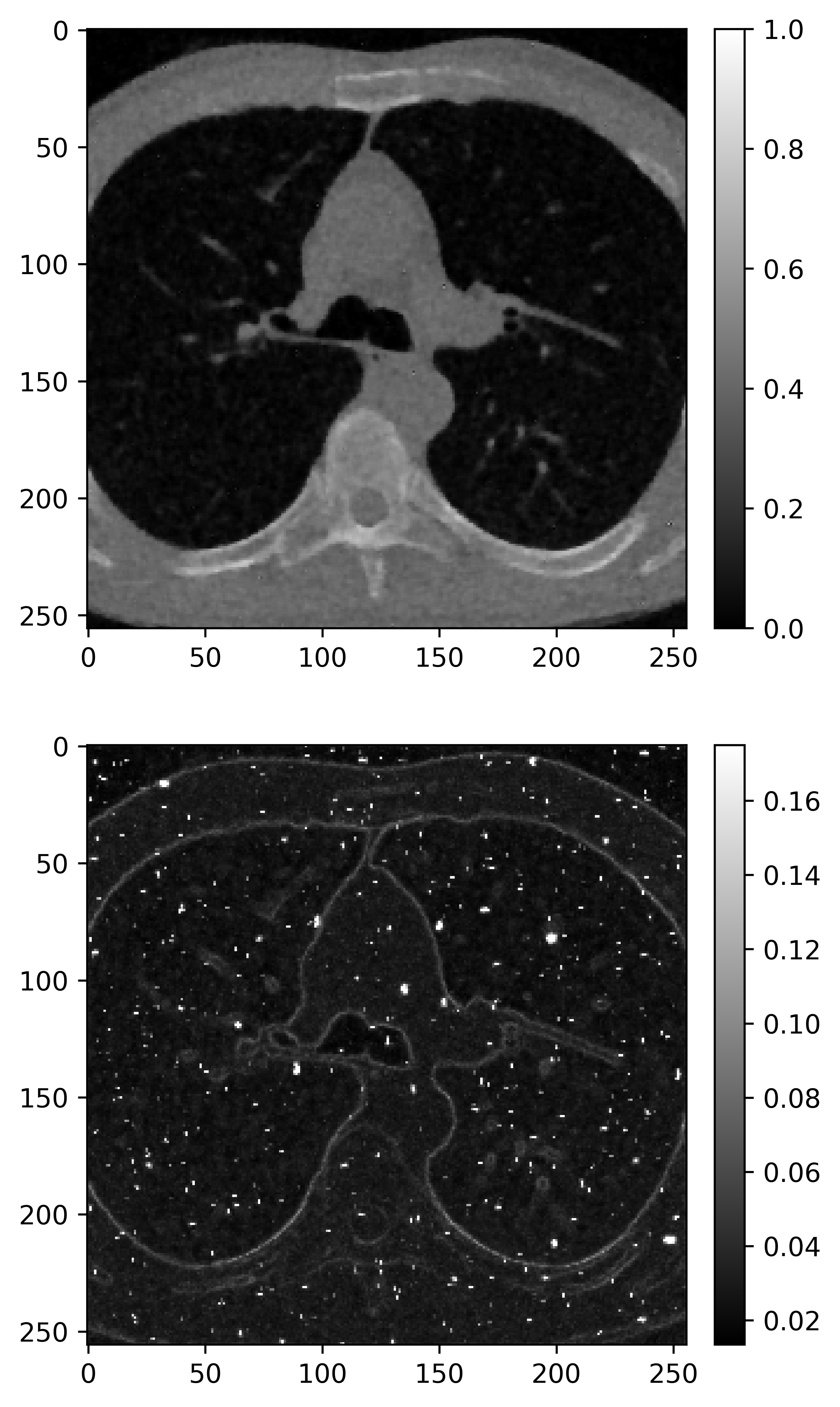}
         \subcaption{Gibbs-BPS}
     \end{subfigure}
     \begin{subfigure}[b]{0.4\textwidth}
         \centering
         \includegraphics[width=\textwidth]
         {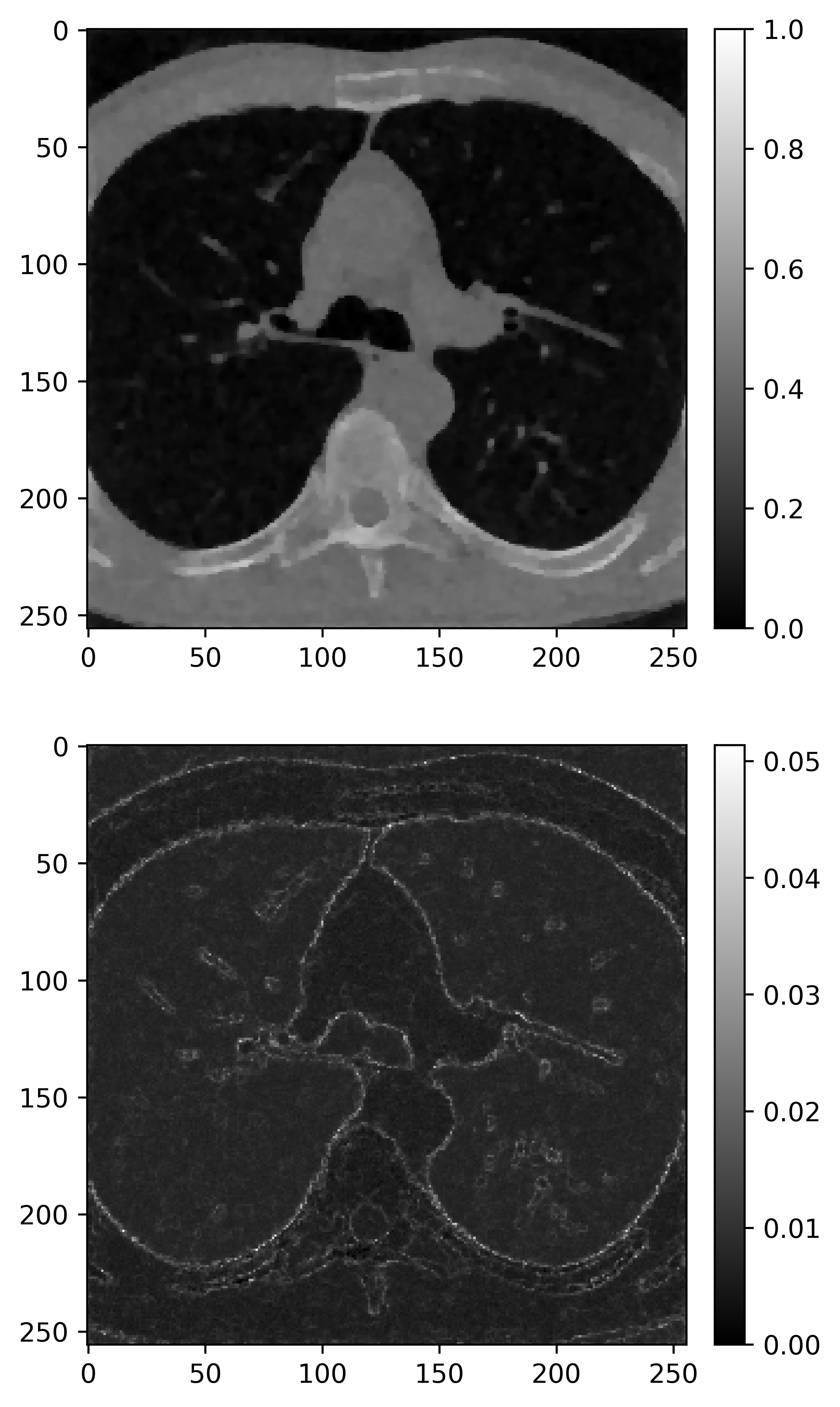}
         \caption{PLD}
     \end{subfigure}
    \caption{Comparison of CT reconstruction for Lung CT Image 2 with fused $L_{1/2}$ prior and fused LASSO prior. The upper images are posterior mean. The bottom images are posterior standard deviations.}
    \label{fig:imageB}
\end{figure}

Table \ref{tab:ESS_large} and Figure \ref{fig:speed_L} compare the convergence speed of MCMC algorithms for large-scale images.  We see that when dealing with high-resolution images, the advantages of Gibbs-BPS become evident. It is the most efficient for Walnut Phantom with $128 \times 128$ pixel, then followed by Gibbs sampler and PLD. The pCN has the slowest convergence speed. For two lung CT images with $256 \times 256$ pixel, the Gibbs sampler does not work due to the inability to sample the Gaussian distribution with dimensions $256^{2}$, and we also found that the pCN suffers from numerical instability. The proximal Langevin dynamic is the only competitor for the problem with size. In this case, the Gibbs-BPS converges slightly faster than PLD. Figures \ref{fig:walnut}-\ref{fig:imageB} show the posterior statistics for the three high resolution images recovered by various methods. For Walnut Phantom, the image recovered by all the methods except for edge-preserving horseshoe prior have the similar quality. For two lung CT images, the fused $L_{1/2}$ prior always did slightly better than the fused LASSO prior in terms of PSNR. The posterior standard deviation estimated by the PLD algorithm is much smaller than Gibbs-BPS. Since this phenomenon consistently holds for all the scenarios for PLD, we suspect that the PLD may underestimate the posterior standard deviations.

\subsection{Hyper parameters setting}\label{sec:hyper-setting}
Finally, we briefly discuss how to tune the hyper parameters in our methods. There are two types of hyper parameters. One is the hyper parameters in the fused $L_{1/2}$ prior, the other is the hyper parameters in the Gibbs-BPS algorithm.
\subsubsection{Hyper parameters in the fused $L_{1/2}$ prior}
\begin{figure}[htbp]
\centering
\subfloat[$\gamma_{1}=1, \gamma_{2}=1$]{\includegraphics[width=0.32\textwidth]{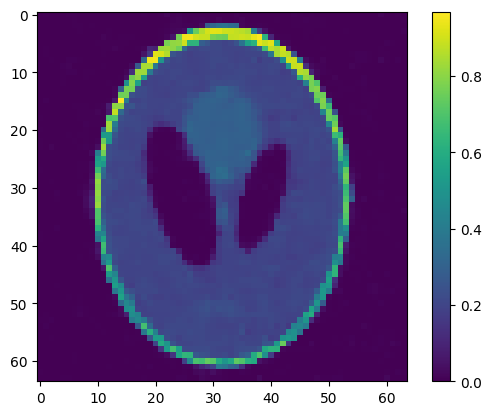}}  \hspace{10pt}
\subfloat[$\gamma_{1}=2, \gamma_{2}=1$]{\includegraphics[width=0.32\textwidth]{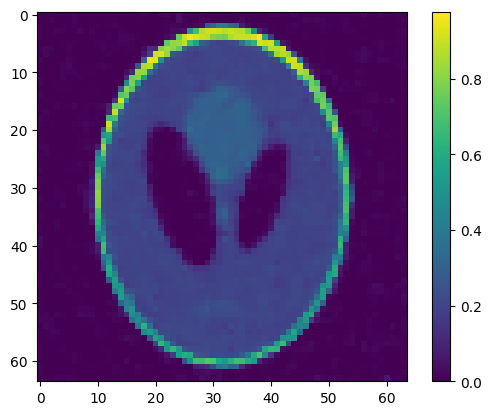}} 
\subfloat[$\gamma_{1}=3, \gamma_{2}=1$]{\includegraphics[width=0.32\textwidth]{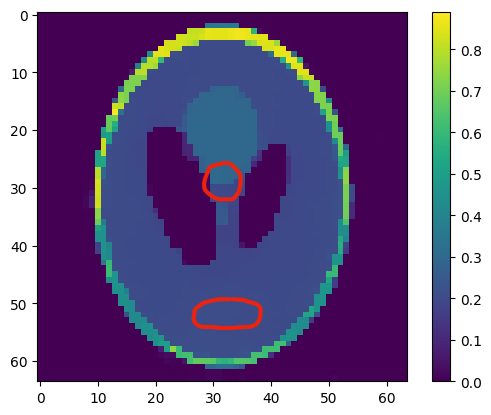}} 

\subfloat[$\gamma_{1}=0, \gamma_{2}=2$]{\includegraphics[width=0.32\textwidth]{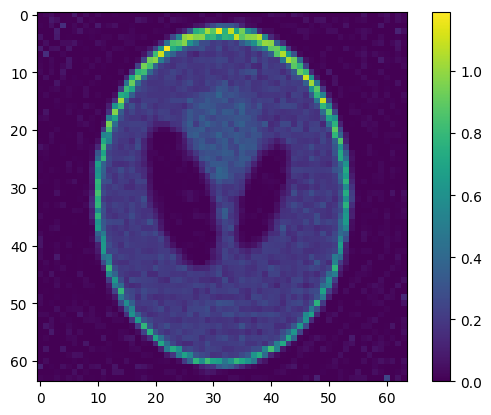}}  \hspace{10pt}
\subfloat[$\gamma_{1}=1, \gamma_{2}=2$]{\includegraphics[width=0.32\textwidth]{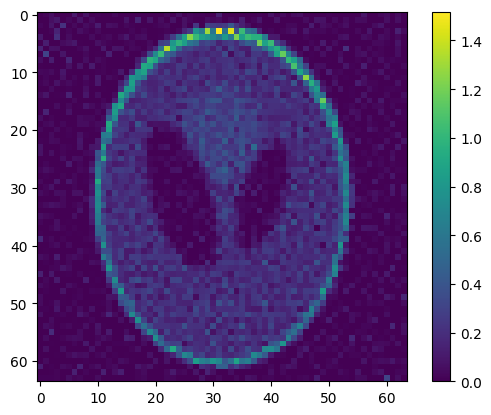}} 
\subfloat[$\gamma_{1}=2, \gamma_{2}=2$]{\includegraphics[width=0.32\textwidth]{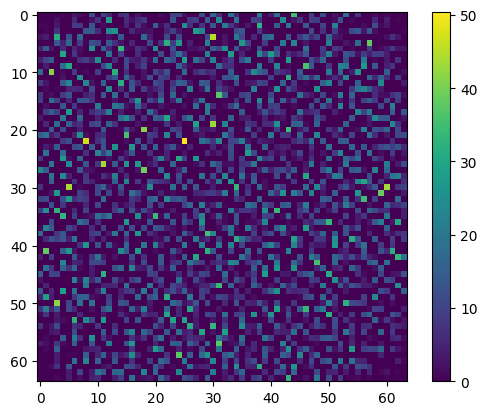}} 
\vspace{-8pt}
\caption{Comparison of CT reconstruction for Shepp–Logan Phantom image using fused $L_{1/2}$ prior with different $\gamma_{1}$ and $\gamma_{2}$. The red cycles in (c) highlight the loss of details.}
\label{fig:comparison}
\end{figure}

We first discuss the choice of $\gamma_1, \gamma_2$. We show that the algorithm can sample the posterior of fused $L_{1/2}$ prior with $\gamma_{1},\gamma_{2} \in \mathrm{N}_{0}$. However, the recovered image is often very blurred, when we set $\gamma_{2} \geq 2$ for edge-preserving terms in equation (\ref{eq:fused_prior}). For the sparsity-promoting term, we found that  $\gamma_{1}=1$ always outperforms $\gamma_{1}=0$. This is within our expectation as it was both theoretically and empirically shown by ~\cite{castillo2015bayesian,song2023nearly} that the LASSO prior is not optimal for high dimensional sparse regression in terms of posterior contraction rate, while recently, ~\cite{ke2021bayesian} showed that the bridge prior with $0 < \alpha  \leq 1/2$ has nearly optimal posterior contraction rate. But, for image problem, we found that when $\gamma_{1} \geq 3\,\,(\text{i.e.} \,\,\alpha_{1} \leq \frac{1}{8} )$, it is quite easy for the recovered image to loose details. Figure~\ref{fig:comparison} demonstrates these phenomenons with Shepp–Logan phantom in CT reconstruction problem. When we try to test them in $256 \times 256$ image, we found that for either $\gamma_{1} \geq 2$ or $\gamma_{2} \geq 2$, the algorithm suffers from numerical instability issues and fails to mix. This is because a large value of $\gamma_{1}$ and $\gamma_{2}$ will lead to the regularize term close to $L_{0}$ norm. In this case, the posterior parameter space is very rugged. This will impact the mixing of the MCMC sampler, who uses the gradient information. Therefore, we recommend set $\gamma_{1} = 1$ and tune $\gamma_{2} \in \left\{0,1\right\}$.  

We also provide a heuristic way to tune $a_{1}, b_{1}, a_{2},b_{2}, a_{3},b_{3}$ in $L_{1/2}$ prior. The hyper parameters $\lambda_{2}$ and $\lambda_{3}$ control the global shrinkage effect of horizontal increments and vertical increments, respectively, and we set their hyper priors identical. Thus, $a_{2}=a_{3}$ and $b_{2}=b_{3}$. Now, we provide a heuristic way to tune their values. Since the conditional posterior of $\lambda_{1}$ is 
$$\lambda_{1} \mid \bm{x} \sim \mathrm{Gamma}\Big(2^{\gamma_{1}}d^{2}+a_{1},\sum_{i,j}\left|x_{ij}\right|^{\frac{1}{2^{\gamma_{1}}}}+b_{1}\Big).$$
Its conditional posterior mean and mode are around $\frac{2^{\gamma_{1}}d^{2}+a_{1}}{\sum_{i,j}\left|x_{ij}\right|^{\frac{1}{2^{\gamma_{1}}}}+b_{1}}$. By default, we set $a_{1}=b_{1}=1$. In this case, the effect of the prior is weak and determining the value of $\lambda_{1}$ in MCMC is fully data driven. To tilt up the value of $\lambda_{1}$, we should increase $a_{1}$ and fixed $b_{1}=1$. To tilt down the value of $\lambda_{1}$, we should fixed $a_{1}=1$ and increase $b_{1}$. The same strategy can also be used to tune $a_{2}$ and $b_{2}$.

\subsubsection{Hyper parameters in the Gibbs-BPS algorithm}
It was shown by ~\cite{bouchard2018bouncy} that for some target distributions, the bouncy particle sampler can be reducible. This implies that there may be parts of the state space that the BPS cannot reach. To address this issue, they introduce a refresh events occur as events of an independent Poisson process of constant rate $\lambda_{\mathrm{ref}}$, and at a refresh event we simulate a new velocity from $\mathcal{N}\left(0_{n},I_{n}\right)$. ~\cite{bouchard2018bouncy} argued that a small value of refresh rate can lead to a failure to visit certain state space, while a large value leads to a random walk behavior, which gives negative impact of the mixing speed of the chain. Table \ref{tab:ESS_ref} confirmed this argument.  But for safety, we still stick with $\lambda_{ref} = 10$ as the default setting. This is because without it, if the initialization is not good, it may be possible that part of the state can not be reached by the Gibbs-BPS.

As for the event rate $\eta$, it determines the frequency of the Gibbs-BPS algorithm to update the global and local shrinkage parameters. From Table \ref{tab:ESS_eta}, we found that increasing the value of $\eta$ from $0$ leads to improve the mixing speed, beyond $\eta=100$ the improvement wear off. Thus, by default we set $\eta=100$ and suggest not tuning it.

\begin{table}[!htb]
\begin{tabular}{ccccc|cccc|cccc}
\hline
                & \multicolumn{4}{c|}{Shepp-Logan} & \multicolumn{4}{c|}{Grains} & \multicolumn{4}{c}{Wallnut} \\ \hline
$\lambda_{ref}$ & Mean   & Median  & Max   & Min   & Mean & Median & Max  & Min  & Mean & Median & Max  & Min  \\ \hline
0               & 3.69   & 3.47    & 5.41  & 0.87  & 4.95 & 5.36   & 7.87 & 1.07 & 2.77 & 3.08   & 3.86 & 0.59 \\ \hline
10              & 3.24   & 3.08    & 5.09  & 0.63  & 4.57 & 4.98   & 7.64 & 0.37 & 2.52 & 2.83   & 3.81 & 0.22 \\ \hline
25              & 2.47   & 2.49    & 4.03  & 0.12  & 3.93 & 4.22   & 7.42 & 0.15 & 2.26 & 2.61   & 3.65 & 0.02 \\ \hline
50              & 2.35   & 2.42    & 4.11  & 0.03  & 2.42 & 2.54   & 6.14 & 0.04 & 0.79 & 0.86   & 1.66 & 0.01 \\ \hline
75              & 1.85   & 1.89    & 3.83  & 0.04  & 1.61 & 1.69   & 4.62 & 0.02 & 0.97 & 1.07   & 1.89 & 0.01 \\ \hline
\end{tabular}
\caption{The mean, median, maximum and minimum  of ESS per second for Gibbs-BPS cross all the pixels with $\eta = 100$ fixed and varies of $\lambda_{ref}$.}
\label{tab:ESS_ref}
\end{table}

\begin{table}[!htb]
\begin{tabular}{ccccc|cccc|cccc}
\hline
       & \multicolumn{4}{c|}{Shepp-Logan} & \multicolumn{4}{c|}{Grains} & \multicolumn{4}{c}{Wallnut} \\ \hline
$\eta$ & Mean   & Median  & Max   & Min   & Mean & Median & Max  & Min  & Mean & Median & Max  & Min  \\ \hline
50     & 2.85   & 2.75    & 4.57  & 0.21  & 3.85 & 4.25   & 6.59 & 0.21 & 2.22 & 2.53   & 3.27 & 0.05 \\
75     & 2.83   & 2.73    & 4.48  & 0.32  & 4.18 & 4.56   & 6.88 & 0.27 & 2.57 & 2.91   & 3.74 & 0.13 \\
100    & 3.21   & 3.08    & 5.09  & 0.61  & 4.57 & 4.98   & 7.61 & 0.37 & 2.52 & 2.83   & 3.81 & 0.22 \\
125    & 2.89   & 2.76    & 4.57  & 0.37  & 3.91 & 4.25   & 6.52 & 0.46 & 2.22 & 2.49   & 3.31 & 0.11 \\
150    & 3.01   & 2.86    & 4.73  & 0.45  & 3.88 & 4.21   & 6.33 & 0.48 & 2.23 & 2.49   & 3.27 & 0.24 \\ \hline
\end{tabular}
\caption{The mean, median, maximum and minimum of ESS per second for Gibbs-BPS cross all the pixels in the image with $\lambda_{ref} = 10$ fixed and varies of $\eta$.}
\label{tab:ESS_eta}
\end{table}

\section{Conclusions}
\label{sec:conclusions}
We proposed the fused $L_{1/2}$ prior for solving Bayesian linear inverse problems, where both preserving edges and sparsity features of the solution are required. Our approach is to put the exponential power prior both on each pixel (sparsity-promoting) and its increment (edge-preserving). We have proved that the fused $L_{1/2}$ prior has an analytical form of Gaussian mixture representation, which allows us to construct the Gibbs sampler with the simple closed form of the conditional posterior. 

We also developed a novel sampler, termed Gibbs-BPS, based on a continuous time Markov chain. This new sampler incorporates the Gibbs type update for the conditional posterior of the global and local shrinkage parameters and uses the piecewise deterministic Markov process to update the conditional posterior of the pixels. The main advantage of this new sampler is that the most heavy computation involved is only the matrix multiplication, making it particularly suitable for large scale linear inverse problems. We have demonstrated the potential of this method using CT reconstruction with various image sizes.

Finally, we discuss some future research directions that can extend our methodology:
\begin{enumerate}[label=(\roman*)]
    \item In Theorem 4.2, we showed that the Gibbs-BPS algorithm is invariant to the target distribution and we demonstrate experimentally that it has good performance. However, the geometric ergodicity results for such scheme has not been established so far. Such theoretical result has been established for BPS algorithm with very restrictive assumptions~\cite{deligiannidis2019exponential} and has been relaxed recently~\cite{durmus2020geometric}. We conjecture that a similar result also holds for the Gibbs-BPS algorithm.
    \item It is also possible to apply the Gibbs-BPS algorithm to the posterior based on the horseshoe prior. However, rather than using the sampling approach from~\cite{makalic2015simple}, we need to develop a more advanced approach to sample the conditional posterior of global and local shrinkage parameters, which allows us to construct the two-block Gibbs sampler similar to algorithm~\ref{alg: Gibbs}. Then the Gibbs-BPS algorithm can be easily applied. In fact, the two-block Gibbs sampler has been constructed in a sparse linear regression setting based on horseshoe prior~\cite{johndrow2020scalable}, but sampling the global shrinkage parameters by their approach is computationally intensive.
    \item Despite our method being tailored for linear inverse problems, it can also be extended to nonlinear ones by using the Poisson thinning~\cite{lewis1979simulation}. This requires us to find a tight upper bound to the gradient of the log-likelihood.
\end{enumerate}

\section*{Acknowledgments}
The work was supported by the Major Scientific and Technological Innovation Platform Project of Human Province (2024JC1003). We are grateful
to the High Performance Computing Center of Central South University for assistance with the computations.

\vspace{10pt}
\appendix 
\section{Proof of Theorem 4.2}

\begin{proof}

It is sufficient to verify that
$$
\iiint \left(\mathcal{L}_{\mathrm{GBPS}} f\right)(\bm{x}, \bm{v},\bm{\phi})\pi(\bm{x},\bm{v},\bm{\phi} \mid \bm{y})d \bm{x}d  \bm{v}d\bm{\phi}=0.
$$
Since $\mathcal{L}_{\mathrm{GBPS}}=\mathcal{L}_{\mathrm{BPS}}+\eta \mathcal{L}_{\mathrm{Gibbs}}$, it is sufficient to verify the following two conditions:\\
Given any fixed value of $\bm{\phi}$,
\begin{equation}\label{eq:condition1}
\iint \left(\mathcal{L}_{\mathrm{BPS}} f\right)(\bm{x}, \bm{v},\bm{\phi})\pi(\bm{v})\pi(\bm{x} \mid \bm{y},\bm{\phi})d \bm{x}d  \bm{v}=0.
\end{equation}
Given any fixed value of $\bm{x}$ and $\bm{v}$, 
\begin{equation}\label{eq:condition2}
\int \left(\mathcal{L}_{\mathrm{Gibbs}} f\right)(\bm{x}, \bm{v},\bm{\phi})\pi( \bm{\phi}\mid \bm{x},\bm{y})d \bm{\phi}=0.
\end{equation}
To verify equation (\ref{eq:condition1}), we plug in equation (\ref{eq:BPS_generator}) on its left-hand side, which given us 
\begin{equation}
\begin{aligned}
& \quad \iint \left(\mathcal{L}_{\mathrm{BPS}} f\right)(\bm{x}, \bm{v},\bm{\phi})\pi(\bm{v})\pi(\bm{x} \mid \bm{y},\bm{\phi})d \bm{x}d  \bm{v}\\
& = \iint \bm{v}^{T} \nabla_{\bm{x}} f(\bm{x}, \bm{v},\bm{\phi})\pi(\bm{v})\pi(\bm{x} \mid \bm{y},\bm{\phi})d \bm{x}d  \bm{v}\\
& \quad +\lambda^{\mathrm{ref}}\iint [f(\bm{x}, \bm{v}^{\prime}, \bm{\phi})-f(\bm{x},\bm{v}, \bm{\phi})]\pi(\bm{v}^{\prime})\pi(\bm{v})\pi(\bm{x} \mid \bm{y},\bm{\phi})d \bm{x}d\bm{v}^{\prime} d  \bm{v}\\
& \quad + \iint \langle\bm{v},  \nabla_{\bm{x}} U(\bm{x},\bm{\phi})\rangle_{+}
[f(\bm{x},R_{\nabla_{\bm{x}} U(\bm{x},\bm{\phi})}(\bm{v}), \bm{\phi})-f(\bm{x},\bm{v}, \bm{\phi})]\pi(\bm{v})\pi(\bm{x} \mid \bm{y},\bm{\phi})d \bm{x}d  \bm{v}.
\end{aligned}
\end{equation}
We see that the second term is trivially equal to zero. For the third term, by change-of-variables, we have
\begin{equation}
\begin{aligned}
& \iint \langle\bm{v},  \nabla_{\bm{x}} U(\bm{x},\bm{\phi})\rangle_{+}
[f(\bm{x},R_{\nabla_{\bm{x}} U(\bm{x},\bm{\phi})}(\bm{v}), \bm{\phi})-f(\bm{x},\bm{v}, \bm{\phi})]\pi(\bm{v})\pi(\bm{x} \mid \bm{y},\bm{\phi})d \bm{x}d  \bm{v}\\
= &  \iint \langle R_{\nabla_{\bm{x}} U(\bm{x},\bm{\phi})}(\bm{v}),  \nabla_{\bm{x}} U(\bm{x},\bm{\phi})\rangle_{+}
f(\bm{x},\bm{v}, \bm{\phi})\pi(\bm{v})\pi(\bm{x} \mid \bm{y},\bm{\phi})d \bm{x}d  \bm{v}\\
& -\iint \langle\bm{v},  \nabla_{\bm{x}} U(\bm{x},\bm{\phi})\rangle_{+} f(\bm{x},\bm{v}, \bm{\phi})\pi(\bm{v})\pi(\bm{x} \mid \bm{y},\bm{\phi})d \bm{x}d  \bm{v}\\
= &  \iint \langle -\bm{v},  \nabla_{\bm{x}} U(\bm{x},\bm{\phi})\rangle_{+}
f(\bm{x},\bm{v}, \bm{\phi})\pi(\bm{v})\pi(\bm{x} \mid \bm{y},\bm{\phi})d \bm{x}d  \bm{v}\\
& -\iint \langle\bm{v},  \nabla_{\bm{x}} U(\bm{x},\bm{\phi})\rangle_{+} f(\bm{x},\bm{v}, \bm{\phi})\pi(\bm{v})\pi(\bm{x} \mid \bm{y},\bm{\phi})d \bm{x}d  \bm{v}\\
= & - \iint  \langle\bm{v},  \nabla_{\bm{x}} U(\bm{x},\bm{\phi})\rangle f(\bm{x}, \bm{v},\bm{\phi})\pi(\bm{v})\pi(\bm{x} \mid \bm{y},\bm{\phi})d \bm{x}d  \bm{v}.
\end{aligned}
\end{equation}
Since $f(\cdot)$ is chosen in the domain of the generator, for any fixed value of $\bm{v}$ and $\bm{\phi}$, we have
$$
\lim_{\|\bm{x}\| \rightarrow +\infty}f(\bm{x},\bm{v},\bm{\phi})\pi(\bm{x} \mid \bm{y},\bm{\phi}) =0
$$
Therefore, using the integration by parts for the first term, we have

\begin{equation}
\begin{aligned}
& \quad  \iint \bm{v}^{T} \nabla_{\bm{x}} f(\bm{x}, \bm{v},\bm{\phi})\pi(\bm{v})\pi(\bm{x} \mid \bm{y},\bm{\phi})d \bm{x}d  \bm{v}\\
& =\iint  \langle\bm{v},  \nabla_{\bm{x}} U(\bm{x},\bm{\phi})\rangle f(\bm{x}, \bm{v},\bm{\phi})\pi(\bm{v})\pi(\bm{x} \mid \bm{y},\bm{\phi})d \bm{x}d  \bm{v}.
\end{aligned}
\end{equation}
Thus, the first and third terms are canceled with each other. We have verified the first condition. To verify the second condition, we plug in equation (\ref{eq:Gibbs_generator}) into left hand side of the equation (\ref{eq:condition2})
\begin{equation}
\begin{aligned}
& \,\quad \iint \left(\mathcal{L}_{\mathrm{Gibbs}} f\right)(\bm{x}, \bm{v},\bm{\phi})\pi( \bm{\phi}\mid \bm{x},\bm{y})d \bm{\phi}\\
& =\iint \left\{f(\bm{x}, \bm{v},\bm{\phi}^{\prime})-f(\bm{x}, \bm{v},\bm{\phi})\right\} \mathcal{Q}(\bm{\phi}, \mathrm{d} \bm{\phi}^{\prime})\pi( \bm{\phi}\mid \bm{x},\bm{y})d \bm{\phi}^{\prime}d \bm{\phi}\\
& =\iint \left\{f(\bm{x}, \bm{v},\bm{\phi}^{\prime})-f(\bm{x}, \bm{v},\bm{\phi})\right\} \pi( \bm{\phi}^{\prime} \mid \bm{x},\bm{y})\pi( \bm{\phi}\mid \bm{x},\bm{y})d \bm{\phi}^{\prime}d \bm{\phi}\\
& = 0.
\end{aligned}
\end{equation}
\end{proof}

\section{Proof of Lemma 4.3}
\begin{proof}
Since 
$\pi(x) \propto \exp \left(-1/2 (\bm{x}-\bm{\mu})^T \bm{\Sigma}^{-1} (\bm{x}-\bm{\mu})\right)$,
we have
$$U(\bm{x})=-\log \pi(\bm{x})=\frac{1}{2} (\bm{x}-\bm{\mu})^T \bm{\Sigma}^{-1}(\bm{x}-\bm{\mu})+C.$$
Then $\nabla U(\bm{x})=\bm{\Sigma}^{-1}(\bm{x}-\bm{\mu})$ and $\lambda(\bm{x}, \bm{v})=\langle \bm{v}, \nabla U(\bm{x})\rangle_{+}=\left(\bm{v}^T \bm{\Sigma}^{-1} (\bm{x}-\bm{\mu})\right)_{+}$.\\\\
Solving equation (\ref{eq:first_arrival_time}) is equivalent to finding $s$ such that
$$
\int_{0}^{s} \Big(\frac{d U(\bm{x}+t\bm{v})}{d t}\Big)_{+}=-\log u
$$
Since the Gaussian distribution has the Log-concave densities, there exists a unique $s^{*}$ such that $s^{*}=\arg \min _{t \geq 0} U(x+t v)$. On $[0,s^{*})$, we have $dU/dt<0$ and $dU/dt \geq 0$ on $[s^{*},\infty)$, so
$$
\int_{s^{*}}^{s} \frac{d U(\bm{x}+t\bm{v})}{d t} d t=U(\bm{x}+s\bm{v})-U\left(\bm{x}+s^{*}\bm{v}\right)=-\log u
$$
For Gaussian distribution, we have
\begin{equation}
\begin{aligned}
s^{*} & =\arg \min _{t \geq 0} U(\bm{x}+t\bm{v}) \\
& =\arg \min _{t \geq 0} \frac{1}{2}(\bm{x}+t \bm{v}-\bm{\mu})^T \bm{\Sigma}^{-1}(\bm{x}+t \bm{v}-\bm{\mu})
\end{aligned}
\end{equation}

which can be solved analytically, such that $s^{*}=\left(-\frac{(\bm{x}-\bm{\mu})^T \bm{\Sigma}^{-1} \bm{v}}{\bm{v}^T \bm{\Sigma}^{-1} \bm{v}}\right)_{+}$.\\

Since equation $U(\bm{x}+s\bm{v})-U\left(\bm{x}+s^{*}\bm{v}\right)=-\log u$ is quadratic in $s$, after inserting the expression of $s^{*}$ inside the equation and making some arrangement, we have
$$
\resizebox{0.95\hsize}{!}{$
s=\left(\bm{v}^T \bm{\Sigma}^{-1} \bm{v}\right)^{-1}\left[-(\bm{x}-\bm{\mu})^T \bm{\Sigma}^{-1} \bm{v}+\sqrt{\left(\left((\bm{x}-\bm{\mu})^T \bm{\Sigma}^{-1} \bm{v}\right)_{+}\right)^2-2 \bm{v}^T \bm{\Sigma}^{-1} \bm{v} \log u}\right]$.}
$$
\end{proof}

\section{The fused horseshoe prior}
\subsection{Prior setting}
To obtain edge-preserving and sparsity-promoting properties, we consider the fused horseshoe prior such that 
\begin{equation}\label{eq:horseshoe}
\resizebox{0.87\hsize}{!}{$\pi(\bm{x} \mid \bm{\lambda},\bm{\tau},\bm{\tau}^{h},\bm{\tau}^{v})  \propto \exp\underbrace{\bigg[-\frac{1}{2 \eta_{1}^{2}}\sum_{i,j}\left(\frac{x_{ij}}{\tau_{ij}}\right)^{2}\bigg.}_{\text{sparsity-promoting}} \underbrace{\bigg.-\frac{1}{2 \eta_{2}^{2}}\sum_{i,j}\left(\frac{\Delta_{i,j}^{h}}{\tau_{ij}^{h}}\right)^{2}-\frac{1}{2\eta_{3}^{2}}\sum_{i,j}\left(\frac{\Delta_{i,j}^{v}}{\tau_{ij}^{v}}\right)^{2}\bigg]}_{\text{edge-preserving}}$}
\end{equation}
with the prior of global shrinkage parameters satisfied:
$$
\eta_{1}\sim t^{+}(v_{1},0,c_{1}) \quad \eta_{2}\sim t^{+}(v_{2},0,c_{2}) \quad \eta_{3} \sim t^{+}(v_{3},0,c_{3})
$$
and the prior of local shrinkage parameters satisfied:
$$
\tau_{ij} \sim  t^{+}(v_{1},0,1) \quad \tau_{ij}^{h} \sim t^{+}(v_{2},0,1) \quad \tau_{ij}^{v} \sim t^{+}(v_{3},0,1)
$$
\textit{\bf Remark:} 
We followed the half-student's $t$ distribution prior setting for global and local shrinkage parameters from \cite{uribe2023horseshoe}, who extended the hierarchical structure of the horseshoe prior \cite{carvalho2010horseshoe}.When $v_{1}=v_{2}=v_{3}=1$, the half-Student's $t$-distribution becomes a half-Cauchy distribution, which resemble to the original one \cite{carvalho2010horseshoe}.The difference between Fused horseshoe prior and the edge-preserving horseshoe prior from \cite{uribe2023horseshoe} is that equation (\ref{eq:horseshoe}) has extra sparsity-promoting term.

\subsection{Gibbs sampling}
If $A \sim t^{+}(v,0,c)$, by using the scale mixture decomposition of a half student t distribution,

\begin{equation*}
\left(A^2 \mid B\right) \sim \operatorname{InvGamma}\left(\frac{\nu}{2}, \frac{\nu}{B}\right),  \quad  B \sim \operatorname{InvGamma}\Big(\frac{1}{2}, \frac{1}{c^2}\Big).
\end{equation*}

Then the prior of global and local shrinkage parameters have hierarchical representation:
{\scriptsize{
\begin{equation}
\begin{aligned}
 \eta_{1} \mid \gamma_{1} &\sim \operatorname{InvGamma}\Big(\frac{v_{1}}{2},\frac{v_{1}}{\gamma_{1}}\Big) 
 &\quad \eta_{2} \mid  \gamma_{2} &\sim \operatorname{InvGamma}\Big(\frac{v_{2}}{2},\frac{v_{2}}{\gamma_{2}}\Big) 
 &\quad  \eta_{3} \mid \gamma_{3} &\sim \operatorname{InvGamma}\Big(\frac{v_{1}}{2}, \frac{v_{3}}{\gamma_{3}}\Big) \\
 \gamma_{1} &\sim \operatorname{InvGamma}\left(\frac{1}{2},1\right) 
 &\quad \gamma_{2} &\sim \operatorname{InvGamma}\Big(\frac{1}{2},1\Big) 
 &\quad \gamma_{3} &\sim \operatorname{InvGamma}\Big(\frac{1}{2},1\Big) \\
 \tau_{ij} \mid w_{ij} &\sim \operatorname{InvGamma}\Big(\frac{v_{1}}{2},\frac{v_{1}}{w_{ij}}\Big) 
 &\quad \tau_{ij}^{h} \mid  w_{ij} &\sim \operatorname{InvGamma}\Big(\frac{v_{2}}{2},\frac{v_{2}}{w_{ij}^{h}}\Big) 
 &\quad  \tau_{ij}^{v} \mid w_{ij}^{v} &\sim \operatorname{InvGamma}\Big(\frac{v_{1}}{2}, \frac{v_{3}}{w_{ij}^{v}}\Big) \\
 w_{ij} &\sim \operatorname{InvGamma}\Big(\frac{1}{2},\frac{1}{c_{1}^{2}}\Big) 
 &\quad w_{ij}^{h} &\sim \operatorname{InvGamma}\Big(\frac{1}{2},\frac{1}{c_{2}^{2}}\Big) 
 &\quad w_{ij}^{v} &\sim \operatorname{InvGamma}\Big(\frac{1}{2},\frac{1}{c_{3}^{2}}\Big).
\end{aligned}
\end{equation}
}}
This hierarchical representation allows a direct application of the Gibbs sampler since the conditional densities for each parameter can be derived in closed form. We denote
$$
\widetilde{\bm{\Lambda}}=\frac{1}{\sigma_{\mathrm{obs}}^2} \bm{A}^{\top} \bm{A}+\bm{\Lambda}+\bm{D}_{h}^{\top} \bm{\Lambda}_{h} \bm{D}_h+\bm{D}_v^{\top} \bm{\Lambda}_v \bm{D}_{v}, \quad \widetilde{\bm{\mu}}=\widetilde{\bm{\Lambda}}^{-1}\Big(\frac{1}{\sigma_{\mathrm{obs}}^2} \bm{A}^{\top} \bm{y}\Big),
$$
where $\bm{D}_{h}=\bm{D} \,\, \otimes \,\,\bm{I}_{d}$, $\bm{D}_{v}=\bm{I}_{d} \,\,\otimes \,\, \bm{D}$, $\bm{I}_{d}$ is $d \times d$ identity matrix and $\bm{D}$ is a $d \times (d-1)$ difference matrix. In addition, we have $\bm{\Lambda}^{\frac{1}{2}}=\operatorname{diag}\left(\operatorname{vec}\left(\eta_{1}/\tau_{i,j}\right)\right)$, $\bm{\Lambda}_{h}^{\frac{1}{2}}=\operatorname{diag}\left(\operatorname{vec}\left(\eta_{2}/\tau_{i,j}^{h}\right)\right)$ and $\bm{\Lambda}_{v}^{\frac{1}{2}}=\operatorname{diag}\left(\operatorname{vec}\left(\eta_{3}/\tau_{i,j}^{v}\right)\right)$. Then the conditional posterior of $\bm{x}$ is
$\mathcal{N}\left(\widetilde{\bm{\Lambda}}^{-1}\left(\frac{1}{\sigma_{\mathrm{obs}}^2} \bm{A}^{\top} \bm{y}\right),\widetilde{\bm{\Lambda}}^{-1}\right)$. Now we can write down the Gibbs sampler below. 

\begin{algorithm}
\begin{algorithmic}[1]
\item[] \textbf{Input:}  $v_{1},v_{2},v_{3},c_{1},c_{2},c_{3} \in \mathrm{R}^{+}$; $T${: Num of iterations};
\item[] \textbf{Output:}  {All the $T$ samples of $\boldsymbol{x}$}
\vspace{4pt}
\FOR{$t \leftarrow 1$ \TO T}
\STATE Sample  $\boldsymbol{x} \mid \boldsymbol{\lambda},\boldsymbol{\tau},\boldsymbol{\tau}^{ h},\boldsymbol{\tau}^{v}, \boldsymbol{y}
\sim \mathcal{N}\left(\widetilde{\boldsymbol{\Lambda}}^{-1}\left(\frac{1}{\sigma_{\mathrm{obs}}^2} \boldsymbol{A}^{\top} \boldsymbol{y}\right),\widetilde{\boldsymbol{\Lambda}}^{-1}\right)$\\ \vspace{3pt}
\STATE Sample $\tau_{ij} \mid w_{ij},x_{ij} \sim \operatorname{InvGamma}\Big(\frac{v_{1}+1}{2},\frac{1}{2}\big(\frac{x_{ij}}{\eta_{1}}\big)^{2}+\frac{v_{1}}{w_{ij}}\Big)$\\ \vspace{3pt}
\STATE Sample $\tau_{ij}^{h} \mid w_{ij},\Delta_{ij}^{h} \sim \operatorname{InvGamma}\Big(\frac{v_{2}+1}{2},\frac{1}{2}\big(\frac{\Delta_{ij}^{h}}{\eta_{2}}\big)^{2}+\frac{v_{2}}{w_{ij}^{h}}\Big)$\\ \vspace{3pt}
\STATE Sample $\tau_{ij}^{v} \mid w_{ij},\Delta_{ij}^{v} \sim \operatorname{InvGamma}\Big(\frac{v_{3}+1}{2},\frac{1}{2}\big(\frac{\Delta_{ij}^{v}}{\eta_{3}}\big)^{2}+\frac{v_{3}}{w_{ij}^{v}}\Big)$\\ \vspace{3pt}
\STATE Sample $\eta_{1} \mid \gamma_{1},\boldsymbol{x} \sim \operatorname{InvGamma}\Big(\frac{n+v_{1}}{2},\frac{1}{2}\sum_{i,j}\frac{x_{ij}^{2}}{\tau_{ij}^{2}}+\frac{v_{1}}{\gamma_{1}}\Big)$\\ \vspace{3pt}
\STATE Sample $\eta_{2} \mid \gamma_{2},\boldsymbol{x} \sim \operatorname{InvGamma}\Big(\frac{n+v_{2}}{2},\frac{1}{2}\sum_{i,j}\big(\frac{\Delta_{ij}^{h}}{\tau_{ij}^{h}}\big)^{2}+\frac{v_{2}}{\gamma_{2}}\Big)$\\ \vspace{3pt}
\STATE Sample $\eta_{3} \mid \gamma_{3},\boldsymbol{x} \sim \operatorname{InvGamma}\Big(\frac{n+v_{3}}{2},\frac{1}{2}\sum_{i,j}\big(\frac{\Delta_{ij}^{v}}{\tau_{ij}^{v}}\big)^{2}+\frac{v_{3}}{\gamma_{3}}\Big)$\\ \vspace{3pt}
\STATE Sample $w_{ij} \mid  \tau_{ij} \sim \operatorname{InvGamma}\left(\frac{v_{1}+1}{2},1+\frac{v_{1}}{\tau_{ij}^{2}}\right)$\\ \vspace{3pt}
\STATE Sample $w_{ij}^{h} \mid  \tau_{ij} \sim \operatorname{InvGamma}\left(\frac{v_{2}+1}{2},1+\frac{v_{2}}{(\tau_{ij}^{h})^{2}}\right)$\\ \vspace{3pt}
\STATE Sample $w_{ij}^{v} \mid  \tau_{ij} \sim \operatorname{InvGamma}\left(\frac{v_{3}+1}{2},1+\frac{v_{3}}{(\tau_{ij}^{v})^{2}}\right)$\\ \vspace{3pt}
\STATE Sample $\gamma_{1} \mid \eta_{1} \sim \operatorname{InvGamma}\left(\frac{v_{1}+1}{2},\frac{1}{c_{1}^{2}}+\frac{v_{1}}{\eta_{1}^{2}}\right)$\\ \vspace{3pt}
\STATE Sample $\gamma_{2} \mid \eta_{2} \sim \operatorname{InvGamma}\left(\frac{v_{2}+1}{2},\frac{1}{c_{2}^{2}}+\frac{v_{2}}{\eta_{2}^{2}}\right)$\\ \vspace{3pt}
\STATE Sample $\gamma_{3} \mid \eta_{3} \sim \operatorname{InvGamma}\left(\frac{v_{3}+1}{2},\frac{1}{c_{3}^{2}}+\frac{v_{3}}{\eta_{3}^{2}}\right)$
\ENDFOR
\end{algorithmic}
\caption{Gibbs sampler}
\end{algorithm}

\section{Proximal Langevin dynamic}

\subsection{Fused lasso prior}
We consider the fused lasso prior
\begin{equation}
\pi(\boldsymbol{x} \mid \lambda_{1},\lambda_{2},\lambda_{3})  \propto \exp \Big(-\lambda_{1}\sum_{i,j}|x_{ij}|-\lambda_{2}\sum_{i,j}|\Delta_{i,j}^{h}|-\lambda_{3}\sum_{i,j}|\Delta_{i,j}^{v}|\Big).
\end{equation}
with the hyper parameter $\lambda_{1}$,$\lambda_{2}$ and $\lambda_{3}$ being tuned manually.

\subsection{Proximal Langevin Dynamic}
Proximal Langevin dynamic is an efficient approximate MCMC approach to perform Bayesian computation for high-dimensional models that are log-concave and non-smooth\cite{durmus2018efficient,durmus2022proximal}. It leverages the unadjusted Langevin dynamics to explore the parameter space and the proximal operator to efficiently handle the non-smooth part of the target distribution. Specifically, given $u>0$ and a step size $\epsilon >0$, we use a Euler-Maruyama approximation to obtain the following discrete-time Markov chain:
$$
\boldsymbol{x}_{k+1}=\left(1-\frac{\epsilon}{u}\right) \boldsymbol{x}_{k}-\epsilon \nabla \log \pi(\boldsymbol{y}\mid \boldsymbol{x}_{k})+\frac{\epsilon}{u} \operatorname{prox}_g^u\left(\boldsymbol{x}_k\right)+\sqrt{2 \epsilon} z_{k+1}
$$
where $z_{k+1}$ is $n$-dimensional standard Gaussian random variables and 
\begin{equation}\label{eq:proximal_map}
\operatorname{prox}_{\mathrm{g}}^{u}(\boldsymbol{x})=\underset{\boldsymbol{t} \in \mathbb{R}^d}{\arg \min }\Big\{\mathrm{g}(\boldsymbol{t})+\frac{1}{2u}\|\boldsymbol{x}-\boldsymbol{t}\|^2\Big\}
\end{equation}
with $g(\boldsymbol{x})=\lambda_{1}\sum_{i,j}|x_{ij}|+\lambda_{2}\sum_{i,j}|\Delta_{i,j}^{h}|+\lambda_{3}\sum_{i,j}|\Delta_{i,j}^{v}|$. The proximal map in (\ref{eq:proximal_map}) can be solved by the ADMM algorithm\cite{boyd2011distributed}. 
\subsection{The proximal map of fused Lasso with the ADMM solver}
Now, we show how to use the ADMM \cite{boyd2011distributed} algorithm to solve the proximal operator.

\begin{equation*}
\begin{aligned}
\operatorname{prox}_{\mathrm{g}}^u(\boldsymbol{x}) & =\underset{\boldsymbol{t} \in \mathbb{R}^d}{\arg \min }\Big\{\mathrm{g}(\boldsymbol{t})+\frac{1}{2u}\|\boldsymbol{x}-\boldsymbol{t}\|_{2}^2\Big\}\\
& = \underset{\boldsymbol{t} \in \mathbb{R}^d}{\arg \min }\Big\{\lambda_{1}\|\boldsymbol{t}\|_{1}+\lambda_{2}\|\boldsymbol{D}_{h}\boldsymbol{t}\|_{1}+\lambda_{3}\|\boldsymbol{D}_{v}\boldsymbol{t}\|_{1}+\frac{1}{2u}\|\boldsymbol{x}-\boldsymbol{t}\|_{2}^2\Big\}
\end{aligned}
\end{equation*}

In ADMM form, this problem can be written as 
\begin{equation}
\begin{array}{ll}
\operatorname{Minimize} & l(\boldsymbol{t})+g_{1}(\boldsymbol{z}_{1})+g_{2}(\boldsymbol{z}_{2})+g_{3}(\boldsymbol{z}_{3}) \\
\text {Subject to } & \quad \boldsymbol{t}-\boldsymbol{z}_{1}=0\\
& \boldsymbol{D}_{h}\boldsymbol{t}-\boldsymbol{z}_{2}=0\\
& \boldsymbol{D}_{v}\boldsymbol{t}-\boldsymbol{z}_{3}=0\\
\end{array}
\end{equation}
where $g_{1}(\boldsymbol{z}_{1})=\lambda_{1}\|\boldsymbol{z}\|_{1}$, $g_{2}(\boldsymbol{z}_{2})=\lambda_{2}\|\boldsymbol{z}_{2}\|_{1}$,$g_{3}(\boldsymbol{z}_{3})=\lambda_{3}\|\boldsymbol{z}_{3}\|_{1}$ and $l(\boldsymbol{t})=\frac{1}{2u}\|\boldsymbol{x}-\boldsymbol{t}\|_{2}^{2}$. With the hyper-parameter $\rho_{1},\rho_{2},\rho_{3},u \in \mathbf{R}^{+}$, we form the augmented Lagrangian 
\begin{equation}
\begin{aligned}
L_{\rho}(\boldsymbol{t},\boldsymbol{z},\boldsymbol{\phi}) & =\frac{1}{2u}\|\boldsymbol{x}-\boldsymbol{t}\|_{2}^{2}+\lambda_{1}\|\boldsymbol{z}\|_{1}+\lambda_{2}\|\boldsymbol{z}_{2}\|_{1}+\lambda_{3}\|\boldsymbol{z}_{3}\|_{1}\\
& \quad +\boldsymbol{\phi}_{1}^{T}(\boldsymbol{t}-\boldsymbol{z}_{2})+\boldsymbol{\phi}_{2}^{T}(\boldsymbol{D}_{h}\boldsymbol{t}-\boldsymbol{z}_{2})+\boldsymbol{\phi}_{3}^{T}(\boldsymbol{D}_{v}\boldsymbol{t}-\boldsymbol{z}_{3})\\
& \quad +\frac{\rho_{1}}{2}\|\boldsymbol{t}-\boldsymbol{z}_{1}\|_{2}^{2}+\frac{\rho_{2}}{2}\|\boldsymbol{D}_{h}\boldsymbol{t}-\boldsymbol{z}_{2}\|_{2}^{2}+\frac{\rho_{3}}{2}\|\boldsymbol{D}_{v}\boldsymbol{t}-\boldsymbol{z}_{3}\|_{2}^{2}\\
\end{aligned}
\end{equation}
where $\boldsymbol{z}=(\boldsymbol{z}_{1},\boldsymbol{z}_{2},\boldsymbol{z}_{3})$ and $\boldsymbol{\phi}=(\boldsymbol{\phi}_{1},\boldsymbol{\phi}_{2},\boldsymbol{\phi}_{3})$.
ADMM consists of the iterations
\begin{equation}
\begin{aligned}
\boldsymbol{t}^{k+1} & :=\underset{\boldsymbol{t}}{\operatorname{argmin}} L_\rho\left(\boldsymbol{t}, \boldsymbol{z}^k, \boldsymbol{\phi}^k\right) \\
\boldsymbol{z}^{k+1} & :=\underset{\boldsymbol{z}}{\operatorname{argmin}} L_\rho\left(\boldsymbol{t}^{k+1}, \boldsymbol{z}, \boldsymbol{\phi}^k\right) \\
\boldsymbol{\phi}_{1}^{k+1} & :=\boldsymbol{\phi}_{1}^{k}+\rho_{1}\left( \boldsymbol{t}^{k+1}- \boldsymbol{z}_{1}^{k+1}\right)\\
\boldsymbol{\phi}_{2}^{k+1} & :=\boldsymbol{\phi}_{2}^{k}+\rho_{2}\left( \boldsymbol{D}_{h}\boldsymbol{t}^{k+1}- \boldsymbol{z}_{2}^{k+1}\right)\\
\boldsymbol{\phi}_{3}^{k+1} & :=\boldsymbol{\phi}_{3}^{k}+\rho_{3}\left( \boldsymbol{D}_{v}\boldsymbol{t}^{k+1}- \boldsymbol{z}_{3}^{k+1}\right)
\end{aligned}
\end{equation}
Then, we have
\begin{equation}
\begin{aligned}
\boldsymbol{t}^{k+1} & =\left((u^{-1}+\rho_{1})I_{n}+\rho_{2} \boldsymbol{D}_{h}^{\top}\boldsymbol{D}_{h}+\rho_{3} \boldsymbol{D}_{v}^{\top}\boldsymbol{D}_{v}\right)^{-1}\Big(\frac{1}{u}\boldsymbol{x}+(\rho_{1} \boldsymbol{z}_{1}-\boldsymbol{\phi}_{1}) + \Big.\\
& \quad \Big.\boldsymbol{D}_{h}^{\top}(\rho_{2} \boldsymbol{z}_{2}-\boldsymbol{\phi}_{2})+\boldsymbol{D}_{v}^{\top}(\rho_{3} \boldsymbol{z}_{3}-\boldsymbol{\phi}_{3})\Big)\\
\boldsymbol{z}_{1}^{k+1} & = \operatorname{Sign}\Big(\boldsymbol{t}+\frac{\boldsymbol{\phi}_{1}}{\rho_{1}}\Big)\Big(\boldsymbol{t}+\frac{\boldsymbol{\phi}_{1}}{\rho_{1}}-\frac{\lambda_{1}}{\rho_{1}}\Big)_{+}\\
\boldsymbol{z}_{2}^{k+1} & = \operatorname{Sign}\Big(\boldsymbol{D}_{h}\boldsymbol{t}+\frac{\boldsymbol{\phi}_{2}}{\rho_{2}}\Big)\Big(\boldsymbol{D}_{h}\boldsymbol{t}+\frac{\boldsymbol{\phi}_{2}}{\rho_{2}}-\frac{\lambda_{2}}{\rho_{2}}\Big)_{+}\\
\boldsymbol{z}_{3}^{k+1} & = \operatorname{Sign}\Big(\boldsymbol{D}_{v}\boldsymbol{t}+\frac{\boldsymbol{\phi}_{3}}{\rho_{3}}\Big)\Big(\boldsymbol{D}_{v}\boldsymbol{t}+\frac{\boldsymbol{\phi}_{3}}{\rho_{3}}-\frac{\lambda_{3}}{\rho_{3}}\Big)_{+}\\
\boldsymbol{\phi}_{1}^{k+1} & :=\boldsymbol{\phi}_{1}^{k}+\rho_{1}\left( \boldsymbol{t}^{k+1}- \boldsymbol{z}_{1}^{k+1}\right)\\
\boldsymbol{\phi}_{2}^{k+1} & :=\boldsymbol{\phi}_{2}^{k}+\rho_{2}\left( \boldsymbol{D}_{h}\boldsymbol{t}^{k+1}- \boldsymbol{z}_{2}^{k+1}\right)\\
\boldsymbol{\phi}_{3}^{k+1} & :=\boldsymbol{\phi}_{3}^{k}+\rho_{3}\left( \boldsymbol{D}_{v}\boldsymbol{t}^{k+1}- \boldsymbol{z}_{3}^{k+1}\right)
\end{aligned}
\end{equation}
By change of variable, we set $\boldsymbol{v}_{i}=\frac{\boldsymbol{\phi}_{i}}{\rho_{i}}$, then
\begin{equation}
\begin{aligned}
\boldsymbol{t}^{k+1} & =\left((1+\rho_{1})I_{n}+u\rho_{2} \boldsymbol{D}_{h}^{\top}\boldsymbol{D}_{h}+u\rho_{3} \boldsymbol{D}_{v}^{\top}\boldsymbol{D}_{v}\right)^{-1}\big(\boldsymbol{x}+u\rho_{1}(\boldsymbol{z}_{1}-\boldsymbol{v}_{1})+\big. \\
& \quad \big. u\boldsymbol{D}_{h}^{\top}\rho_{2}(\boldsymbol{z}_{2}-\boldsymbol{v}_{2})+u\boldsymbol{D}_{v}^{\top}\rho_{3}(\boldsymbol{z}_{3}-\boldsymbol{v}_{3})\big)\\
\boldsymbol{z}_{1}^{k+1} & = \operatorname{Sign}\left(\boldsymbol{t}+\boldsymbol{v}_{1}\right)\Big(\boldsymbol{t}+\boldsymbol{v}_{1}-\frac{\lambda_{1}}{\rho_{1}}\Big)_{+}\\
\boldsymbol{z}_{2}^{k+1} & = \operatorname{Sign}\left(\boldsymbol{D}_{h}\boldsymbol{t}+\boldsymbol{v}_{2}\right)\Big(\boldsymbol{D}_{h}\boldsymbol{t}+\boldsymbol{v}_{2}-\frac{\lambda_{2}}{\rho_{2}}\Big)_{+}\\
\boldsymbol{z}_{3}^{k+1} & = \operatorname{Sign}\left(\boldsymbol{D}_{v}\boldsymbol{t}+\boldsymbol{v}_{3}\right)\Big(\boldsymbol{D}_{v}\boldsymbol{t}+\boldsymbol{v}_{3}-\frac{\lambda_{3}}{\rho_{3}}\Big)_{+}\\
\boldsymbol{v}_{1}^{k+1} & =\boldsymbol{v}_{1}^k+\left( \boldsymbol{t}^{k+1}- \boldsymbol{z}_{1}^{k+1}\right)\\
\boldsymbol{v}_{2}^{k+1} & =\boldsymbol{v}_{2}^k+\left( \boldsymbol{D}_{h}\boldsymbol{t}^{k+1}- \boldsymbol{z}_{2}^{k+1}\right)\\
\boldsymbol{v}_{3}^{k+1} & =\boldsymbol{v}_{3}^k+\left( \boldsymbol{D}_{v}\boldsymbol{t}^{k+1}- \boldsymbol{z}_{3}^{k+1}\right)
\end{aligned}
\end{equation}
\textit{\bf Remark:}
Since updating $\boldsymbol{t}^{k+1}$ explicitly is expensive, we consider the gradient descent algorithm, with the gradient.
$$
\frac{\partial L_{\rho}(\boldsymbol{t},\boldsymbol{z},\boldsymbol{\phi})  }{\partial \boldsymbol{t}}=\frac{1}{u}\left(\boldsymbol{t}-\boldsymbol{x}\right)+\rho_{1} (\boldsymbol{v}_{1}+\boldsymbol{t}-\boldsymbol{z}_{1})+\rho_{2} \boldsymbol{D}_{h}^{\top}(\boldsymbol{v}_{2}+\boldsymbol{D}_{h}\boldsymbol{t}-\boldsymbol{z}_{2})+\rho_{3} \boldsymbol{D}_{v}^{\top}(\boldsymbol{v}_{3}+\boldsymbol{D}_{v}\boldsymbol{t}-\boldsymbol{z}_{3})
$$

\section{Extra numerical studies: Full Bayesian vs Empirical Bayes}

In this section, we provide extra numerical studies for the comparison of full Bayesian approach and empirical Bayes approach.  Figure E1, E2, E3 and Table E1 show that the Empirical Bayesian approach has the similar performance as the Full Bayesian approach. Table E2 shows that the estimators for $\lambda_{1}$ and $\lambda_{2}$ from Empirical Bayesian approach is always slightly smaller than the posterior mean reported from Full Bayesian approach. For $\lambda_{3}$, their difference is negligible.

\begin{figure}[htbp]
\centering
\subfloat[Empirical Bayesian]{\includegraphics[width=0.48\textwidth]{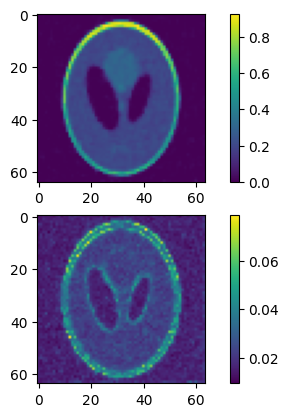}}  \hspace{10pt}
\subfloat[Full Bayesian]{\includegraphics[width=0.48\textwidth]{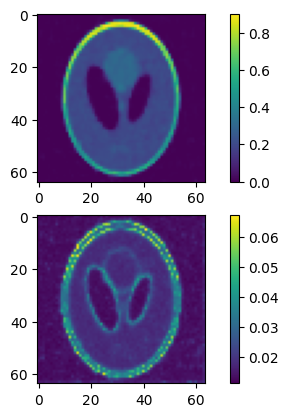}} 
\vspace{-8pt}
\caption{Empirical Bayesian vs  Full Bayesian for Shepp–Logan Phantom}
\end{figure}

\begin{figure}[htbp]
\centering
\subfloat[Empirical Bayesian]{\includegraphics[width=0.48\textwidth]{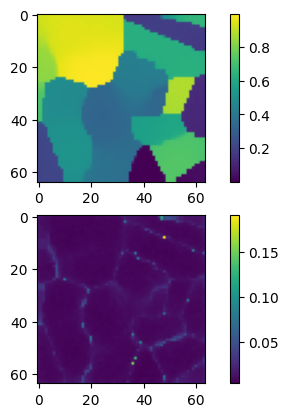}}  \hspace{10pt}
\subfloat[Full Bayesian]{\includegraphics[width=0.47\textwidth]{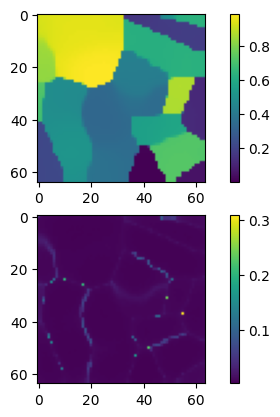}} 
\vspace{-8pt}
\caption{Empirical Bayesian vs  Full Bayesian for Grains Phantom}
\end{figure}

\begin{figure}[htbp]
\centering
\subfloat[Empirical Bayesian]{\includegraphics[width=0.48\textwidth]{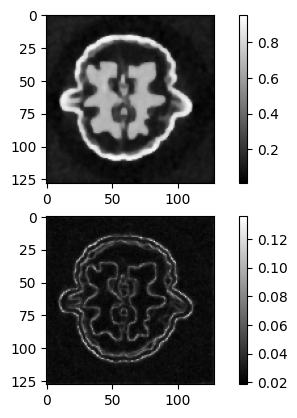}}  \hspace{10pt}
\subfloat[Full Bayesian]{\includegraphics[width=0.47\textwidth]{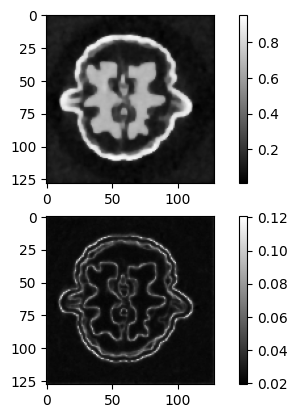}} 
\vspace{-8pt}
\caption{Empirical Bayesian vs  Full Bayesian for Walnut Phantom}
\end{figure}

\begin{table}[!htb]
\centering
\begin{tabular}{ccc}
\hline
\multicolumn{1}{l}{} & \multicolumn{2}{c}{PSNR(SSIM)}     \\ \cline{2-3} 
                     & Empirical Bayesian & Full Bayesian \\ \hline
Shepp-Logan          & 30.80(0.95)        & 31.20(0.96)   \\ \hline
Grains               & 27.01(0.90)        & 27.11(0.90)   \\ \hline
Wullnat              & 27.85(0.91)        & 27.91(0.92)   \\ \hline
\end{tabular}
\caption{Empirical Bayesian approach vs Full Bayesian approach.}
\end{table}

\begin{table}[!htb]
\centering
\begin{tabular}{ccc}
\hline
$\boldsymbol{\lambda}$ & Empirical Bayesian & Full Bayesian \\ \hline
                       & \multicolumn{2}{c}{Shepp Logan}    \\ \hline
$\lambda_1$            & 19.66              & 22.53         \\ \hline
$\lambda_2$            & 18.23              & 20.37         \\ \hline
$\lambda_3$            & 7.01               & 7.02          \\ \hline
                       & \multicolumn{2}{c}{Grains}         \\ \hline
$\lambda_1$            & 18.54              & 20.67         \\ \hline
$\lambda_2$            & 19.34              & 20.41         \\ \hline
$\lambda_3$            & 2.86               & 2.85          \\ \hline
                       & \multicolumn{2}{c}{Wullnat}        \\ \hline
$\lambda_1$            & 21.23              & 21.75         \\ \hline
$\lambda_2$            & 21.83              & 22.46         \\ \hline
$\lambda_3$            & 4.65               & 4.64          \\ \hline
\end{tabular}
\caption{Comparison of the hyper-parameters estimation between Empirical Bayesian approach and full Bayesian approach. For full Bayesian approach, we report the posterior mean of $\boldsymbol{\lambda}$.}
\end{table}

\clearpage

\section*{References}
\bibliographystyle{plain}
\bibliography{references}

\end{document}